\documentclass[aps,prb,reprint,showpacs,floatfix,superscriptaddress]{revtex4-1}
\usepackage{amsmath,amssymb,graphics,epsfig,color,verbatim,amsfonts,tensor}
\usepackage{variables}
\usepackage{tensor}
\usepackage{multirow}
\begin{document}

\title{A group theoretical approach to computing phonons and their interactions}
\author{Lyuwen Fu}
\email{lyuwen.fu@columbia.edu}
\affiliation{Department of Applied Physics and Applied Mathematics, Columbia University, New York, NY 10027}
\author{Mordechai Kornbluth}
\affiliation{Robert Bosch LLC, Research and Technology Center North America, 255 Main St., Cambridge, Massachusetts 02142, USA }
\email{mckornbluth@gmail.com}
\author{Zhengqian Cheng}
\email{chengzhengqian@gmail.com}
\affiliation{Department of Applied Physics and Applied Mathematics, Columbia University, New York, NY 10027}
\author{Chris A. Marianetti}
\email{chris.marianetti@columbia.edu}
\affiliation{Department of Applied Physics and Applied Mathematics, Columbia University, New York, NY 10027}

\begin{abstract}
Phonons and their interactions are necessary for determining a wide range of materials properties.
Here we present four 
independent advances which facilitate the computation of phonons and their interactions
from first-principles.  First, we implement a group-theoretical approach to
construct the order $\order$ Taylor series of a $d$-dimensional crystal purely in terms of space group
irreducible derivatives (ID), which guarantees symmetry by construction
and allows for a practical means
of communicating and storing phonons and their interactions.  
Second, we prove that the smallest possible supercell which accommodates $\order$ given wavevectors in a $d$-dimensional crystal
is determined 
 using the Smith Normal Form
of the matrix formed from the corresponding wavevectors; resulting in negligible computational cost to find said supercell,
in addition to providing the maximum required multiplicity for uniform supercells at arbitrary $\order$ and $\cdim$.
Third, we develop 
a series of finite displacement methodologies to compute phonons and their
interactions which exploit the first two developments: lone and bundled irreducible derivative (LID and BID) approaches. LID
computes a single ID, or as few as possible, at a time in the smallest supercell possible, while BID exploits perturbative derivatives for some order
less than $\order$ (e.g. Hellman-Feynman forces) in order to extract all ID in the smallest possible supercells using the fewest possible computations.
Finally, we derive an equation for the order $N$ volume derivatives of the phonons in terms of the order $\order=N+2$ ID. Given that
the former are easily computed, they can be used as a stringent, infinite ranged test of the ID.
Our general framework is illustrated on graphene, 
yielding irreducible phonon interactions to fifth order. Additionally, we provide a cost analysis for the rock-salt structure at $\order=3$, demonstrating a massive speedup compared 
to popular finite displacement methods in the literature.
\end{abstract}

\date{\today}
\maketitle

\section{Introduction}
\subsection{General Background}
\label{sec:genback}
Phonons and phonon interactions dictate a wide array of materials properties,
often including thermal conductivity, thermal expansion, linear and nonlinear elasticity, structural phase stability, and many
other properties\cite{Dove1993book,Srivastava19900852741537,Reissland19730471715859,Fultz2010247,Walle200211}. Even when studying purely electronic phenomena, knowledge of
phonons and their interactions can be critical to interpreting experimental measurements.
While computing phonons from first-principles is largely considered a solved
problem, practical shortcomings of existing methods still preclude their use on
a broad swath of materials with select first-principles approaches. Furthermore, computing phonon interactions from
first-principles is still a rapidly evolving field, and the basic form of
phonon interactions is not well known even in classic materials systems.

The problem of computing phonons and their interactions from first-principles
is equivalent to computing the Taylor series expansion of the
Born-Oppenheimer energy surface with respect to the nuclear displacements of
the crystal. The second order terms (i.e.
harmonic) dictate the phonons, while higher order terms (i.e. anharmonic)
dictate phonon interactions.  Given that a crystal is infinite in extent, the
computed Taylor series at each order will necessarily be truncated at some maximum resolution.
 An important task is to
obtain a sufficiently high resolution such that the expansion is converged at a
given order, meaning that a higher resolution will have no appreciable
influence on relevant observables.  

There are two basic approaches to computing phonons\cite{Martin2008,Baroni2001515} and the same can be claimed regarding their
interactions: perturbation theory and finite
displacements; where the latter encompasses usual finite difference approaches
or more complicated fitting procedures based on finite displacements.
Furthermore, these approaches are naturally combined, using perturbation theory
to obtain some low order derivative (e.g. Hellman-Feynman forces) and finite
displacements for a higher derivative. For an early example at second order within density functional theory (DFT), Ihm \emph{et. al} used the Hellman-Feynman forces
and finite difference to compute phonons\cite{Ihm1981491}. In this same spirit, Bonini \emph{et. al} used density functional perturbation theory (DFPT) to compute second order
terms and then used finite difference to compute third and fourth order terms in graphene\cite{Bonini2007176802}. 

In any case, whether it be perturbation
theory or finite displacement or a combination thereof, the \emph{de facto} standard is to compute all
derivatives associated with displacements that transform as irreducible representations of some
finite translation group (FTG) (i.e. $q$-points commensurate with a Born-von-Karman supercell, described further in Section \ref{sec:ugb}).
This set of derivatives  allows for a Fourier interpolation,
which exactly preserves the derivatives at a $q$-point that is an
irreducible representation of the FTG while providing a smooth interpolation
for all other $q$-points (see Refs. \onlinecite{Giannozzi19917231,Parlinski19974063} for early examples at second order). 
Assuming that the
discretization errors of finite difference calculations are properly
extrapolated to zero; and that spurious behavior is properly handled when
encountered via perturbation theory (e.g. see reference \onlinecite{Zhang2004167102});
and that the underlying first-principles approach is properly
converged with respect to its own discretizations (e.g. plane wave cutoff,
$k$-point density, etc); and that the energy function is analytic; then all approaches must agree on derivatives with respect to displacements which transform
as irreducible representations of the FTG.

Within DFT at second order, it should be noted that a distinct advantage of perturbation theory
(i.e. density functional perturbation theory) is that an arbitrary point within
the Brillouin Zone may be computed with a cost on the order of a standard DFT
calculation of the primitive unit cell\cite{Baroni2001515}, while finite
displacement approaches are limited to supercells for which a DFT calculation
can be tractably performed. 
However, not all mainstream DFT codes have fully implemented DFPT yet.
Moreover, perturbation theory is not ubiquitous for 
techniques which go beyond DFT, and even simple approaches like DFT+$U$ only have a few
demonstrations to date where perturbation theory has been executed at second order\cite{Floris2011161102,Dorado2017245402}.
Therefore, both perturbation theory and finite displacement approaches will continue to play
an important role for the foreseeable future in the context of computing phonons and their interactions.

This paper describes several novel approaches, applicable to a broad variety of phonon and phonon-interaction methodologies.
First, we write the Taylor series purely in terms of space group irreducible derivatives, building in all possible symmetry
by construction; we are not aware of existing studies that employ this in complex scenarios beyond second order (i.e. for a sufficiently large FTG to describe generic observables).
Aside from computational efficiency, symmetry is essential for characterizing, storing, and 
disseminating the vibrational Hamiltonian. Second, we devise two finite displacement approaches, which focus on getting the most precise answer or getting
a robust answer as efficiently as allowed by group theory. Finally, we evaluate various approaches to assessing the integrity of the Taylor series.

The remainder of the paper is organized as follows. Sections \ref{sec:irrderivs}-\ref{sec:fdanharm} review the relevant
literature with respect to group theory, perturbation theory, and previous approaches using finite displacements. Section
\ref{sec:gtmethod} outlines our group theoretical methodology, which is illustrated throughout with examples from graphene
for the sake of clarity. Additionally, a glossary of all key variables can be found in Supplementary Information, Table S\ref{si:table:glossary}.
Section \ref{sec:finitedisp} puts forward our finite displacement approaches. Section \ref{sec:hierarchy} solves the minimal supercell problem using the
Smith Normal Form, resulting in the Minimum Supercell Multiplicity equation; while Sections \ref{sec:LID}-\ref{sec:HSBID} introduce our LID and 
BID finite displacement approaches. Finally, Section \ref{sec:assessing} focusses on how to asses the quality of the extracted irreducible derivatives,
including the $N$-th order strain derivatives of the phonons. 

Applications are presented throughout the manuscript, and DFT calculations were executed as follows (unless otherwise noted).
DFT calculations within the local density approximation (LDA)\cite{Perdew19815048}%
were performed using the
Projector Augmented Wave (PAW) method \cite{Blochl199417953,Kresse19991758},
as implemented in the Vienna Ab-initio Simulation Package (VASP) \cite{Kresse1993558,Kresse199414251,Kresse199615,Kresse199611169}. 
A plane wave basis with a kinetic energy cutoff of 625 eV was employed. We used a $\Gamma$-centered \textbf{k}-point 
mesh of 100$\times$100$\times$1. 
All $k$-point integrations were done using Gaussian smearing with a smearing width of 0.2eV. 
The DFT energies were converged to within 10$^{-8}$ eV, while ionic relaxations were converged to within 10$^{-7}$ eV.
The relaxed lattice parameter in graphene was found to be $a_0$=2.44994\AA.

\subsection{Symmetry and Irreducible Derivatives}
\label{sec:irrderivs}
Group theory is a central tenet of physics\cite{Cornwell,Zee20160691162697}, and it should 
characterize phonons and their interactions; regardless of how these
quantities are computed (i.e. finite displacement or perturbation theory). In
the context of atomic physics, for example, where continuous groups
characterize the invariance of the Hamiltonian, the notion of an ``irreducible
matrix element" as given by the Wigner-Eckhardt theorem is textbook material\cite{Sakurai19930201539292}; 
and the same could be said for nuclear physics\cite{Greiner1996354059180X}.  The beauty of
irreducible matrix elements is that absolutely no excess information needs to
be provided, beyond the chosen phase conventions, to characterize any possible
matrix element; and one is guaranteed that the theory satisfies
symmetry by construction. Generically, we refer to this type of symmetrization as ``intrinsic symmetrization",
because it begins with basis functions that transform like irreducible representations of the group,
and determines the  existence of an arbitrary matrix element \emph{a priori}.

In the context of lattice vibrations, the corresponding quantities are ``space
group irreducible derivatives" of the Born-Oppenheimer potential.  Such an
approach will automatically satisfy all space group symmetry by construction,
in addition to homogeneity of free space and permutation symmetry of each
derivative.  At second order, space group irreducible derivatives are
constructed using standard tools of solid state physics\cite{Cornwell,Tinkham}:
the irreducible Brillouin zone and the little group of a given $q$-point.
Beyond second order, the use of space group irreducible derivatives is far less
common, most likely because the group theory is more complex. Nonetheless,
constructing the symmetric products of irreducible representations of space
groups was essentially a solved problem by the year 1980, and the history of
this saga is well described in Cracknell \emph{et. al} (see Vol. 1 of Ref.
\onlinecite{Cracknell19790306651750}).  There are two complimentary
approaches\cite{Birman19621093,Lax1965793,Birman1966771,Zak1966464,Lewis1973125,Gard19731807,Gard19731829,Birman1974354013395X}:
the full group approach and the subgroup approach.  While both approaches have
their respective merits, Cracknell \emph{et. al} used the subgroup formulation
of Gard\cite{Gard19731807,Gard19731829} to completely automate the process,
resulting in a code which could be executed at an arbitrary order
$\order$\cite{Cracknell19790306651750}; only limited by the computers of their
time period.  They produced printouts containing the selection rules for third
order symmetric products within all crystallographic space groups, and
therefore the composition of the third order Taylor series in terms of space group
irreducible derivatives can be obtained for any possible crystal. They
also report that they produced an archived volume with quartic symmetric
products for all space groups.

Despite the power of intrinsic symmetrization in the context of lattice dynamics, which works with basis functions that
transform as irreducible representations of the space group and obey clear selection rules which can be determined once and for all \emph{a priori}, it
remains highly underutilized; with applications beyond second order often
involving Landau expansions, where a phase transition may be associated with a
single star of wavevectors\cite{Toledano19879971500264}; or optical transitions\cite{Birman1974354013395X}.  However, we are not
aware of any systematic approach which utilizes intrinsic symmetrization to
construct phonon interactions in general, which is an intent of this paper. 

The major alternative to utilizing intrinsic symmetrization is
to start with the order $\order$ Taylor series in real space (i.e. with displacements labeled by a lattice translation) 
and then impose invariance with respect to
the space group operations, permutation symmetry of the derivative, and homogeneity and isotropy of free space\cite{BornHuang,Leibfried1961275,Horton19740720402786};
and this results in a system of linear equations that the real space derivatives must obey.
This approach is the \emph{direct opposite} of intrinsic symmetrization: instead of starting with symmetry and only creating allowed polynomials,
one starts with the most general polynomials and then determines their relations. We refer to this alternate procedure as ``extrinsic symmetrization". 
While extrinsic symmetrization is most naturally associated with a real space basis, we note that it can be used for an arbitrary basis.
Extrinsic symmetrization can be straightforwardly implemented in scenarios that are sufficiently low order and short range, allowing one to solve for
a set of irreducible real space derivatives. However, this approach quickly becomes challenging as the size of the initial unsymmetrized polynomial
will grow rapidly with order and range. 

Practitioners typically numerically implement extrinsic symmetrization while simultaneously fitting the real space derivatives, 
resulting in a procedure where it is unclear to the outside observer if symmetry is actually being fulfilled. This even happens regularly at second order.
For example, in the well known paper of Parlinski \emph{et. al}, which puts forward an approach to compute phonons using finite difference\cite{Parlinski19974063}, they implement
point symmetry using extrinsic symmetrization and apply this to the case of ZrO$_2$. For a $2\times2\times2$ supercell relative to the conventional cubic cell, their symmetry analysis finds
that there are 68 independent parameters, though they report that only 59 of these 68 are nonzero. Nonetheless, group theory
dictates that there are precisely 52 irreducible derivatives, all of which can be chosen as real numbers (see Appendix \ref{app:zro2} for details). 
Strictly speaking, their Born-Oppenheimer potential will not satisfy symmetry, though their results are clearly robust and not affected by this inefficacy.
However, it is also worth noting that group theory dictates that all irreducible derivatives can be extracted 
with a single central finite difference calculation instead of two, which is used in their study (see Appendix \ref{app:zro2} for details).
Clearly, it is much easier to employ intrinsic symmetrization instead of a numerical implementation of extrinsic symmetrization where the answer is not obvious.
While the aforementioned paper is relatively old, extrinsic symmetrization still persists at second order\cite{Alfe20092622,Wang20142950,Togo20151}
and is commonplace beyond second order\cite{Esfarjani2008144112,Chaput2011094302,Hellman2013144301,Zhou2014185501}.
More importantly, we demonstrate that the practical inefficacy of extrinsic symmetrization is dramatically worse in
some popular approaches for computing cubic interactions (See Section \ref{sec:SSBID}).

An important point to consider is how the Taylor series is truncated at a given
order, and there are two natural approaches to doing this.  First, one can
create a homomorphic mapping between the infinite translation group and  a
finite translation group (FTG)  via a Born von Karman (BvK)
supercell\cite{Tinkham,Cornwell}; and this type of truncation is naturally
compatible with the irreducible representations of the space group and the
accompanying intrinsic symmetrization.  Second, one can retain the infinite crystal, or a
sufficiently large BvK supercell, and define a range in real space via nearest
neighbor shells or some cluster size beyond which all derivatives are zero; and
this type of truncation is naturally compatible with a
real space basis and extrinsic symmetrization. We refer to these two types of truncation as reciprocal
space truncation and real space truncation, respectively, given that the former
restricts to some finite number of $q$-points and the latter restricts to some
neighbor shell in real space.  An important point to realize is that these two
truncations do not have a direct correspondence in general, and it is often not
clear which truncation a practitioner is using.

An additional important point is that translation group irreducible derivatives, and therefore space group irreducible derivatives as well, 
\emph{are invariant to the supercell in which they are computed},
whereas real space derivatives are only exact in the infinite crystal. Of course, a real space basis can
always be used in any supercell, even very small supercells, but in such situations the real space derivatives are simply containers and interpolants for the
space group irreducible derivatives. Under normal circumstances, the real space derivatives will converge when taken in a sufficiently large supercell,
but space group irreducible derivatives are always converged with respect to supercell size by construction. However, a sufficient number of 
space group irreducible derivatives must be resolved in order to precisely interpolate to an arbitrary $q$-point,
which is equivalent to the real space derivatives being sufficiently diminished within the truncation range. 

Finally, we point out that there is a middle ground between intrinsic and extrinsic symmetrization, which can be convenient if
a real-space truncation is needed. 
One can consider the crystal to be an infinite array of overlapping clusters, and the local modes of each cluster can then be used
as the new degrees of freedom subject to a constraint. Such a program was originally put
forth and implemented for model Hamiltonians in two
dimensions\cite{Ahn2003092101,Ahn2004401,Seman20125}. The same type of framework, called
the Slave Mode Expansion, was put forward purely for the purpose of symmetrizing the lattice potential\cite{Ai2014014308,Kornbluth2017}.
The basic idea is to perform intrinsic symmetrization with respect to the point group, and then to perform extrinsic symmetrization
with respect to the translation group; assuming that the clusters overlap (see Ref. \onlinecite{Thomas2013214111} for an approach similar in spirit, yet distinct). 

\subsection{Perturbation theory}
Perturbation theory is normally the preferred method for computing derivatives, and should be used when possible.
The Hellman-Feynman theorem provides first derivatives of the energy (i.e. the force) at a very
small computational cost\cite{Martin2008}, and have become standard in density functional theory codes. 
Perturbative forces are often implemented in static approaches like DFT+$U$, and a few studies have succeeded in computing forces
in more advanced methods such as DFT plus dynamical mean-field theory\cite{Leonov2014146401,Haule2016195146}.

For second order derivatives, 
density functional perturbation theory (DFPT)\cite{Zein19841825,Baroni19871861,Gonze198913120,Gonze19951096,Baroni2001515} may be executed at an arbitrary reciprocal space point,
with a cost which is on the order of a primitive cell self-consistent DFT calculation\cite{Baroni2001515};
and there is a large literature of such calculations. 
DFPT is not as widely available as the ubiquitous Hellman-Feynman forces,
and therefore DFPT may not be available for all codes or basis sets in practice. 
Furthermore, DFPT often does not support
even simple beyond DFT methods such as DFT+U, and at present we are only aware of several examples in the literature\cite{Floris2011161102,Dorado2017245402}. 
Therefore, DFPT is not always an option for second order derivatives.

DFPT may be extended to third order\cite{Gonze198913120,Gonze19951096,Debernardi19951819},
and this has been implemented for the most general case (i.e. arbitrary $q$ vectors, metals and insulators)\cite{Lazzeri2002245402,Paulatto2013214303}.
A small number of applications can be found in the literature thus far\cite{Deinzer2003144304,Fugallo2013045430,Paulatto2015054304,Campi2017024311,Markov2016064301,Giannozzi2017465901}.
We are not aware of any studies using third order DFPT within DFT+$U$.

DFPT naturally works with irreducible derivatives of the translation group, and at least some implementations
at second order work with irreducible representations of the space group when performing perturbation theory\cite{Giannozzi2009395502}. It is unclear to what extent
point symmetry, or full space group symmetry, is exploited for third order. In any case, it would be ideal if DFPT studies reported space group irreducible derivatives,
as this would allow for a direct comparison with competing methods.

\subsection{Finite displacement phonon approaches}
Finite displacement approaches are those which explicitly move the atoms in a series of different displacement fields and perform
a full, self-consistent first-principles calculation in each case. This could range from performing a first-principles molecular dynamics trajectory, to a more
standard central or forward finite difference calculation; and we focus on the latter.
We begin by reviewing  the earliest papers in the literature,
and discuss them in terms of the framework we will be presenting.
Perhaps the
earliest study performed
second and third order finite difference derivatives of the energy using
a displacement which transforms as an irreducible representation of the space
group\cite{Wendel1978950}, and this came to be known as a ``frozen-phonon" calculation. In terms of our categorization, the preceding paper falls under LID with
$\pd[0]$ (see Section \ref{sec:LID}). 

Several similar studies followed soon after on various materials\cite{Yin19801004,Kunc19812311,Ihm1981491}, 
and Ihm \emph{et. al} used the Hellman-Feynman forces in this same context\cite{Ihm1981491}, which we categorize as LID using $\pd[1]$.  
Martin subsequently announced a major advance which further exploited Hellman-Feynman forces\cite{Martin1981617}, whereby the displacement field
was intentionally chosen not to transform as an irreducible representation of the space group such that many independent force constants could
be independently measured. This general philosophy falls under the category of SS-BID using $\pd[1]$ (see Section \ref{sec:SSBID}).
A subsequent  study then executed Martin's previous announcement with an application to GaAs\cite{Kunc1982406}, showing the power of this approach.
However, several additional steps would be needed to satisfy all the conditions of SS-BID. First, 
the force constants should be extracted in a manner which preserves the irreducible derivatives of the translation group. 
Second, the approach for displacing the atoms could be optimized.

In order to better exploit the forces, displacement should be constructed so as to sample
as many irreducible derivatives as possible in a single calculation.
Frank \emph{et. al} made another step forward, performing finite difference calculations where they displaced a single atom at a time\cite{Frank19951791}.
This approach goes a long way towards
achieving the goal, given that a local displacement in real space is guaranteed
to sample all $q$-points in the supercell; though a shortcoming is that point symmetry 
is not
explicitly dealt with in any way. More problematic is that care is needed to
ensure the translation group irreducible derivatives are extracted properly. 

Parlinski \emph{et. al} resolved a main shortcoming of the preceding studies\cite{Parlinski19974063},
introducing  a proper weighting of the real
space force constants on the boundary of the Wigner-Seitz supercell, which ensures their Fourier interpolation yields the
numerically exact irreducible derivatives of the finite translation group for the supercell being used. 
The authors also directly account for point symmetry, determining the minimum number of calculations 
required to extract all force constant in conjunction with the forces (though there were some inefficacies in their analysis, see discussion in Section \ref{sec:irrderivs}). 
We categorize this method as a $\order=2$ SS-BID approach using $\pd[1]$ (see Section \ref{sec:SSBID}).

A relevant factor which had not been considered by the aforementioned approaches is that they extract all force constants from a single supercell, 
and we refer to these as \emph{single supercell} (SS) approaches.
An important development occurred relatively recently with the work of Monserrat \emph{et al.}\cite{Lloyd-williams2015184301},
which recognized the importance of using so-called non-diagonal supercells.
They show that given a three dimensional crystal, 
all $q$-points within a $n_1 \times n_2 \times n_3$ supercell can always be probed in a multiplicity $\textrm{lcm}(n_1, n_2, n_3)$ supercell. 
This result has far reaching implications for computing phonons, offering a massive speedup for first-principles approaches which scale in a super-linear manner. 
We note that their result is a special case of our Minimum Supercell Multiplicity equation (Eq. \ref{eq:ssmult}), and equation \ref{eq:msmt} which follows.

\subsection{Finite displacement anharmonic approaches}
\label{sec:fdanharm}
Finite displacement approaches have also been employed to compute anharmonic terms. As mentioned, the very first frozen phonon calculation by Wendel and Martin
computed a third order derivative using finite difference of the energy\cite{Wendel1978950}. More systematic approaches began to appear thereafter, such as when 
Vanderbilt \emph{et. al} used the forces and finite displacement calculations to fit an assortment of cubic and quartic phonon interactions at products of the $\Gamma$ and $X$ points 
in diamond\cite{Vanderbilt19895657,Narasimhan19914541}. These interactions were then fit to a modified Keating model
which was then used to extrapolate throughout the Brillouin zone; and this approach provided reasonable results for the
phonon lifetimes in Si.  %

As time progressed and computing resources increased, new efforts emerged to systematically compute more interactions. Esfarjani 
and Stokes employed a extrinsic symmetrization approach with a real space truncation (see Section \ref{sec:irrderivs})  in order to compute the real space force constants
up to fourth order\cite{Esfarjani2008144112}. 
They suggested that a data set of forces could be obtained from DFT calculations on a sufficiently large supercell by generating a first-principles molecular
dynamics trajectory, random displacements, or symmetrically displacing one atom at a time; and they opted for the latter in a test on Si.
Using this data set and the aforementioned symmetrization constraints, they fit the real space force constants up to fourth order. 
Applications of this method in a wide range of materials
soon followed, all in the context of thermal conductivity\cite{Esfarjani2011085204,Shiomi2011104302,Tian2011053122,Lee2014085206}. 
Many approaches similar to the aforementioned approach, yet distinct in various ways, soon 
followed\cite{Chaput2011094302,Lindsay2012095901,Thomas2013214111,Li20141747,Zhou2014185501,Ai2014014308,Plata201745},
with each producing the real space force constants up to some order and within some real space truncation range.
In section \ref{sec:SSBID}, we compare our SS-BID approach to several of the aforementioned approaches\cite{Chaput2011094302,Li20141747,Plata201745}, 
demonstrating that we can extract all space group
irreducible derivatives far more efficiently for a given test case in rock salt. Furthermore, all of the above approaches could
benefit from our hierarchical supercell approach (see Sections \ref{sec:hierarchy} and \ref{sec:HSBID}).

While all the preceding studies relied upon forces (i.e. $\pd[1]$), 
third and fourth order phonon interactions have been
computed using finite difference of second order DFPT calculations in graphene and graphite\cite{Bonini2007176802}.  
In that study, a FTG is used for truncation (i.e. with graphene, they used $\supa[BZ]=4\hat1$ for $\order=3$ and $\order=4$ with $\pd[2]$) instead of a real space truncation, and the cubic and quartic
derivatives appear to be translation group irreducible. A study of this sort could fully exploit both the SS-BID and the 
HS-BID approaches we outline in this paper, which would yield a major increase in efficiency (see Section \ref{sec:SSBID} and \ref{sec:HSBID}).

\section{Group Theoretical Methodology}
\label{sec:gtmethod}
\subsection{Crystalline potential and its derivatives} 
We begin by discussing the Born-Oppenheimer potential energy, $\pot$, defined over the set of all nuclear displacements in the
crystal, $\{\rdispvec[\ctrans][b][\beta]\}$; where $\ctrans$ labels a unit cell in lattice coordinates, $b$ labels one of the $\natom$ different atoms within the unit cell, 
and $\beta$ labels 
one of the $\npol$ possible displacements of the atom. 
The function $\pot$ is invariant to all operations of some space
group and conserves total linear and angular momentum; and $\pot$ is presumed to be analytic. 
Our convention is to define $\pot$ as the energy of the crystal \emph{per unit cell}, so it is an intensive quantity.
While we focus on $\pot$ in this work, any function defined over the lattice could be considered.
Due to the large number of variables defined in this paper, a glossary is provided in Supplementary Information, Table S\ref{si:table:glossary}. 
Additionally, our application to graphene is distributed throughout the manuscript, which should aid in understanding all definitions.

If the crystal is $d$-dimensional, the translation group is
defined via $\cdim$ linearly independent vectors $\lvec$ in $\mathbb{R}^d$, and 
stored as row-stacked vectors in the rank-$\cdim$ matrix $\lmat$.
Basis atoms are specified as $\natom$ distinct  Cartesian vectors $\cbvec$. 
A corresponding set of reciprocal lattice vectors are defined as $\hat b=2\pi \hat a^{-1}$,
where $\rmat$ gives the \emph{column stacked} vectors $\rvec[i]$. The continuum of $q$-points within the first Brillouin zone
can be used to form basis functions $\{\qdispvec[\qq][b][\beta]\}$ that transform like irreducible representations of the translation group.
For the case of graphene (see schematic in Figure \ref{fig:uniform_grid}a), we have:
\begin{align}\label{}
\lmat = \frac{a_o}{2} \begin{bmatrix}
  \sqrt 3 &  1 \\[0.5em]
  \sqrt 3 & -1 
\end{bmatrix}  
  &&
  \rmat = \frac{2\pi}{a_o\sqrt 3} \begin{bmatrix}
   1      &   1        \\[0.5em]
   \sqrt 3& - \sqrt 3 
\end{bmatrix}  
 \\ 
   \cbvec[1] = a_0\frac{\sqrt 3}{3}\vec i && \cbvec[2]  = a_0\frac{2\sqrt 3}{3}\vec i
\end{align}
where $a_0=2.44994$\AA, as computed within DFT (see Section \ref{sec:genback} for computational details), and $\vec i$ is the unit vector of $x$-axis. 
An arbitrary lattice point may be expressed as $\ctrans \lmat$, where $\ctrans$ is a $d$-dimensional vector of integers: $\ctrans\in\mathbb{Z}^d$.

An $\order$-th order derivative of $\pot$ is denoted as $\pot_{i_1\dots
i_\order}^{j_1\dots j_\order}$, where $i$ labels either some linear combination
of reciprocal lattice vectors or real lattice vectors, and $j$ labels some linear
combination of the $\natom\npol$ degrees of freedom within the unit cell.
For the specific case of derivatives taken with respect to the displacements in the real lattice basis,
we define the force tensor
    \begin{align}\label{}
      \phinsym\indices*{*_{\vec 0}^{(a_1, \alpha_1)}_{\ctrans[2]-\ctrans[1]}^{(a_2,\alpha_2)}_{\dots}^{\dots}_{\ctrans[\order]-\ctrans[1]}^{(a_\order,\alpha_\order)}}=
      \frac{\partial^\order \pot }{\prod_i^\order\partial\qdispvec[\ctrans[i]-\ctrans[1]][a_i][\alpha_i] }
    \end{align}
where we explicitly retain the identity translation $\vec 0$ as the first index (i.e. $\ctrans[1]-\ctrans[1]=\vec 0$), and $\qdispvec[\ctrans[i]][a_i][\alpha_i]$ follows the same convention as previously defined; with $\alpha$ labeling a displacement and $a$ labeling an atom within the primitive cell.
For the specific case of derivatives taken with respect to displacements that transform as irreducible representations of the translation group,
we define the dynamical tensor:
\begin{align}\label{}
\dqnsym\indices*{*_{\qq[1]}^{(a_1,\alpha_1)}_{\dots}^{\dots}_{\qq[\order]}^{(a_{\order},\alpha_{\order})}}
= 
\frac{\partial^\order V }{\prod_i^\order\partial\qdispvec[\qq[i]][a_i][\alpha_i] }
\end{align}
where $\sum_{i=1}^\order\qq[i]$ is a reciprocal lattice vector, and $a_i, \alpha_i$ follows the same convention as previously defined.

\subsection{Finite Translation Group}
Here we define the familiar notion
of the finite translation group (FTG), which is a homomorphic mapping with the infinite translation group via periodic boundary conditions\cite{Tinkham,Cornwell};
though we consider the most general case. 
The FTG is equivalently defined using a supercell of the real
space lattice or a subcell of the reciprocal lattice; which we refer to as the Born-von Karman (BvK) supercell and Brillouin Zone (BZ) subcell, respectively. 
We note that non-diagonal BvK supercells are considered in this work.
Mathematically, we define 
the BvK supercell lattice vectors of the real space lattice and the corresponding 
BZ subcell vectors of the reciprocal
lattice using the matrix $\supa[BZ]$:
\begin{align}\label{}
  \lmat[BZ] = \supa[BZ]\lmat
  &&  
  \rmat[BZ] = \rmat \supa[BZ]^{-1} 
\end{align}
where
\begin{align}\label{eq:supa}
\nonumber \supa[BZ] \in \{& \hat n \in \mathbb{Z}^{(d,d)}:\det(\hat n)\ne 0, \\ & \hat n\lmat\symop^\tp(\hat{n}\lmat)^{-1} \in \mathbb{Z}^{(d,d)} \forall \hat R\in\pg \}
\end{align}
where $\cdim$ is the dimension of the crystal and $\pg$ is the point group of the space group.
In words, $\supa[BZ]$ is an invertible $\cdim\times\cdim$ matrix of integers, with a real (reciprocal) space Wigner-Seitz super (sub) cell that is invariant to $\pg$.
The translation vectors of the FTG are equivalently either all of the real space lattice points that fit within $\lmat[BZ]$ or all of the 
subcell reciprocal lattice points that fit within $\rmat$, where the lattice points can be expressed in lattice
coordinates of $\lmat$ and $\rmat[BZ]$, respectively.
Mathematically, a translation vector of the FTG is represented as $\ctrans\lmat$, where  $\ctrans$ is vector of integers $\ctrans\in\mathbb{Z}^d$ constrained to
\begin{align}\label{eq:tpoints}
 0 \le \ctrans\supa[BZ]^{-1}\cdot \vec e_j < 1   && \textrm{ for }&& j=1,\dots,\cdim
\end{align}
where $\vec e_j$ is a unit vector in $\mathbb{Z}^d$.
These integer vectors $\ctrans$ are lattice coordinates of FTG points, and we refer to them as ``$t$-points"; though it should be emphasized that $t$-points can either be used to specify a lattice translation
in lattice coordinates of $\lmat$ or reciprocal space points in lattice coordinates of $\rmat[BZ]$. Therefore, these points characterize both the FTG and its irreducible representations.
The set of all $t$-points is defined as $\bctrans[BZ]$ (sets are always denoted with a tilde), and $|\bctrans[BZ]|=\det(\supa[BZ] )$. Given the importance
of the total number of $t$-points, we define the variable $\nq=|\bctrans[BZ]|=\det(\supa[BZ] )$.

When a $t$-point $\ctrans$ is denoted in lattice coordinates of $\rmat$, it will be a
fraction that is less than one; and it is simply denoted as $\qq$ (i.e.
$\qq[i]=\supa[BZ]^{-1}\ctrans[i]$). We naturally refer to these points as ``$q$-points", and the characters of 
the irreducible representations of the FTG are then $e^{\imag2\pi \ctrans\qq}$. The set of all $\qq$ is denoted as $\bqq[BZ]$, and this stores all $q$-points within the first Brillouin Zone.
Clearly, we have $|\bqq[BZ]|=\nq$.
Another key property of the FTG is the largest least common denominator of all components of any $\qq\in\bqq[BZ]$,  denoted $\lcmdm$.
In the common case of $\supa[BZ]=n\hat 1$, where $n\in\mathbb{Z_+}$,
we simply have $\lcmdm=n$; and we refer to this as a \emph{uniform} supercell. 

The physical meaning of the real space supercell and reciprocal space subcell is
then clear: the allowed subcell points $\qq\in\bqq[BZ]$ correspond to waves that are commensurate with the supercell $\lmat[BZ]$.
Stated differently,
each allowed subcell $q$-point transforms like the identity representation of the infinite supercell
translation group.  
It is clear that the reciprocal lattice sub-cell volume is reduced by a factor of $\nq$ while the corresponding real space supercell is increased
by this same factor.

The symmetrized displacement amplitudes are obtained with the projection operator, recovering the usual discrete Fourier transform, 
though we use a normalization such that the $q$-space amplitudes are intensive quantities:
\begin{align}\label{}
\qdispvec &= \frac{1}{\nq}\sum_{\ctrans\in \bctrans[BZ]} \rdispvec e^{-2\pi\imag \ctrans \cdot \qq }
\\[0.5em]
\rdispvec &= \sum_{\qq\in \bqq[BZ]} \qdispvec e^{2\pi\imag \ctrans \cdot \qq }
\end{align}
where these modes are imparted on some supercell $\supa[BZ]$.

\begin{figure}[htpb]
  \centering
  \includegraphics[width=0.99\linewidth]{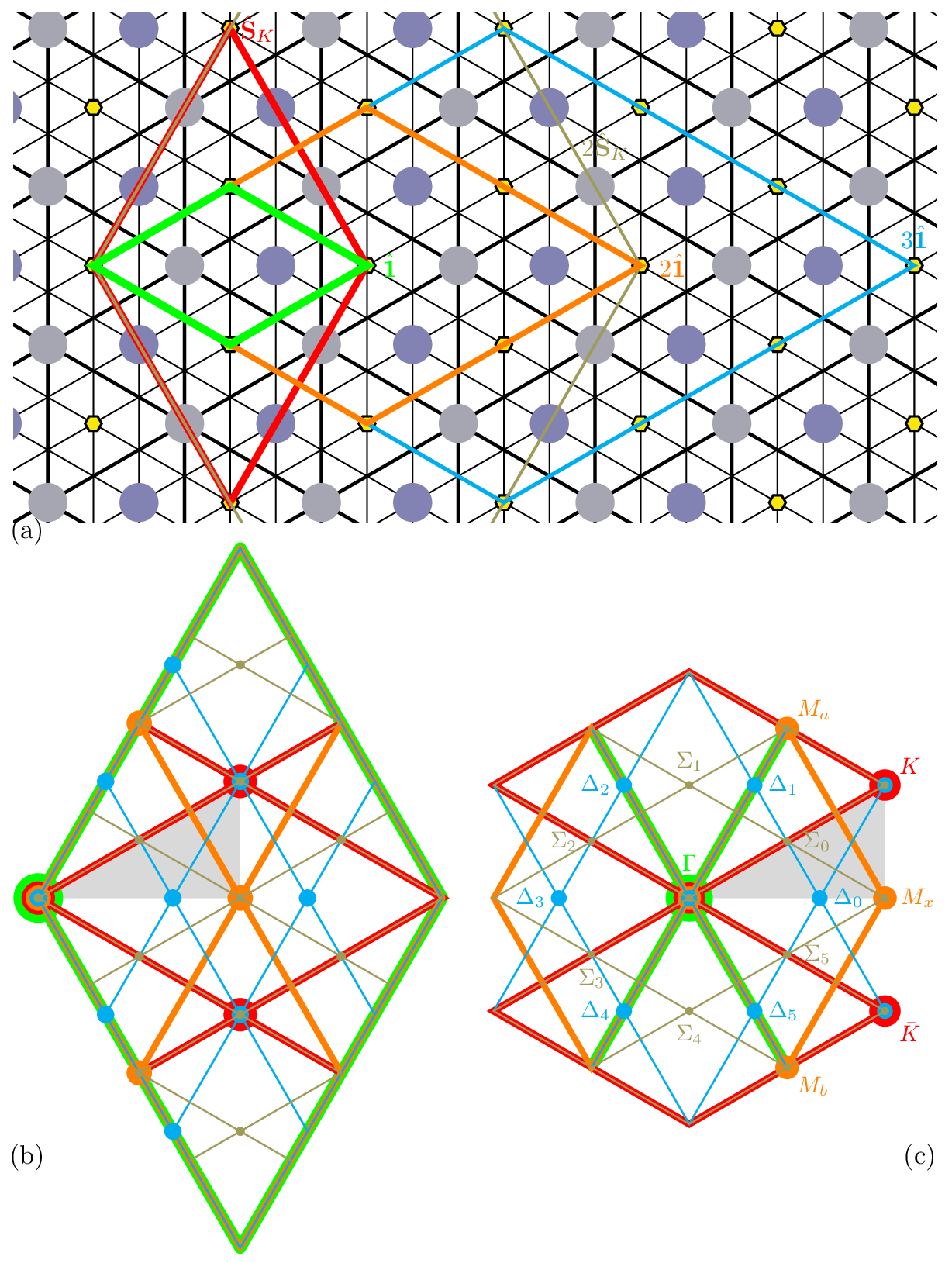}
  \caption{(a) A schematic of the structure of graphene. 
  Yellow hexagons are lattice points, while circles represent carbon atoms.
  The first five smallest BvK supercells are denoted, and the lattice points of a corresponding FTG is given by Eq. \ref{eq:tpoints}. 
  (b) The corresponding (color coded) five reciprocal lattice subcells which
  are repeated to tile the First Brillouin Zone; irreducible Brillouin zone is
  shaded grey. (c) Same as (b), but with the FBZ in the Wigner-Seitz cell
  convention. All $q$-points are labeled according to their star.
  }
  \label{fig:uniform_grid}
\end{figure}

In the case of graphene, it is straightforward to find that all FTG can be obtained as integer multiples
of the rank 2 identity matrix $\hat 1$ or the $K$-supercell, $\supa[K]=2\hat 1-\pauli[x]$, where $\pauli[x]$ is a Pauli matrix;
and this results in 
grid densities of $\nq=\{n^2 | n\in \mathbb{Z}_+\}$  and $\nq=\{3n^2 | n\in \mathbb{Z}_+\}$
 points per Brillouin zone, respectively (See schematic in Figures \ref{fig:uniform_grid}a-c for FTG corresponding to 
 $\supa[BZ]=\hat 1$, $\supa[K]$, $2\hat 1$, $3\hat 1$, and $2\supa[K]$). 
\label{sec:ugb}
While the FTG $\hat 1$ is already nontrivial given that $\natom>1$ in graphene,  it is pedagogically instructive to consider the next largest FTG $\supa[BZ]=\supa[K]$;
which corresponds to the order 3 cyclic group. For this FTG, $\nq=3$ and we have
\begin{align}\label{}
  \bctrans[BZ]&=\left\{ (0,0), (1,0), (0,1) \right\}
               &&= \left\{ \vec 0 , \ctrans[a],\ctrans[b] \right\} \\
  \bqq[BZ]&=\left\{
\left(0,0 \right),
\left(\frac{2}{3}, \frac{1}{3} \right),
\left(\frac{1}{3}, \frac{2}{3} \right)
                   \right\} &&= \{\Gamma,K, \bar K \}
\end{align}
which can be deduced from the diagrams in Figure \ref{fig:uniform_grid}.

\subsection{Point Symmetry of Finite Translation Group}
The point symmetry of the FTG must also be considered, and several additional definitions are needed. 
First, $\bctrans[IBZ]$ and $\bqq[IBZ]$ are irreducible sets which can generate all elements of $\bctrans[BZ]$ and $\bqq[BZ]$, respectively,
in conjunction with some point operation $\hat R\in\pg$. Furthermore, it is important to identify the so-called ``little group" $\pg_{\qq}$
for each $\qq\in\bqq[IBZ]$, which is the subgroup of $\pg$ that leaves $\qq$ invariant to within a shift in $\mathbb{Z}^d$.
Finally, we must introduce the ``star" of the $q$-point, which is the set of points generate by $\pg$: 
$\qorbit[\qq]=\{\rmat^{-1}\hat R\rmat\qq | \forall\hat R\in\pg\}$, where $|\qorbit[\qq]|\le \pgorder$; and there will be one star
for each $\qq\in\bqq[IBZ]$. 
The set of all stars is then denoted as $\qorbit[BZ]=\{\qorbit[\qq] | \forall \qq \in \bqq[IBZ]\}$. 
A given star may be used to create a $|\qorbit[\qq]|$-dimensional representation of star vectors.

In the case of graphene, we have $\pg=D_{6h}$, using Schoenflies notation. For $\supa[BZ]=\supa[K]$, $\bqq[IBZ]=\{\Gamma,K\}$;
the little groups are $\pg_\Gamma=D_{6h}$ and $\pg_K=D_{3h}$; the two stars are $\qorbit[\Gamma]=\{\Gamma\}$ and $\qorbit[K]=\{K,\bar K\}$,
and the corresponding representations of the star vectors decompose to $A_{1g}$ and $A_{1g}\oplus B_{2u}$, respectively. The explicit 
representation are:
\begin{align}\label{}
\qorbitvec[\Gamma][A_{1g}]=\Gamma && 
\qorbitvec[K][A_{1g}]=\frac{1}{\sqrt 2}(K+\bar K) &&
\qorbitvec[K][B_{2u}]=\frac{1}{\sqrt 2}(K-\bar K)
\end{align}

\subsection{Order $\order$ identity representations of FTG and permutation symmetry}
Having defined the FTG, the resolution of the problem has been set.  We  proceed by creating all
of the order $\order$ direct product representations of $\bqq[BZ]$  which transform like the identity under the translation group. 
Each identity representation is given 
by a variable $\qset$, which is a row stacked matrix of $\order$ vectors $\{\qq[i]\}$ in a particular order;
and the notation $(\qset)$ implies a $\order$-tuple of the $\{\qq[i]\}$ with the same ordering. 
The translation group demands that the identity representation satisfy $(\sum_{\qq\in(\qset)}\qq)\in\mathbb{Z}^\cdim$.
Clearly, one of the $\qq\in(\qset)$ is not independent, and there must be 
$\nq^{\order-1}$ distinct identity representations. Therefore, we can identify
each order $\order$ identity irreducible representation with a corresponding irreducible representation of the 
$\order -1$ direct product group formed from $\bctrans[BZ]$; with the set of all product translations collected in the set $\btset[BZ]$. 
The set of all identity representations 
$\{\qset[i]\}$ within the BvK supercell is denoted as $\bqset[\supa[BZ]]$ (abbreviated as  $\bqset[BZ]$).
For the case of graphene with $\order=3$ and $\supa[BZ]=\supa[K]$, we have:
\begin{align}\label{}
  \bqset[BZ]= \{ & (\Gamma,\Gamma,\Gamma),  (\Gamma,\bar K,K),  (\Gamma,K, \bar K),  (K,\Gamma,\bar K)  \nonumber \\
              & (\bar K,\Gamma,K),        (K,\bar K, \Gamma),  (\bar K,K, \Gamma),  (K,K,K),  \nonumber \\
              & (\bar K,\bar K,\bar K)                                                   
   \} \nonumber \\
   =\{& \qset[1],  \qset[2],  \qset[3],  \qset[4],  \qset[5],  \qset[6],  \qset[7],  \qset[8],  \qset[9] \}
   \\
  \bqset[IBZ]= \{ & (\Gamma,\Gamma,\Gamma) , (\Gamma,\bar K,  K), (K,\Gamma,\bar K), (K,\bar K, \Gamma),
              \nonumber \\ & (K,K,K) 
   \} \nonumber \\
   =\{& \qset[1],  \qset[2],  \qset[4],  \qset[6],  \qset[8] \}
   \\
  \btset[BZ]=\{&
    (\vec 0,\vec 0,\vec 0), 
    (\vec 0,\vec 0,\ctrans[a]), 
    (\vec 0,\vec 0,\ctrans[b]), 
    (\vec 0,\ctrans[a],\vec 0), 
    (\vec 0,\ctrans[a],\ctrans[a]),  
    \nonumber \\ & 
    (\vec 0,\ctrans[a],\ctrans[b]), 
    (\vec 0,\ctrans[b],\vec 0), 
    (\vec 0,\ctrans[b],\ctrans[a]), 
    (\vec 0,\ctrans[b],\ctrans[b]) 
    \}
\end{align}
We will also need to form the stars for each $\qset\in\bqset$, which is straightforwardly constructed. 
\begin{align}\label{}
\orbit[\qset[1]]&=\{\qset[1]\}
&&
\orbit[\qset[2]]=\{\qset[2], \qset[3]\}
&&
\orbit[\qset[4]]=\{\qset[4], \qset[5]\}
\nonumber \\
\orbit[\qset[6]]&=\{\qset[6], \qset[7]\}
&&
\orbit[\qset[8]]=\{\qset[8], \qset[9]\}
\end{align}

Given that any derivative
is invariant to permutation symmetry, it is necessary to define a \emph{multiset}  $\qsetp=\{\qq|\qq\in(\qset)\}$ (i.e. there is no ordering and repeating $\qq[i]$ are allowed).
We can immediately reduce $\bqset[BZ]$ to the identity representations of the symmetric product group
by retaining only the unique $\qsetp$ generated from $\bqset[BZ]$; and this is denoted $\bqsetpbz$ . 
Finally, we can create the point irreducible set of $\bqsetpbz$, denoted as $\bqsetpibz$. 
For the case of graphene with $\order=3$ and $\supa[BZ]=\supa[K]$, we have:
\begin{align}\label{}
  \bqsetpbz&= \{(\Gamma,\Gamma,\Gamma) , (\Gamma,\bar K,  K), (K,K,K), (\bar K,\bar K, \bar K)  \} 
  \nonumber \\ &= \{ \qsetp[1], \qsetp[2], \qsetp[3], \qsetp[4] \}
  \\
  \bqsetpibz&=\{(\Gamma,\Gamma,\Gamma) , (\Gamma,\bar K,  K), (K,K,K)  \} 
  \nonumber \\&=\{\qsetp[1], \qsetp[2], \qsetp[3]  \} 
\end{align}
For each $\qsetp\in \bqsetpibz$,  all distinct $\qsetp[i]$ that are generated from point operations form a star,
denoted as $\orbit[\qsetp]$ (where $1\le|\orbit|\le\pgorder$). 
\begin{align}\label{}
\orbit[\qsetp[1]]&=\{\qsetp[1]\}
&&
\orbit[\qsetp[2]]=\{\qsetp[2]\}
&&
\orbit[\qsetp[3]]=\{\qsetp[3], \qsetp[4]\}
\end{align}
The set  composed of all such stars is denoted $\borbit[BZ]$ (where $|\borbit[BZ]|=|\bqsetpibz|$).

\subsection{Point symmetry including the basis}

Having accounted for translation, permutation, and point symmetry of the order $\order$ identity representations of the pure lattice,
point symmetry of the atoms and their corresponding displacements vectors must now be incorporated. 
First, one must symmetrize the $\natom\npol$ displacements $\{\qdispvec\}$ for all $\qq\in\bqq[IBZ]$ according the little group of each respective $\qq$; as these are
the building blocks for a given $\qq\in\qset$. The representation of displacements at a given $\qq$ is then 
given as $\hat \Gamma(\dispsym_{\qq})=\bigoplus_\alpha \hat \Gamma_\alpha$, where $\alpha$ labels an irreducible representation; and the irreducible representation labels can be stored in the set $\tilde \Gamma(\dispsym_{\qq})$.
For the case of $\qq=M_x$, for example, we have the following six displacement amplitudes:
\begin{align}\label{}
\hat \Gamma(\dispsym_{M_x})=A_{1g}\oplus B_{2g}\oplus B_{3g}\oplus B_{1u}\oplus B_{2u}\oplus B_{3u}
\end{align}
We tabulate the explicit form of all symmetrized displacements of graphene, for an arbitrary $\qq\in\bqq[IBZ]$, in Supplementary Information, Table S\ref{si:table:qbasis}. 
All point group conventions in this study follow Cornwell\cite{Cornwell}.

Given some $\qsetp\in\bqsetpibz$ at order $\order$, where $\qsetp=\{\qq[1]\dots\qq[\order]\}$,
the task at hand is to determine if a given derivative with respect to
$u_{\qq[1]}^{\alpha_1}\dots u_{\qq[\order]}^{\alpha_{\order}}$, where
$\alpha_i$ is an irreducible representation of the
little group of $\qq[i]$,  is symmetry allowed;
and if so, to determine how many irreducible derivatives it yields (in the case where  multidimensional irreducible representations are present in the set).
Each displacement $u_{\qq[i]}^{\alpha_i}$ will be associated with a set of star displacement vectors $\{\dispsym_{\qorbit[\qq[i]]}^{\beta_j} : j\in[1,|\qorbit[\qq[i]]|]\}$,
which form full space group irreducible representations\cite{Birman19621093}.
Therefore, existence of derivatives with respect to $u_{\qq[1]}^{\alpha_1}\dots u_{\qq[\order]}^{\alpha_{\order}}$ can be determined 
from evaluating the corresponding derivatives with respect to the stars. Group theoretically, one is left with the problem of forming symmetric
direct products\cite{Zhou1989935,Lyubarskii19601483212556} of a set of stars\cite{Cracknell19790306651750,Birman1974354013395X}.
As discussed in Section \ref{sec:irrderivs}, this is a solved problem, though it is still nontrivial to execute at arbitrary order $\order$, as we have. 
Explicit results are illustrated for graphene (see Table \ref{table:irrD} and S\ref{si:table:irr_deriv_n3}) and rock salt (see Appendix \ref{app:rocksalt}).

We now define relevant variables to count the total number of identity representations.
The resulting number of identity representations for a given $\qsetp$ is denoted $\nirrep[\qsetp]$, and $\nirrep[\qsetp]=\nirrep[\orbit[\qsetp]]$ for all $\qsetp\in\orbit$. 
The total number of irreducible derivatives
can then be found as:
\begin{align}\label{}
\nirreptotal[\supa[BZ]] =
\sum_{\qsetp\in\bqsetp[IBZ]} \nirrep[\qsetp]
=
\sum_{\orbit\in\borbit[BZ]} \nirrep[\orbit]
\end{align}

A final point is that time reversal symmetry can be employed in conjunction
with space group symmetry to determine if space group irreducible derivatives
can have a phase convention which ensures that they are purely real numbers (or
purely imaginary); and this is realized in all applications in this paper.

\subsection{Homogeneity and isotropy of space}
In addition to space group symmetry and permutation of derivative indices, the potential will also conserve total linear and angular momentum. The former implies that
an arbitrary shift of the system will leave all derivatives of the Born-Oppenheimer surface invariant\cite{BornHuang,Leibfried1961275}; and this is referred to as the acoustic sum
rule in the context of a Taylor series in the real space basis. The acoustic
sum rules can be quite challenging for real space Taylor series approaches
to enforce\cite{Plata201745,Li20141747,Giannozzi2009395502}.
To the contrary, when working with space group irreducible derivatives, and even
simply translation group irreducible derivatives, the acoustic sum rules are
automatically satisfied to all order by construction. Moreover, each irreducible derivative will individually satisfy
the acoustic sum rule, and therefore the acoustic sum rule does not redistribute error among different irreducible derivatives.
This is true for any translation group (see Eq. \ref{eq:supa}), irrespective of its size.
The only care that is needed occurs when the acoustic modes, at the $\Gamma$ point, are a repeating irreducible representation,
and then one should ensure that they are orthogonalized to the modes of the same symmetry; which is trivial to enforce by construction.
Given that space group irreducible derivatives are invariant to supercell size, and that the acoustic sum rules are automatically satisfied,
there are major incentives to work purely with space group irreducible derivatives.

In the case of conservation of total angular momentum, an arbitrary global
rotation will leave the potential unchanged; and enforcing this in the limit of small rotations will link a given order
of real space derivatives to infinite range, in addition to linking them to the next highest order\cite{Leibfried1961275}. 
However, this does not impart any constraints on the space group irreducible derivatives within a FTG, given that the basis of the FTG does not
describe pure rotation. However, the constraint may be placed within the method of Fourier interpolation (see Section \ref{sec:FI}), which 
interpolates the irreducible derivatives to the infinite lattice; here, free infinitesimal rotation can be enforced. In summary, isotropy of free space is 
not a consideration when extracting space group irreducible derivatives.

\subsection{Taylor series of $V$ in symmetrized variables}
Having accounted for all symmetries, 
we are now in a position to write the Taylor series purely in terms of space group irreducible derivatives.
We will label a given irreducible derivative at order $\order$ as ${}^{j}\dqnirrsym\indices*{*_{\qq[1]}^{\alpha_1}_{\cdots}^{\cdots}_{\qq[\order]}^{\alpha_\order}}$,
where $\qq[i]\in\qsetp$ and $\alpha_i\in\tilde\Gamma(\dispsym_{\qq[i]})$ and $j$ labels repeating instances of a particular identity representation. A given 
derivative of the Born-Oppenheimer surface can be written in terms of the irreducible derivative as:
\begin{align}
\frac{\partial^\order V }{\prod_i^\order\partial\qdispvec[\qq[i]][\alpha_i][a_i] }
&= 
\sum_j {}^{j}\hspace{-0.5mm}\theta_{a_1\dots a_\order}^{\alpha_1\dots\alpha_\order}[\qset]
\hspace{0.5mm}{}^{j}\hspace*{-0.7mm}\dqnirrsym_{\qq[1]\dots\qq[\order]}^{\alpha_1\dots\alpha_\order}
\nonumber \\&
=
\dqnsym_{\qset}^{(\alpha_1,a_1)\dots(\alpha_\order,a_\order)}
\end{align}
where $\qq[i]\in(\qset)$, $a_i$ is a given row of the $\alpha_i$ irreducible representation, 
$\theta_{a_1\dots a_\order}^{\alpha_1\dots\alpha_\order}[\qset]$ are the Clebsch-Gordon (CG) coefficients of the direct product representation, 
the left superscript $j$ is a label for repeating instances of a given irreducible derivative,
and the symbol $\dqnsym$ is used for the derivative of the potential with 
respect to irreducible representations of the displacements. 
The distinction between $\dqnsym$ and $\dqnirrsym$ should be appreciated, as
the latter only depends on irreducible representations and \emph{not} the rows
of the irreducible representations. Our convention for the CG coefficients is to start with the normalized product representation,
and then rescale by $\sqrt n$ where $n$ is the smallest positive integer that produces the smallest number of radical CG coefficients.

The Taylor series of the potential energy, per unit cell, is then written for a given FTG and order as:
\begin{align}\label{}
V^{(\order)}_{\supa[BZ]} &=\frac{1}{\order!} \sum_{\qset\in\bqset[BZ]}\sum_{\substack{\alpha_1\dots\alpha_\order\\[0.3em]a_1\dots a_\order}} 
\dqnsym_{\qset}^{(\alpha_1,a_1)\dots(\alpha_\order,a_\order)}
\prod_{i=1}^\order \qdispvec[\qq[i]][a_i][\alpha_i]
\nonumber \\
&=\frac{1}{\order!}
\sum_{\substack{\qsetp[i] \in \bqsetp[IBZ]\\[0.3em]\alpha_1\dots\alpha_\order,j}}
\hspace{-4mm}
^{j}\hspace{-0.7mm}\dqnirrsym_{\qsetp[i]}^{\alpha_1\dots\alpha_\order}
\hspace*{-3mm}
\sum_{\qset \in \orbit[\qset[i]]} 
\hspace*{-1mm}
\sum_{a_1\dots a_\order } 
\hspace*{-2mm}
{}^{j}\hspace{-0.5mm}\theta_{a_1\dots a_\order}^{\alpha_1\dots\alpha_\order}[\qset]
\prod_{i=1}^\order \qdispvec[\qq[i]][a_i][\alpha_i]
\end{align}
For the specific case of in-plane displacements in graphene at $\order=3$ with $\supa[BZ]=\supa[K]$, we have:
\begin{widetext}
\begin{align}\label{eq:potKinplane}
&V^{(3)}_{\supa[K]} = 
\frac{1}{6}\dqnirrsym\indices*{*_\Gamma^{E_2}_\Gamma^{E_2}_\Gamma^{E_2}} 
\left(3 u_\Gamma^{E^0_2}u_\Gamma^{E^0_2}u_\Gamma^{E^1_2} - u_\Gamma^{E^1_2}u_\Gamma^{E^1_2}u_\Gamma^{E^1_2} \right)
+
\dqnirrsym\indices*{*_\Gamma^E _{\bar K}^E_K^E} 
  \left( u_\Gamma^{E^0}u_{\bar K}^{E^0}u_{ K}^{E^1}  +   u_\Gamma^{E^0}u_{\bar K}^{E^1}u_K^{E^0} +  
    u_\Gamma^{E^1}u_{\bar K}^{E^0}u_{ K}^{E^0}  -   u_\Gamma^{E^1}u_{\bar K}^{E^1}u_{ K}^{E^1}  \right)  
+ \nonumber \\[0.5em]& 
\dqnirrsym\indices*{*_\Gamma^E _{\bar K}^E_K^{A_1}}
  \left( u_{K}^{A_1} (u_{\Gamma}^{E^0}u_{\bar K}^{E^0}+u_{\Gamma}^{E^1}u_{\bar K}^{E^1})  + \textrm{cc} \right)
+
\dqnirrsym\indices*{*_\Gamma^E _{\bar K}^E_K^{A_2}}
  \left( u_{K}^{A_2} (u_{\Gamma}^{E^0}u_{\bar K}^{E^1}-u_{\Gamma}^{E^1}u_{\bar K}^{E^0}) + \textrm{cc}\right)
+ 
\frac{1}{6}\dqnirrsym\indices*{*_K^{A_1}_K^{A_1}_K^{A_1}} \left( u_{K}^{A_1} u_{K}^{A_1}u_{K}^{A_1} + \textrm{cc}\right)
+ \nonumber \\[0.5em]& 
\frac{1}{6}\dqnirrsym\indices*{*_K^{A_1}_K^{A_2}_K^{A_2}} \left(  u_{K}^{A_1} u_{K}^{A_2}u_{K}^{A_2} + \textrm{cc} \right)
+ 
\frac{1}{2} \dqnirrsym\indices*{*_K^E _{ K}^E_K^{A_1}}
  \left( u_{K}^{A_1} (u_{K}^{E^0}u_{ K}^{E^0}+u_{K}^{E^1}u_{ K}^{E^1}) +\textrm{cc} \right)
+ 
\frac{1}{6}\dqnirrsym\indices*{*_K^{E}_K^{E}_K^{E}}   
\left(
(3 u_K^{E^0}u_K^{E^0}u_K^{E^1} - u_K^{E^1}u_K^{E^1}u_K^{E^1})+\textrm{cc}
\right)
\end{align}
\end{widetext}
where cc indicates the complex conjugate of the preceding term, superscripts of irreducible representations indicate a given row of a multidimensional irreducible representation, and we have used $C_{3v}$ labels for the little group of $K$ for convenience; as opposed to 
$D_{3h}$, which is needed when including out-of-plane displacements. 
The values of the above derivatives can be found in Table \ref{table:irrD}, and the approaches to computing them are discussed in Section \ref{sec:finitedisp}.
We emphasize that to third order, \emph{any possible} in-plane displacement within $\supa[K]$ is purely characterized by the eight real irreducible derivatives shown in Eq. \ref{eq:potKinplane} in addition to the four in-plane irreducible derivatives at second order (see Table \ref{table:irrD}).

\subsection{Fourier Interpolation}
\label{sec:FI}
Given a set of irreducible derivatives defined over some FTG, one may interpolate to a different FTG or the infinite lattice; and
this can be achieved using Fourier Interpolation (FI)\cite{Giannozzi19917231,Parlinski19974063}. 
Such
trigonometric interpolations have a long history in physics, dating back to the
beginning of classical mechanics\cite{Heideman198414}.  
We emphasize that FI is not unique, and
one could supply additional information, such as the elastic constants, to improve the FI.
Beyond second order, 
the only description of FI we are aware of is the treatment of third order in Ref. \onlinecite{Paulatto2013214303}. 
In this work, we need a FI scheme for arbitrary order,
and therefore we implement the most straightforward generalization of the usual FI at second order\cite{Parlinski19974063,Paulatto2013214303};
which amounts repacking the force tensor into the Wigner-Seitz cell. 

Here we outline
the various steps in our FI approach.
First, the dynamical tensor needs to be rotated to a common basis at each $\qset\in\bqset[BZ]$,
which is chosen as the naive basis labeled by each atom and cartesian displacement:
\begin{align}\label{}
\dqnsym_{\qset}^{'i_1,\dots,i_{\order}} = \sum_{\ell_1,\dots,\ell_{\order}}\prod_{j=1}^\order  U^{i_j\ell_j}_{\qq[j]} \dqnsym_{\qset}^{\ell_1,\dots,\ell_{\order}}
\end{align}
where $\hat U_{\qq}$ are the matrices that transform from the symmetrized basis under the little group 
of $\qq$ to the naive basis (provided for graphene in Supplementary Material, Table S\ref{si:table:qbasis}), and
the index $i_j$ is  a two tuple containing both an atom and displacement label, while $\ell_j$ labels an irreducible representation of the little group of $\qq[j]$.
Subsequently, the dynamical tensor can be Fourier transformed to obtain the force tensor:
    \begin{align}\label{}
      \phin&=\frac{1}{\nq^{\order-1}}\sum_{\qset \in \bqset[BZ]} \dqn'
      e^{\imag 2\pi \mathrm{Tr}\left(\qset\cdot\tset^\tp\right) }
    \end{align}
where $\ctrans[i]\in(\tset)$ and $\tset\in\btset[BZ]$.
At this point, $\{\phin|\tset\in\btset[BZ]\}$ can then be used to predict $\dqn$ at an arbitrary $\qset$ point.  
However, such an interpolation does not guarantee point symmetry for $\qset\notin\bqset[BZ]$, and therefore an additional transformation is needed.
The basic approach is to repack $\phin$, defined over $\btset[BZ]$, into the corresponding Wigner-Seitz cell. To do so, a map $\mathcal{M}^{a_1\cdots a_\order}_{\tset}$,
where $a_i$ label one of the $\natom$ basis atoms in the primitive unit cell, must be created from the translation points
$\btset[BZ]^{WS}$ defined over the WS BvK supercell to the conventional BvK supercell $\btset[BZ]$. 

\begin{figure}[htbp]
\begin{center}
\includegraphics[width=1.0\linewidth]{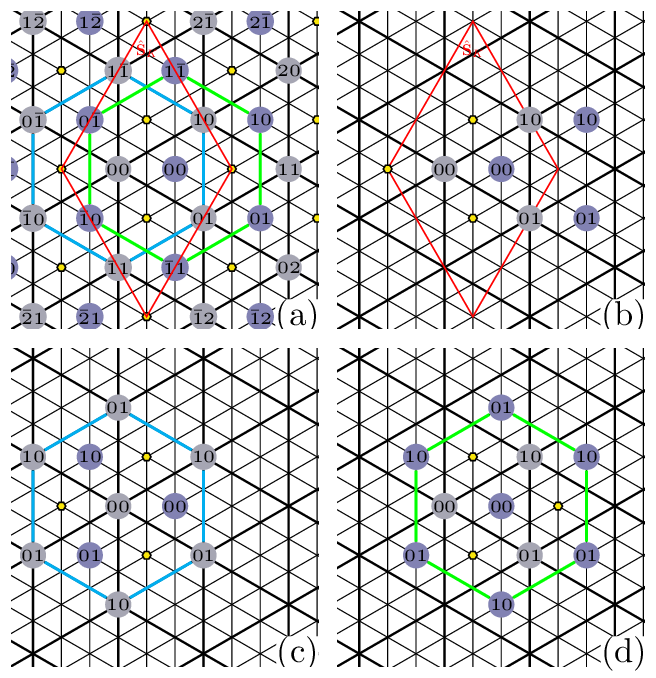}
\end{center}
\caption{
  (a) Schematic of the graphene crystal structure, where yellow hexagons represent lattice points and circles represent carbon atoms; and each
  carbon atom is labeled by two integers which correspond to a translation in lattice coordinates. The $\supa[K]$ supercell is shown in red, and the
  corresponding WS cell is shown in  blue and green for a centering on the first and second carbon atom, respectively. 
  (b) Schematic showing $\ctrans\in\bctrans[BZ]$ for $\supa[BZ]=\supa[K]$ along with the corresponding basis atoms.
  (c) Schematic showing how the basis atoms are translated back into the WS cell using some vector $\vec t\supa[K]\lmat $, with $\vec t\in\mathbb{Z}^d$,
  where the centering of the WS cell is on the first carbon atom. (d) Same as (c) but with the WS cell centered on the second carbon atom.
}
  \label{fig:wscell}
\end{figure}

In order to build $\mathcal{M}^{a_1\cdots a_\order}_{\tset}$, we begin by building $m^{ij}_{\ctrans}$,
which is the corresponding map from $\bctrans[BZ]$ to $\bctrans[BZ]^{WS}$. The process of deducing this map is related to finding the Wigner-Seitz cell associated
with $\supa[BZ]$, and this is illustrated
in the case of $\supa[BZ]=\supa[K]$ in graphene (see Figure \ref{fig:wscell}). Figure \ref{fig:wscell}a
contains a schematic of the graphene lattice, with each basis atom labeled by the translation, in lattice coordinates, of the infinite lattice.
The FTG $\supa[K]$ is illustrated in red, while the corresponding WS cell is illustrated in blue and green for centerings on the first
and second carbon atom, respectively. Figure \ref{fig:wscell}b retains only the six carbon atoms associated with $\supa[K]$, and the 
task is to shift all of these atoms by any translation $\{\supa[K]\ctrans | \ctrans \in  \mathbb{Z}^d\}$ that maps the atom
into or onto the boundary of the WS cell; and each atom may be shifted by more than one translation.
Figures \ref{fig:wscell}c-d show the result of this for the two different WS cells, and the resulting map can be deduced by comparing
to Figure \ref{fig:wscell}a:
\begin{align}\label{}
m^{i,i}_{\vec 0}&=\{\vec 0\}
&
m^{i,j}_{\vec 0}&=\{\vec 0\}
\nonumber \\
m^{i,i}_{\ctrans[a]}&=\{\ctrans[a],(0,\bar 1),(\bar 1 ,1)\}
&
m^{0,1}_{\ctrans[a]}&=\{(0,\bar 1)\}
\nonumber \\
m^{i,i}_{\ctrans[b]}&=\{\ctrans[b],(\bar 1,0),(1,\bar 1)\}
&
m^{0,1}_{\ctrans[b]}&=\{(\bar 1,0)\}
\nonumber \\
m^{1,0}_{\ctrans[b]}&=\{\ctrans[b]\}
&
m^{1,0}_{\ctrans[a]}&=\{\ctrans[a]\}
\end{align}
Clearly, $m^{i,i}_{\ctrans}$ is purely a property of the lattice, with $\bctrans[BZ]^{WS}=\bigcup_{\ctrans\in\bctrans[bz]}m^{i,i}_{\ctrans}$,  while $m^{i,j}_{\ctrans}$ ($i\ne j$) will depend on the relative positions of the basis atoms.
Having deduced $\{m^{i,j}_{\ctrans}| \ctrans \in\bctrans[BZ]\}$, any element $\mathcal{M}^{a_1\cdots a_\order}_{\tset}$ can now straightforwardly be constructed at an arbitrary
$\order$ within $\supa[K]$.
For example, at $\order=3$ one case is:
\begin{align}\label{}
\mathcal{M}^{0,0,1}_{(\vec 0,\ctrans[a],\ctrans[b])}=\{ &
  (\vec 0,\ctrans[a],(\bar 1, 0)), 
  \nonumber \\
    & %
  (\vec 0,(0,\bar 1),(\bar 1, 0)),
  (\vec 0,(\bar 1,1),(\bar 1, 0))
  \}
\end{align}

Once the map is obtained, $\phin^{WS}$ can be constructed:
\begin{align}\label{}
  \phin[\tset']^{WS,(a_1,\alpha_1)\cdots(a_\order,\alpha_\order)} = 
  |\mathcal{M}^{a_1\cdots a_\order}_{\tset}|^{-1}
  \phin[\tset]^{(a_1,\alpha_1)\cdots(a_\order,\alpha_\order)}
\end{align}
where $\tset'\in\mathcal{M}^{a_1\cdots a_\order}_{\tset}$. Finally, an arbitrary $\qset$ can be constructed as
\begin{align}
  \dqn=\sum_{\tset \in \btset[BZ]^{WS}} \phin^{WS}
      e^{-\imag 2\pi \mathrm{Tr}\left(\qset\cdot\tset^\tp\right)}
\end{align}
This procedure has been straightforwardly executed on graphene up to $\order=5$ (see Section \ref{sec:assessing}).

\begin{figure}[htbp]
\begin{center}
\includegraphics[width=1.0\linewidth]{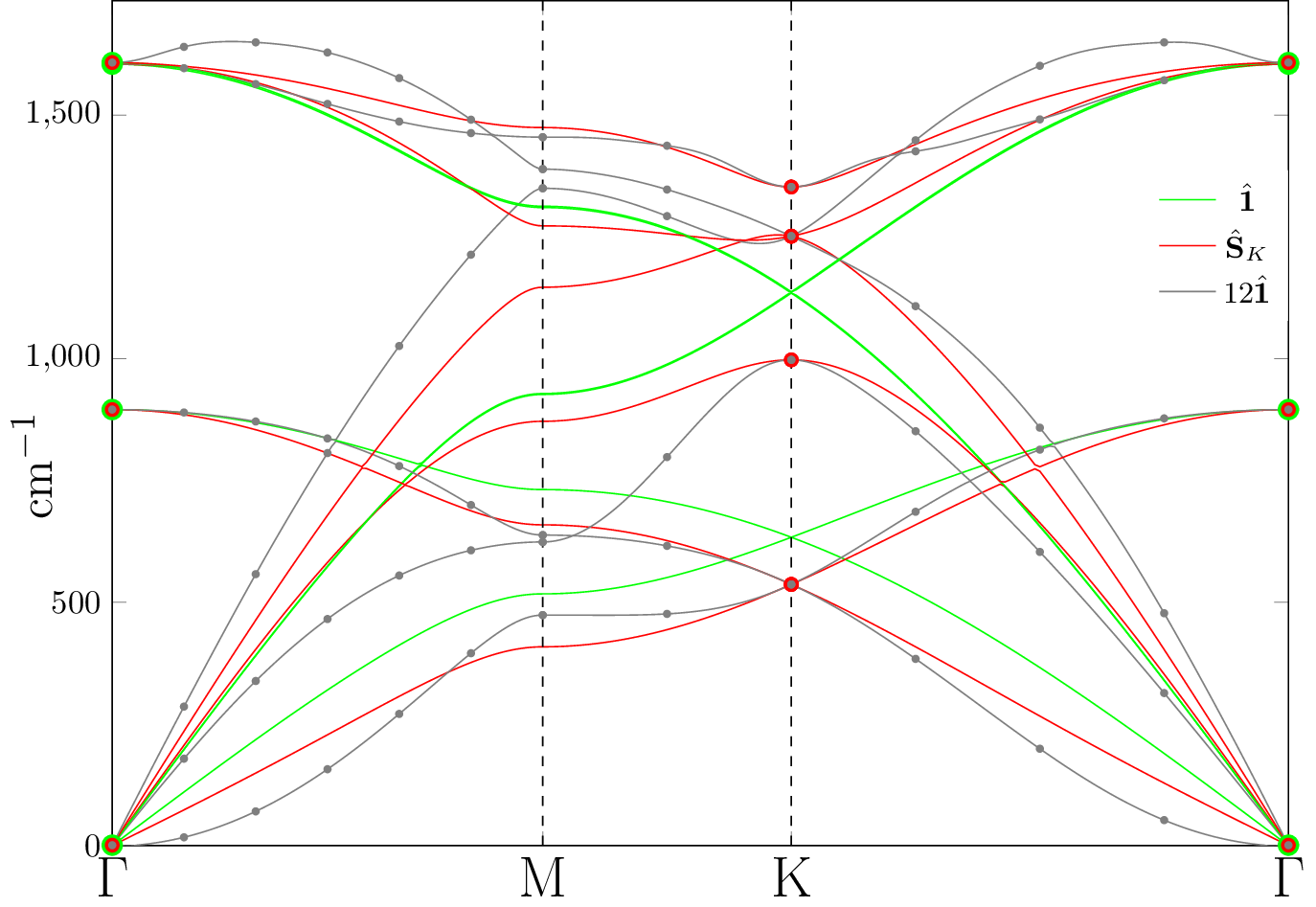}
\end{center}
\caption{
  Phonons of graphene within DFT for $\supa[BZ]=\hat 1$, $\supa[K]$, and $12\hat 1$, where data points are direct computational measurements and lines are Fourier interpolation. 
  The irreducible derivatives for $\hat 1$ and $\supa[K]$ are shown in Table \ref{table:irrD}; and the corresponding frequencies are obtained, in units of s$^{-1}$,
  as $\omega_{\qq}^{\alpha}=\sqrt{\dqnirrsym\indices*{*_{\bar \qq}^{\alpha}_{\qq}^{\alpha}}/m}$, where $m=12.011\cdot 1.0364\times 10^{-28}\textrm{eV}\cdot \textrm{s}^2/\AA^2 $ for carbon.
  The $y$-axis plots $\omega_{\qq}^{\alpha}/(2\pi c)$, where $c$ is the speed of light in units of cm/s.
}
  \label{fig:phonons}
\end{figure}

As an illustration, we provide the Fourier Interpolation of graphene at second order for $\supa[BZ]=\hat 1$, $\supa[BZ]=\supa[K]$, and $\supa[BZ]=12\hat 1$
(see Figure \ref{fig:phonons}). The data points denote frequencies at specific $\qq$ which result from direct measurement, while the lines
are the result of the Fourier interpolation, and it is clear that all symmetries are satisfied. We emphasize that only the data points are robust, and
the lines are only reliable for a sufficiently large FTG. Given that there are no repeating irreducible representation for $\supa[BZ]=\hat 1$ and $\supa[BZ]=\supa[K]$,
the phonon frequencies at the irreducible representations of the FTG can be obtained without any matrix diagonalization (see caption of Figure \ref{fig:phonons}).

\label{sec:fi}
\section{Finite Displacement Methodology}
\label{sec:finitedisp}
\subsection{Statement of problem}
Having developed a Taylor series purely in terms of space group irreducible derivatives at
order $\order$, we now turn to the problem of how to compute these derivatives
using finite displacements; while exploiting perturbative derivatives (e.g. Hellman-Feynman forces) up to order $n$,
with  $n<\order$, that the first-principles approach may provide.
We refer to the order $n$ perturbative derivatives as $\pd[n]$.

Generically speaking, we define a finite displacement method as any method which explicitly moves the nuclei and fully computes the electronic structure.
There are now many techniques which use a first-principles molecular dynamics trajectory as a source of data from which to fit\cite{Hellman2013144301,Zhang2014058501,Zhou2014185501},
and this would fall under the category of a finite displacement approach. Furthermore, those approaches extracting third order derivatives from a molecular dynamics trajectory
could obviously exploit our Hierarchical Supercell approach outlined in Section \ref{sec:hierarchy}, though we do not pursue such a program in this work because
we believe fitting tens to thousands of parameters simultaneously should always be a method of last resort. Instead, we seek to use 
central finite difference, where the only simultaneous fitting involved is that of a quadratic function which has two parameters, and 
order $\order$ derivatives are isolated from all other orders.

We define
two finite difference based approaches at competing extremes: the lone irreducible derivative (LID)
and  the bundled irreducible derivative (BID) approach.  The LID approach
measures the smallest possible number of irreducible derivatives simultaneously, sacrificing 
efficiency for accuracy, while BID simultaneously measures the maximum number of irreducible 
derivatives that the perturbative derivatives will allow, prioritizing efficiency over accuracy. A spectrum possibilities exists between these two approaches,
though we focus on these two extremes. Both LID and BID can be executed in a single-supercell approach, performing all calculations within
the BvK supercell $\supa[BZ]$, or a hierarchical supercell approach; whereby all irreducible derivatives are measured in the smallest supercell allowed by group theory.
We proceed by first outlining how to derive the smallest supercell that will accommodate an arbitrary set of $\order$ waves $\qq[i]\in\qset[bz]$.
We emphasize that this question is generic to any sort of waves within the lattice.

\subsection{ Minimal supercell problem}
\label{sec:hierarchy}
The following unresolved problem is of utmost importance in any finite displacement approach:  given $\qset$, find the smallest 
possible supercell, denoted $\supa[\qset]$, that accommodates all $\order$ vectors $\qq\in(\qset)$. 
Mathematically, we demand that all $\qq\in(\qset)$ are identity representations of the supercell, $\qset\supa[\qset]^\tp \in\mathbb{Z}^{(\order,\cdim)}$, with
the constraint that $|\det(\supa[\qset])|$ is a minimum. 
Recall that $\sum_{\qq\in(\qset)}\qq\in\mathbb{Z}^\cdim$, which demands that a supercell which accommodates any $\order-1$ of the $\qq\in(\qset)$ 
will automatically accommodate the remaining $\qq$. Therefore, we are free to remove any row in $\qset$.
Furthermore, it is useful to express
$\qset$ in units of $\rmat[BZ]$, so we
we define a new matrix $\qset[][']$, which is obtained by removing any row from $\qset$ and converting to lattice coordinates of $\rmat[BZ]$.
The matrix $\qset[][']$ is a $(\order-1)\times\cdim$ matrix of integers, and the commensuration requirement becomes $\qset[][']\supa[\qset]^\tp  (\textrm{mod} ~\lcmd)= \vec 0$,
where $\lcmd$ is the least common denominator for all components of $\qq\in(\qset)$ and $\vec 0$
is a $(\order-1)\times\cdim$ dimensional zero matrix.
Finally, the mathematical requirement for a valid $\supa[\qset]$ is 
\begin{align}\label{}
\supa[\qset]\in \argmin_{\hat n \in \mathbb{Z}^{(d,d)}}\{ |\hat n| : \qset[][']\hat{n}^\tp  (\textrm{mod} ~\lcmd)= \vec 0, |\hat n|\ge 1  \}
\end{align}
Performing this minimization is achieved by constructing the modulo $\lcmd$
kernel of $\qset[][']$, which is obtained by bringing $\qset[][']$ into Smith Normal Form (SNF)\cite{Norman20121447127293}, denoted $\snf$;
and this is achieved via elementary row and column operations:
\begin{align}\label{}
  \snf = \hat R\qset[]['] \hat C
\end{align}
where $\snf$ is a $(\order-1) \times \cdim$ diagonal matrix of integers, 
$\hat R$ is a $(\order-1)\times (\order-1)$ unimodular matrix  of integers obtained from  a sequence
of elementary row transformations, and $\hat C$ is a $\cdim\times\cdim$ unimodular
matrix of integers obtained from a  sequence of elementary column transformations.

The modulo $\lcmd$ kernel of $\snf$ can be formed as
\begin{align}\label{eq:kernn}
\ker(\snf)_i=\frac{\lcmd }{\gcd (\lcmd,G_{ii})} \vec e_i
  &&  &&
  \begin{matrix}
    G_{ii}= \snfsym_{ii}  ~ \textrm{if}~  i\le\order-1 
    \\[0.5em]
    G_{ii}= \lcmd ~ \textrm{if}~  i>\order-1 
  \end{matrix}
\end{align}
where $\vec e_i$ is a unit vector in $\mathbb{Z}^\cdim$. Therefore, we have
\begin{align}\label{}
 \hat R\qset[]['] \hat C \ker(\snf)_i (\textrm{mod} ~\lcmd) =\vec 0 
\end{align}
Finally, we can then define a given basis vector of the kernel of $\qset[][']$ and the resulting supercell
\begin{align}\label{}
\ker(\qset[]['])_i = \hat C \ker(\snf)_i && \supa[\qset]=\ker(\qset[]['])^\tp 
\end{align}
though it should be emphasized that this supercell is not unique and may be reshaped.
Most importantly, the multiplicity of the supercell is 
\begin{align}\label{eq:ssmult}
\det(\supa[\qset]) =  \frac{\lcmd^\cdim}{\prod_{i=1}^\cdim\gcd (\lcmd,G_{ii})} %
\end{align}
We refer to Eq. \ref{eq:ssmult} as the Minimum Supercell Multiplicity (MSM) equation. Given that calculating the  Smith Normal Form
is computationally inexpensive for $\cdim\le3$ at any realistic $\order$, the MSM equation can be efficiently evaluated.
In order to clearly illustrate this approach, a worked example is provided in Appendix \ref{app:SNF}.

Under certain restrictions, the largest necessary supercell multiplicity of a given FTG can be determined from Eq. \ref{eq:ssmult} \emph{a priori}.
For any FTG of an arbitrary $\cdim$-dimensional crystal  at $\order=2$ (i.e. phonons), in addition to FTG's corresponding to $\supa[BZ]=n\hat1$, where $n\in\mathbb{Z}^+$, %
at arbitrary order $\order$,
the largest necessary supercell multiplicity is $\lcmdm^{\min(\order-1,\cdim)}$.
Restated in equations, we have
\begin{align}\label{eq:msmt}
\order=2 \vee \supa[bz]=n\hat1 
\Rightarrow 
\max_{\qset\in\bqset[BZ]}|\supa[\qset]|= \lcmdm^{\min(\order-1,\cdim)}
\end{align}
The above can be proven in two parts. 
For $\order=2$, 
an arbitrary $\qq=(q_1/\lcmdm,q_2/\lcmdm,\dots,q_\cdim/\lcmdm)$, where $0\le q_i <\lcmdm$,
and we have $N_{11}=\gcd(q_1,\dots,q_\cdim)$.
Since $\qset[][']$ is a single row, 
the multiplicity is %
\begin{align}\label{}
\det(\supa[\qset])=\frac{\lcmdm}{\gcd (\lcmdm,N_{11})} =\frac{\lcmdm}{\gcd (\lcmdm,q_1,\dots,q_\cdim)} 
\end{align}
Given that the minimum of the denominator is 1, the maximum multiplicity is $\lcmdm$.
For the case of $\supa[BZ]=n\hat1$ at arbitrary $\order$, we have $\lcmdm=n$ and therefore 
$\qq[i]=(q_{i,1}/n,\dots,q_{i,\cdim}/n)$; where $\forall q_{i,j}\in[0,n-1]$. Then, we have
\begin{align}
\det(\supa[\qset]) =  \frac{n^{\min(\order-1,\cdim)}}{\prod_{i=1}^{\min(\order-1,\cdim)}\gcd (n,G_{ii})} %
\end{align}
The worst case is $\gcd (n,G_{ii})=1$, yielding a  maximum multiplicity  $n^{\min(\order-1,\cdim)}$.

Eq. \ref{eq:msmt} has far reaching implications which should be appreciated.
For typical materials systems (i.e. $\cdim=1,2,3$), Eq. \ref{eq:msmt} dictates that 
phonons can always be obtained from a collection of supercells of multiplicity $\lcmdm$, as was only recently
realized\cite{Lloyd-williams2015184301}. Moreover, for three dimensional materials with FTG $\supa[BZ]=n\hat1$, 
cubic terms can always be obtained from a collection of supercells of maximum multiplicity $n^2$, proving
that the BvK supercell can always be avoided for cubic terms in this common scenario.

\subsection{Central finite difference}
\label{sec:cfd}
Central finite difference (CFD) is the method of choice in this study for computing an arbitrary derivative. 
The main virtue of CFD is that the error is a quadratic function of the discretization parameter $\Delta$.
A given derivative
is  the intercept  of the following function (the indices of $\rdispvec$, $\qdispvec$ are compressed to $\rdisp[i]$ for brevity):
 \begin{align}\label{finitediff}
   \nonumber
\pot_{\rdisp[1]\dots\rdisp[\order]}(\Delta) &=
   \sum_{n_{1},\cdots,n_{\order}=(-1,1)} \frac{ \left(\prod_{i=1}^{\order} n_{i}\right) \pot(\{n_i\Delta\})}%
   {2^\order\Delta^\order  } \\
  & =  \frac{\partial^\order \pot}{\prod_{i=1}^{\order} \partial \rdisp[i] } + O( \Delta^2) + \cdots
 \end{align}
where $\order$ is the order of the derivative and $\Delta$ is a positive real number. 
Higher order derivatives of a
given variable are obtained by repeating the same variable. 
A given $\Delta$ for an order $\order$ derivative will require up to $2^\order$ evaluations of $\pot$.
The intercept of the Eq. \ref{finitediff} gives the value of the derivative,
and CFD 
guarantees that the leading order correction of an order $\order$ derivative is comprised of the order $\order+2$ derivatives;
which dictate the strength of the quadratic error tail (see Ref. \onlinecite{Kornbluth2017} for additional details).
Every evaluation of $\pot$ requires the numerical solution of a differential
equation (e.g. Kohn-Sham equation of DFT) which is subject to it's own
discretization errors (e.g. plane-wave cutoff, etc). Therefore, for
sufficiently small $\Delta$, the finite difference will be dominated by errors;
while if $\Delta$ is too large, then the results will be beyond the quadratic
regime. One needs to ensure that the quadratic regime is obtained such that a
valid extrapolation $\Delta\rightarrow 0$ can be obtained: a practical but
essential point.  
We will demonstrate that this quadratic extrapolation can typically be achieved
even for fourth order derivatives within DFT (i.e. fifth derivatives of the energy if the forces are being used). 

Choosing the discretization grid is an interesting optimization problem in its own right, and we aim for simplicity
in this work; given that the current status quo at second and even sometimes third order is simply choosing a single delta based on experience.
At least three $\Delta$ would be needed to compute an error associated with fitting a quadratic.
In this work, we typically compute up to fifteen $\Delta$ for a given derivative, which is normally excessive, but it allowed for the testing of various schemes for optimizing the quadratic fit.
Typical ranges of $\Delta$ for force derivatives were $\Delta=0.005-0.05\AA$ for first order; $\Delta=0.01-0.1\AA$ for second order; and $\Delta=0.01-0.15\AA$ for third and fourth order.
Given $\pot_{\rdisp[1]\dots\rdisp[\order]}(\Delta)$ evaluated over some set of $N$ different $\Delta$, we need to choose which points to 
use in the least squares fit of the quadratic error tail. To do so, we construct the least squares fit for all sets of $\Delta$ obtained from
choosing $n$ from $N$, where $n\in[4,N]$. Clearly, the smallest number of points will always deliver the smallest error, so we choose our metric to be the standard
error of the fit divided by the number of points used in the fit. We reiterate that there are many different schemes one can choose, and 
in some situations it will suffice to choose a single $\Delta$, such as most first order derivatives, but it is difficult to know \emph{a priori}. 
An illustration of the result of choosing the quadratic error tail can be seen in Figure \ref{fig:centralfd}, which will be discussed in Section \ref{sec:LID}.

Hereafter, we refer to the determination of a given derivative via finite difference as a single ``measurement",
and this should not be confused with a single calculation; as the number of calculations is determined by the number
of $\Delta$ one chooses. Given that different practitioners will choose different numbers of $\Delta$, the number
of measurements is what should be compared when contrasting different methods of extracting all derivatives. 
Finally, it should be noted that the cost of obtaining $n$ distinct $\Delta$ may be
considered to be far less than performing $n$ calculations, given that the wave function of the $(n-1)$-th $\Delta$ can be used to
seed the $n$-th $\Delta$ at a great reduction in computational cost; and we exploit this.

\subsection{Individually resolving irreducible derivatives: lone irreducible derivative approach }
\label{sec:LID}
The first procedure we outline involves measuring a single irreducible
derivative at a time, or as few as group theoretically possible, which we call
the lone irreducible derivative (LID) approach. This approach encompasses the
original frozen phonon approach\cite{Wendel1978950}, but we apply it under the most general conditions. 
We emphasize that LID specifically refers to irreducible derivatives of the space
group, and not simply irreducible derivatives of the translation group. 

If the first-principles method to evaluate $\pot$ does not provide any perturbative derivatives, then LID is 
a natural choice. While any complete basis can be employed at the same cost, 
directly probing a given irreducible derivative could help circumvent potential numerical problems.
If perturbative derivatives $\pd[n]$ are available, where $n<\order$, LID becomes an inefficient choice,
as the most efficient possibility is to simultaneously measure a maximum number of irreducible derivatives at
once (see Section \ref{sec:SSBID} and \ref{sec:HSBID} for the Bundled Irreducible Derivative approaches). 
However, LID is still essential in that
it should be the method of choice for the most accurate measurement of a \emph{given} irreducible derivative. 
For example, when constructing a Taylor series of a particular mode associated with a structural phase transition, LID is the method of choice
to ensure that each irreducible derivative is resolved as precisely as possible. 

Given that the irreducible representations of the translation group are inherently complex numbers,  $\dispsym_{\qq}$ are in general complex.
Therefore, a unitary transformation to a real representation is needed:
\begin{align}\label{eq:realq}
\dispsym_{\qq[][c]}=\frac{1}{\sqrt 2}(\dispsym_{\qq}+\dispsym_{\bar \qq}) &&
\dispsym_{\qq[][s]}=\frac{\imag}{\sqrt 2}(\dispsym_{\qq}-\dispsym_{\bar \qq})
\end{align}
We refer to this basis as the ``real-$q$" representation, and it should be emphasized that these functions do not
transform like irreducible representations of the translation group, though this is easily accounted for.

Given some irreducible derivative
$\dqnirrsym_{\qq[1]\dots\qq[\order]}^{\alpha_1\dots\alpha_\order}$, one needs
to determine which corresponding real-$q$ derivatives need to be measured. The
first point to appreciate is that an irreducible derivative will in general be
a complex number; though specific cases may be purely real due to the combination of time reversal and inversion
symmetry, or if all $\{\dispsym_{\qq}|\qq\in{\qsetp}\}$ are purely real (e.g.
$\Gamma$-point).  We begin by considering the simplest case of $\pd[0]$.  A
complex derivative will require at least two measurements, in order to recover
both the real and imaginary parts.  For example, in order to determine the second order
complex derivative $\dqnirrsym_{\bar \qq\qq}^{\alpha_1\alpha_2}$, where
$\alpha_1$ and $\alpha_2$ are different instances of the same irreducible
representation, then the chain rule in conjunction with Eq. \ref{eq:realq} indicates that two derivatives must be
measured, such as $\pot_{\qq[][c]\qq[][c]}^{\alpha_1\alpha_2}$ and
$\pot_{\qq[][c]\qq[][s]}^{\alpha_1\alpha_2}$. If inversion symmetry is present, then 
a pre-determined phase convention exists such that $\dqnirrsym_{\bar \qq\qq}^{\alpha_1\alpha_2}$ can be
chosen to be real, and only $\pot_{\qq[][c]\qq[][c]}^{\alpha_1\alpha_2}$ would need to be measured,
as $\pot_{\qq[][c]\qq[][s]}^{\alpha_1\alpha_2}$ would be zero by symmetry. 

The same logic applies at higher order, though there are differences to consider. When using the
real-$q$ representation at higher order, it is possible that multiple irreducible derivatives will 
inherently be probed simultaneously. For example, consider the fourth order derivative
$\dqnirrsym\indices*{*_{\vec{\bar{q}}}^{\alpha_1}  _{\vec{\bar{q}}}^{\alpha_2}  _{\qq}^{\alpha_3}  _{\qq}^{\alpha_4}}$,
where $\alpha_i$ are all distinct irreducible representations. In this case, any possible derivative
$\pot\indices*{*_{\qq[][r_1]}^{\alpha_1}  _{\qq[][r_2]}^{\alpha_2}  _{\qq[][r_3]}^{\alpha_3}  _{\qq[][r_4]}^{\alpha_4}}$,
where $r_i\in\{c,s\}$, will inherently probe six complex irreducible derivatives:
\begin{align}\label{}
\dqnirrsym\indices*{*_{\vec{\bar{q}}}^{\alpha_1}  _{\vec{\bar{q}}}^{\alpha_2}  _{\qq}^{\alpha_3}  _{\qq}^{\alpha_4}} &&
\dqnirrsym\indices*{*_{\vec{\bar{q}}}^{\alpha_1}  _{\vec{\bar{q}}}^{\alpha_3}  _{\qq}^{\alpha_2}  _{\qq}^{\alpha_4}} &&
\dqnirrsym\indices*{*_{\vec{\bar{q}}}^{\alpha_1}  _{\vec{\bar{q}}}^{\alpha_4}  _{\qq}^{\alpha_2}  _{\qq}^{\alpha_3}} \nonumber \\
\dqnirrsym\indices*{*_{\vec{\bar{q}}}^{\alpha_3}  _{\vec{\bar{q}}}^{\alpha_4}  _{\qq}^{\alpha_1}  _{\qq}^{\alpha_2}} &&
\dqnirrsym\indices*{*_{\vec{\bar{q}}}^{\alpha_2}  _{\vec{\bar{q}}}^{\alpha_4}  _{\qq}^{\alpha_1}  _{\qq}^{\alpha_3}} &&
\dqnirrsym\indices*{*_{\vec{\bar{q}}}^{\alpha_2}  _{\vec{\bar{q}}}^{\alpha_3}  _{\qq}^{\alpha_1}  _{\qq}^{\alpha_4}}
\end{align}
Consequently, the chain rule dictates that six real-$q$ derivatives must be measured, such as: 
\begin{align}\label{}
\pot\indices*{*_{\qq[][c]}^{\alpha_1}  _{\qq[][c]}^{\alpha_2}  _{\qq[][c]}^{\alpha_3}  _{\qq[][c]}^{\alpha_4}} &&
\pot\indices*{*_{\qq[][c]}^{\alpha_1}  _{\qq[][c]}^{\alpha_2}  _{\qq[][c]}^{\alpha_3}  _{\qq[][s]}^{\alpha_4}} &&
\pot\indices*{*_{\qq[][c]}^{\alpha_1}  _{\qq[][c]}^{\alpha_2}  _{\qq[][s]}^{\alpha_3}  _{\qq[][c]}^{\alpha_4}} \nonumber \\
\pot\indices*{*_{\qq[][c]}^{\alpha_1}  _{\qq[][s]}^{\alpha_2}  _{\qq[][c]}^{\alpha_3}  _{\qq[][c]}^{\alpha_4}} &&
\pot\indices*{*_{\qq[][s]}^{\alpha_1}  _{\qq[][c]}^{\alpha_2}  _{\qq[][c]}^{\alpha_3}  _{\qq[][c]}^{\alpha_4}} &&
\pot\indices*{*_{\qq[][c]}^{\alpha_1}  _{\qq[][c]}^{\alpha_2}  _{\qq[][s]}^{\alpha_3}  _{\qq[][s]}^{\alpha_4}} 
\end{align}
Therefore, in the most general case, multiple irreducible derivatives must be simultaneously considered even in the LID approach,
though in many cases a single irreducible derivative can be probed. 

Now we consider LID in the case where there are perturbative derivatives, and we focus on the common scenario of $\pd[1]$ (i.e. Hellman-Feynman forces). We can now
reexamine the previous two examples. In the case of the complex derivative $\dqnirrsym_{\bar \qq\qq}^{\alpha_1\alpha_2}$,
both the real and imaginary parts can be simultaneously measured, given that a derivative along $\dispsym_{\qq[][c]}^{\alpha_2}$
will generate $\pot_{\qq[][c]\qq[][c]}^{\alpha_1\alpha_2}$ and $\pot_{\qq[][s]\qq[][c]}^{\alpha_1\alpha_2}$, in addition
to $\pot_{\qq[][c]\qq[][c]}^{\alpha_2\alpha_2}$ and $\pot_{\qq[][s]\qq[][c]}^{\alpha_2\alpha_2}$. Therefore, even though our
intent was to measure a single irreducible derivative, we immediately obtain a second one given that we have repeating irreducible representations
in this example. In the simpler case of $\dqnirrsym_{\bar \qq\qq}^{\alpha\alpha}$, $\pd[1]$ has precisely the same cost as $\pd[0]$ given that 
both cases require one measurement (assuming the undistorted energy is known); though $\pd[1]$ has the possibility of performing forward finite difference
which would save a factor of two.

For the case of 
$\dqnirrsym\indices*{*_{\vec{\bar{q}}}^{\alpha_1}  _{\vec{\bar{q}}}^{\alpha_2}  _{\qq}^{\alpha_3}  _{\qq}^{\alpha_4}}$,
using $\pd[1]$, all six real-$q$ derivatives
can be obtained from three measurements of the  $\order-1$ derivatives of the forces:
$\{\dispsym_{\qq[][c]}^{\alpha_2},\dispsym_{\qq[][c]}^{\alpha_3},\dispsym_{\qq[][c]}^{\alpha_4}\}$,
$\{\dispsym_{\qq[][c]}^{\alpha_2},\dispsym_{\qq[][c]}^{\alpha_3},\dispsym_{\qq[][s]}^{\alpha_4}\}$, and
$\{\dispsym_{\qq[][c]}^{\alpha_2},\dispsym_{\qq[][s]}^{\alpha_3},\dispsym_{\qq[][c]}^{\alpha_4}\}$.
Therefore, $\pd[1]$ will save a factor of two in this case. 

\begin{figure}[htbp]
\begin{center}
\includegraphics[width=1.0\linewidth]{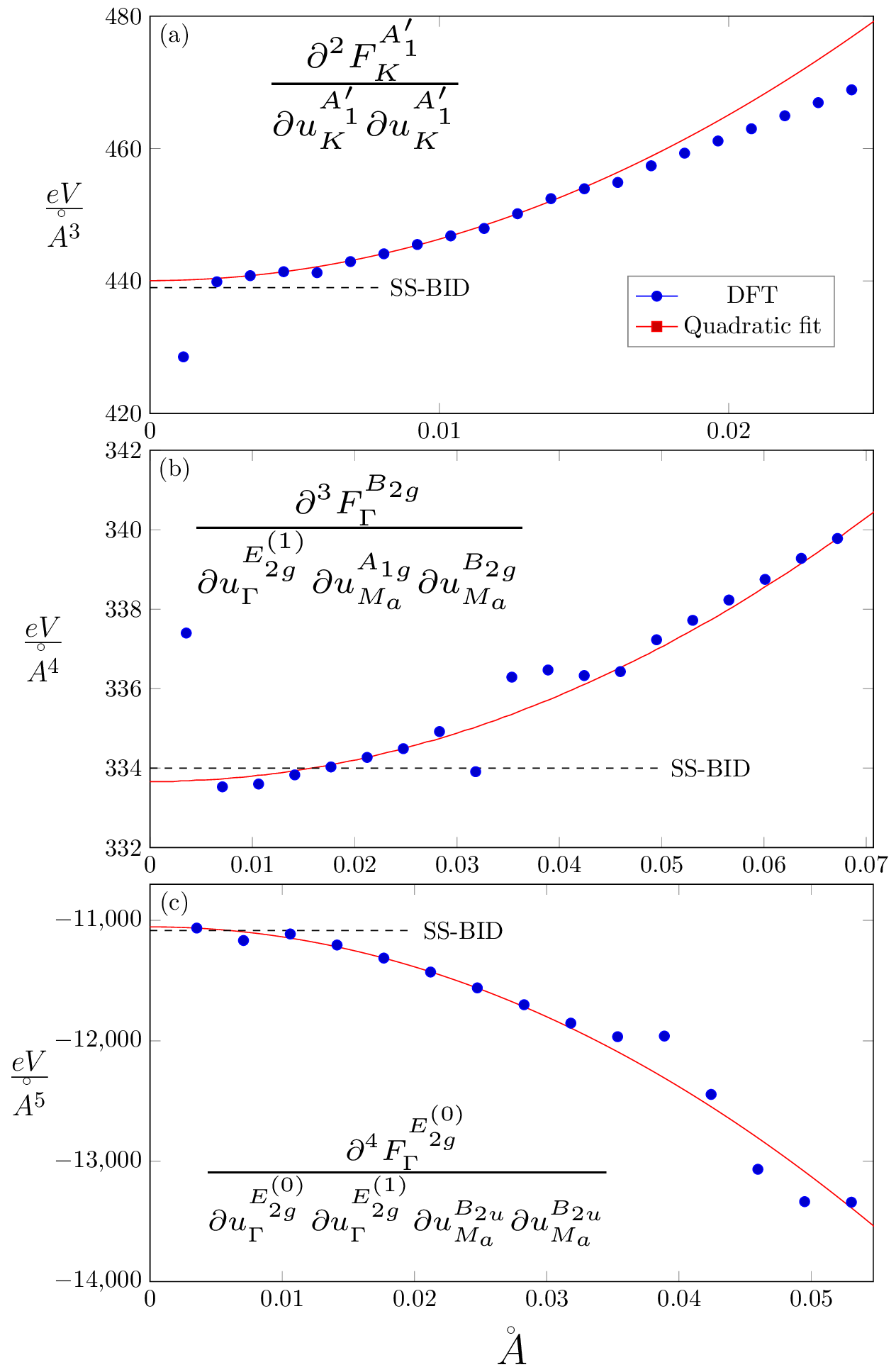}
\end{center}
\caption{
All panels display central finite difference calculations as a function of
$\Delta$; points are calculated values while the red line
is a quadratic fit to a subset of points chosen by the algorithm defined in
Section \ref{sec:cfd}.  Panels (a), (b), and (c) display a particular third,
fourth, and fifth order space group irreducible derivative, respectively, as
obtained using the LID method with $\pd[1]$.  The horizontal dashed lines
show the result of the SS-BID method, where many irreducible derivative are
simultaneously extracted.
}
  \label{fig:centralfd}
\end{figure}

We executed the LID approach using $\pd[1]$ for graphene at $\order$=2, 3, 4, and 5, with FTG up to $\supa[bz]$=12$\hat1$, 3$\hat1$, 2$\hat1$, and 2$\hat1$, respectively.
In Figure \ref{fig:centralfd}, we provide an example for $\order$=3, 4, and 5,
where each data point corresponds to a single evaluation of
$\pot_{\rdisp[1]\dots\rdisp[\order]}(\Delta)$  (i.e. up to 2$^\order$ DFT calculations for a given $\Delta$).
The red line is a quadratic fit to a subset of the points, as described in Section \ref{sec:cfd}, and the intercept of this curve is the value of the
indicated irreducible derivative. 
The values of all irreducible derivatives for $\supa[BZ]=\supa[K]$ at $\order=2$ and $\order=3$ are given in Table \ref{table:irrD}, while the values for $3\hat 1$ and $\supa[2K]$ at $\order=3$
are given in Supplementary Information in Table S\ref{si:table:irr_deriv_n3}.

\begin{table}
\caption{A table of the irreducible derivatives for graphene at $\order=2$ and $\order=3$ with $\supa[BZ]=\supa[K]$. %
  Units are eV/\AA$^{\order}$.%
}
\label{table:irrD}
\begin{tabular}{rrrr}
\hline\hline
Derivative & Value & Derivative & Value \\
\hline\\ [-0.9em]
        $\dqnirrsym\indices*{*_{\Gamma}^{B_{2g}^{}}_{\Gamma}^{B_{2g}^{}}}$ & $35.417$ &                 $\dqnirrsym\indices*{*_{\Gamma}^{E_{2g}^{}}_{\Gamma}^{E_{2g}^{}}}$ & $113.986$ \\[0.4em]
  $\dqnirrsym\indices*{*_{\bar{K}}^{A_{1}^{\prime}}_{K}^{A_{1}^{\prime}}}$ & $80.806$ &           $\dqnirrsym\indices*{*_{\bar{K}}^{A_{2}^{\prime}}_{K}^{A_{2}^{\prime}}}$ &  $43.951$ \\[0.4em]
    $\dqnirrsym\indices*{*_{\bar{K}}^{E_{}^{\prime}}_{K}^{E_{}^{\prime}}}$ & $69.174$ & $\dqnirrsym\indices*{*_{\bar{K}}^{E_{}^{\prime\prime}}_{K}^{E_{}^{\prime\prime}}}$ &  $12.708$  \\[0.4em]
\hline\\ [-0.9em]
            $\dqnirrsym\indices*{*_{\Gamma}^{E_{2g}^{}}_{\Gamma}^{E_{2g}^{}}_{\Gamma}^{E_{2g}^{}}}$ &   $425.751$ &                 $\dqnirrsym\indices*{*_{K}^{A_{1}^{\prime}}_{K}^{A_{1}^{\prime}}_{K}^{A_{1}^{\prime}}}$ &    $440.064$ \\[0.4em]
            $\dqnirrsym\indices*{*_{K}^{A_{1}^{\prime}}_{K}^{A_{2}^{\prime}}_{K}^{A_{2}^{\prime}}}$ &    $11.138$ &                   $\dqnirrsym\indices*{*_{K}^{A_{1}^{\prime}}_{K}^{E_{}^{\prime}}_{K}^{E_{}^{\prime}}}$ &    $289.379$ \\[0.4em]
  $\dqnirrsym\indices*{*_{K}^{A_{1}^{\prime}}_{K}^{E_{}^{\prime\prime}}_{K}^{E_{}^{\prime\prime}}}$ &  $- 53.543$ &                    $\dqnirrsym\indices*{*_{K}^{E_{}^{\prime}}_{K}^{E_{}^{\prime}}_{K}^{E_{}^{\prime}}}$ &  $- 239.640$ \\[0.4em]
   $\dqnirrsym\indices*{*_{K}^{E_{}^{\prime\prime}}_{K}^{E_{}^{\prime\prime}}_{K}^{E_{}^{\prime}}}$ &    $24.732$ &       $\dqnirrsym\indices*{*_{\Gamma}^{B_{2g}^{}}_{\bar{K}}^{E_{}^{\prime}}_{K}^{E_{}^{\prime\prime}}}$ &   $- 41.699$ \\[0.4em]
       $\dqnirrsym\indices*{*_{\Gamma}^{E_{2g}^{}}_{\bar{K}}^{A_{1}^{\prime}}_{K}^{E_{}^{\prime}}}$ & $- 455.829$ &            $\dqnirrsym\indices*{*_{\Gamma}^{E_{2g}^{}}_{\bar{K}}^{A_{2}^{\prime}}_{K}^{E_{}^{\prime}}}$ &   $- 50.356$ \\[0.4em]
        $\dqnirrsym\indices*{*_{\Gamma}^{E_{2g}^{}}_{\bar{K}}^{E_{}^{\prime}}_{K}^{E_{}^{\prime}}}$ &   $204.211$ & $\dqnirrsym\indices*{*_{\Gamma}^{E_{2g}^{}}_{\bar{K}}^{E_{}^{\prime\prime}}_{K}^{E_{}^{\prime\prime}}}$ &   $- 32.416$  \\[0.2em]
\hline\hline
\end{tabular}
 \end{table}

\subsection{Maximally exploiting perturbative derivatives: bundled irreducible derivative approach}
\label{sec:SSBID}
Here we consider the most efficient approach for extracting order $\order$ derivatives given $\pd[n]$, where $n<\order$, while restricting
all calculations to the BvK supercell $\supa[BZ]$; and this latter constraint will be removed in the next section.  The intent is to determine
as many irreducible derivatives as possible in a given measurement, and therefore we refer to this approach
as bundled irreducible derivative (BID) approach; and given the use of the BvK supercell,
we refer to this as the single-supercell bundled irreducible derivative (SS-BID) approach. 
In any BID approach, a basis is explicitly chosen to maximally avoid the block diagonal structure of the dynamical tensor.
For simplicity, we focus on the most common case where only forces are \emph{a priori}  known, $\pd[1]$, though generalizing to other cases is straightforward.
We have already defined the total number of unknowns which must be computed in the BvK supercell $\supa[BZ]$ as $\nirreptotal[\supa[BZ]]$.
We now must determine the total number of measurements, denoted $\nmeasure[\supa[BZ]]$,
in some specifically chosen basis which is yet to be determined. 
It is straightforward to \emph{a priori} determine the upper bound of $\nmeasure[\supa[BZ]]$ using group theory alone for $\order= 2$. One needs to count
the largest number of repeats for a given irreducible representation at a given $q$-point, and then divide by the length of the corresponding star:
\begin{align}\label{eq:nmphonons}
\nmeasure[\supa[BZ]] \le \max_{\qq\in\bqq[BZ],\alpha\in\tilde\Gamma_{\dispsym_{\qq}}}
\ceil*{
\frac{a^\alpha_{\qq}}{|\qorbit[q]|}
}
\end{align}
where $a^\alpha_{\qq}$ is the number of times the $\alpha$ irreducible representation repeats at $\qq$, and the outer bracket denotes the ceiling function.
For a detailed example illustrating this procedure at second order in ZrO$_2$, see Appendix \ref{app:zro2}.

Beyond second order, it is straightforward
to determine the lower bound using a counting argument:
\begin{align}\label{eq:nmeasure}
\nmeasure[\supa[BZ]]\ge\ceil*{\frac{\nirreptotal[\supa[BZ]]}{(\nq\natom\npol-\cdim)}}
\end{align}
where the numerator is the number of irreducible derivatives and the denominator is the number of nonzero force equations $\nforce[\supa[BZ]]=\nq\natom\npol-\cdim$.
The exact $\nmeasure[\supa[BZ]]$ can straightforwardly be determined by explicit calculation, and we have found that the equality in Eq. \ref{eq:nmeasure}
holds for $\order\ge 3$ in every case we examined in graphene and rock salt.

Having determined $\nmeasure[\supa[BZ]]$, the specific
choice of basis, which we call the ``bundled basis", must be constructed for all measurement; being a set of
real displacement vectors $\{b_1^i,\dots,b_{\order-1}^i\}$, where there is one set $i$ for each $\nmeasure[\supa[BZ]]$ measurements. 
It is useful to store the $\nmeasure[\supa[BZ]]$ measured derivatives stacked into a vector  $\mderivv[\supa[BZ]]$, and all $\nirreptotal[\supa[BZ]]$
irreducible derivatives  which are contained within $\supa[BZ]$ are stored in the vector  $\derivv[\supa[BZ]]$.
The order $\order$ chain rule generates a linear system of equations which
relates the derivatives in the bundled basis $\mderivv[\supa[BZ]]$ to the irreducible basis $\derivv[\supa[BZ]]$: 
$\mderivv[\supa[BZ]] = \crulemat[\supa[BZ]] \derivv[\supa[BZ]]$;
where $\crulemat[\supa[BZ]]$ is the $(\nq\natom\npol-\cdim)\nmeasure[\supa[BZ]]\times\nirreptotal[\supa[BZ]]$
complex chain rule matrix. A necessary condition for the bundled basis is that $\textrm{rank}(\crulemat[\supa[BZ]])=\nirreptotal[\supa[BZ]]$.
The choice of bundled basis is not unique, but an obvious criterion is to minimize the condition number of $\crulemat[\supa[BZ]]$,
which will ensure a minimal propagation of error upon solving for $\derivv[\supa[BZ]]$. We explored this possibility by generating thousands of random
bundled basis sets and choosing the one with the smallest condition number. We refer to this as the condition number optimized (CNO)
bundled basis. The only downside to this is that it is inconvenient to disseminate the choices that we made. 

A simple option is
to create a sequence of rational numbers using the FTG's of a one dimensional lattice: 
\begin{align}\label{}
j=\bigcup_{n=1}^{\infty}\bqq[BZ]^n= \{0,\frac{1}{2},\frac{1}{3},\frac{2}{3},\frac{1}{4},\frac{3}{4},\frac{1}{5},\frac{2}{5},\frac{3}{5},\frac{4}{5},\dots\}
\end{align}
where
$\bqq[BZ]^n$ corresponds to supercell $n$. 
The first bundled vector is obtained by iterating over every displacement within $\supa[BZ]$ and imparting an amplitude of $\cos(2\pi j_n n)$,
where $j_n$ is the $n$th element of the set $j$ and $n$ has an inner loop running over the
$\npol$ displacements and an outer loop running over all $\natom\nq$ atoms in $\supa[BZ]$. The remaining $(2^\order\nmeasure[\supa[BZ]]-1)$ bundled
basis vectors are generated by continuing along the sequence $j$. We simply refer to this as the simple bundled basis (SBB), and
in all cases we tested the $2^\order\nmeasure[\supa[BZ]]$ vectors generated in this manner did fulfill $\textrm{rank}(\crulemat[\supa[BZ]])=\nirreptotal[\supa[BZ]]$. 
While the condition number of the resulting $\crulemat[\supa[BZ]]$ for SBB will generally be larger than the CNO basis, the differences in the resulting irreducible derivatives
were typically very small (direct comparisons are made in Supplementary Information, Figure S\ref{si:fig:cno_v_rc}). 
All BID results in this paper were generated using the SBB basis unless otherwise noted. 

We illustrate some specific results using BID in Figure \ref{fig:centralfd},
indicated by a dashed line. As shown, the results agree with the LID approach
to within fractions of a percent. This excellent agreement signifies that we
successfully resolved the quadratic error tails within the SBB bundled basis,
indicating that the Hellman-Feynman forces were \emph{maximally} harnessed
without any appreciable loss in precision. 

Given that our method works purely in terms of irreducible derivatives, we are guaranteed to satisfy all possible symmetries
of the order $\order$ Taylor series by construction; and our BID approach allows them to be extracted in the smallest number of measurements. 
Therefore, it is useful to compare with competing approaches which implement symmetry using extrinsic real space symmetry approaches, and 
we focus on the example of the rock salt structure at $\order=3$. 

\begin{figure}[htbp]
\begin{center}
\includegraphics[width=1.0\linewidth]{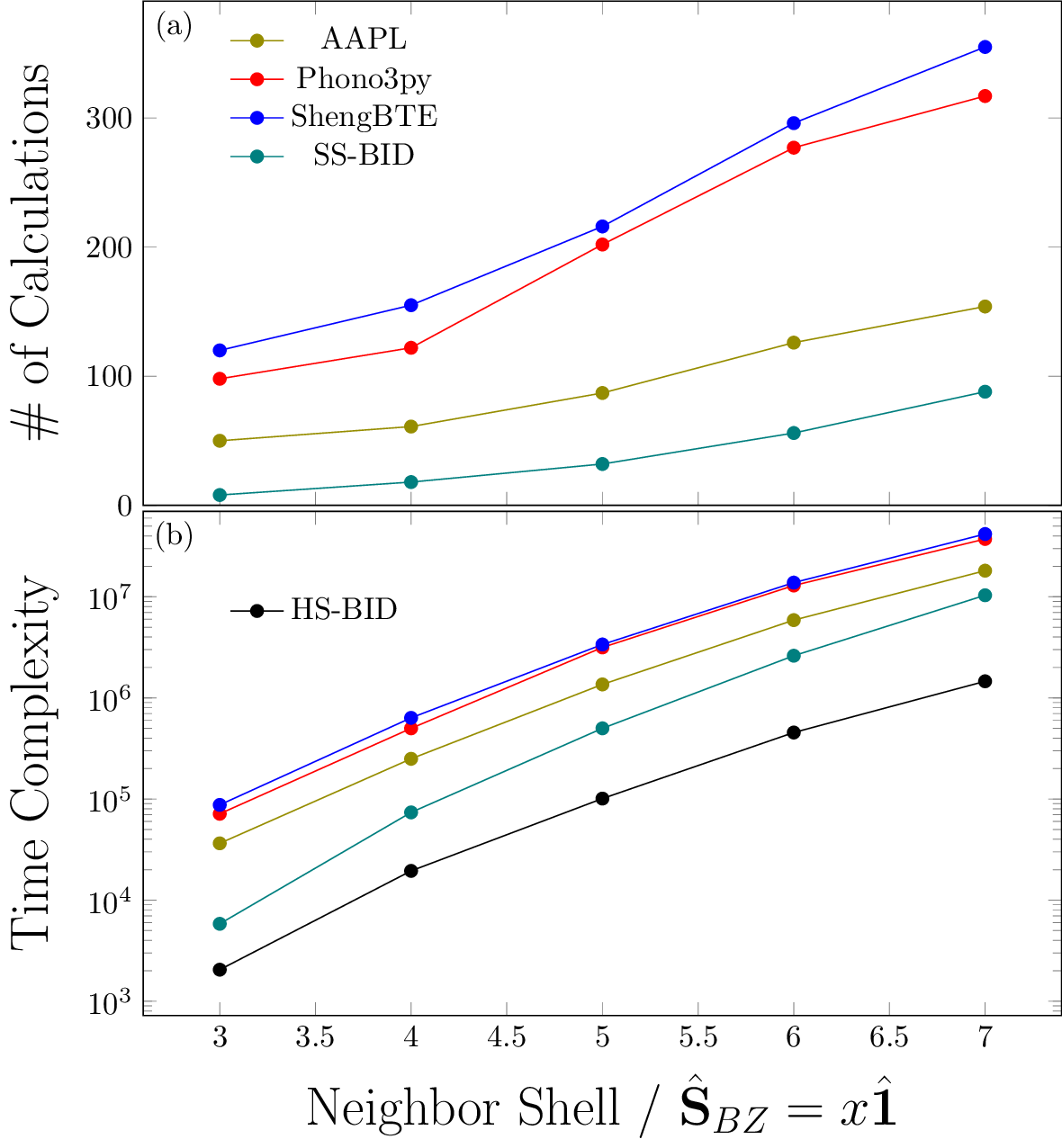}
\end{center}
\caption{
  Complexity analysis for rocksalt structure at $\order=3$, comparing existing published methods (ShengBTE, Phono3py, and AAPL, taken from Ref. \onlinecite{Plata201745}) 
  with our approaches (SS-BID and HS-BID);
  including (a) number of required DFT calculations and (b) time complexity assuming that the DFT calculations
    scale quadratically with system size.
  Existing methods only calculate the derivatives out to the $x$ neighbor shell
  (where $x$ is the horizontal axis)
  within supercell $\supa[BZ]=x \hat{1}$, while our methods computes \emph{all} derivatives within the corresponding supercell.
}
  \label{fig:countruns}
\end{figure}

A recent
paper compared the efficiency of three popular approaches to compute cubic
terms using finite displacements\cite{Plata201745}, 
which we shall label by the
codes which implement them: AAPL\cite{Plata201745},
Phono3py\cite{Togo2015094306}, and ShengBTE\cite{Li20141747}. Figure
\ref{fig:countruns}a replots the results that were presented in reference \onlinecite{Plata201745},
which determines the number of DFT calculations required to determine all cubic derivatives within some real space cutoff shell within a given supercell, 
and serves as a measure of the extent to which symmetry has been
accounted for. We have reproduced the results for the case of Phono3py,
which ensures we have properly understood the conventions and assumptions when using Phono3py in reference \onlinecite{Plata201745}; and we assume that
the analogous procedures were applied for AAPL and ShengBTE, as we did not attempt to interpret the choices made in executing these latter codes. 

It is important to first clarify the x-axis of Figure \ref{fig:countruns}a, which we labeled as being both
the ``Neighbor Shell" and $\supa[BZ]=x\hat 1$. For the competing approaches (i.e. AAPL, Phono3py, and ShengBTE), 
this means that a $\supa[BZ]=x\hat 1$ supercell is constructed and only derivatives within a $x$-neighbor shell
are retained. Alternatively, when we used our SS-BID method for comparison, 
we compute \emph{all possible derivatives} which exist within $\supa[BZ]=x\hat 1$. 
Therefore, this is not not a fair comparison with respect to our space group irreducible approach. 
It is worth noting that if one does not include a real space truncation in the Phono3py code, allowing it to compute
all derivatives within the supercell, the numbers are substantially larger. For example, if one execute $\supa[BZ]=3\hat 1$ in Phono3py without any truncation,
the number of DFT runs increases to 194; nearly doubling as compared to the truncated case (i.e. $x=3$ in \ref{fig:countruns}a).  

Figure \ref{fig:countruns}a shows that
AAPL, Phono3py, and ShengBTE  all overestimate the actual number of calculations which are required to extract \emph{all} irreducible derivatives within the supercell. 
To give an idea of the computational speedup, we assume that the first-principles
method will scale as the square of the number of atoms and plot the total time in Figure \ref{fig:countruns}b, demonstrating a substantial gain over all competing approaches.

In order to clearly demonstrate the group theoretical nature of our results, 
we explicitly list all irreducible derivatives for the case of $\supa[BZ]=2\hat 1$ in Appendix \ref{app:rocksalt}. 
As shown, there are 33 real irreducible derivatives 
and these can all be obtained within a single measurement according to Eq. \ref{eq:nmeasure}. %
We emphasize that the result of our group theoretical analysis is not original in this case, as Birman \emph{et. al} first derived all possible results for a third order product, 
symmetric or otherwise, in Fm$\bar3$m\cite{Chen1968639}.

\subsection{Bundled irreducible derivatives with hierarchical supercells}
\label{sec:HSBID}
Here we consider an alternative BID approach which demands that each irreducible derivative is computed within the smallest
possible supercell in which it fits; and we refer to this as the hierarchical supercell bundled irreducible derivative (HS-BID) approach.
Our Minimum Supercell Multiplicity equation dictates that for three dimensional materials having uniform supercells, the BvK supercell
can be completely avoided for $\order\le 3$, and therefore HS-BID will yield a substantial increase in computational efficiency
for first-principles approaches which scale in a super-linear fashion, as most do, despite the fact that more total calculations are required. 

The first step is to categorize the smallest supercell into which each irreducible derivatives fits.
Therefore, for all $\orbit\in\borbit[BZ]$, we must determine the smallest supercell $\supa[\orbit]$ which contains at least one $\qsetp\in\orbit$;
and the set of all supercells is denoted as $\bsupastar$,
where $|\bsupastar|\le|\bqsetp[IBZ]|$. 
Additionally, we construct a set $\borbit[\supa]$ which contains all $\{\orbit\}$ commensurate with $\supa$. Furthermore,
for every $\borbit$, we create a subset denoted $\borbitup[\supa]$, which consists of
all orbits  $\orbit[i]$ that are contained by $\supa$ and \emph{not} contained by any $\supa[j]$ where $\det(\supa[j])<\det(\supa)$.
Now, the number of irreducible derivatives which must be computed in a given supercell is $\nirreptotalup=\sum_{\orbit\in\borbitup}\nirrep$.
We will also define corresponding quantities $\borbitdn$ and $\nirreptotaldn$ to characterize the irreducible derivatives
in $\supa$ which are contained in a smaller supercell; where $\nirreptotal=\nirreptotalup+\nirreptotaldn$.

In order to illustrate the definitions in the preceding paragraph, we consider graphene at $\order=3$ and $\supa[BZ]=3\hat 1$,
where 
\begin{align}\bqset[IBZ]=\{&
\left(\Gamma,\Gamma,\Gamma\right),
\left(\Gamma,\bar K,K\right)
\left(K,K,K\right),
\left(\Gamma,\Delta_{0},\Delta_{3}\right),
\nonumber \\&
\left(\Delta_{0},\Delta_{0},\Delta_{0}\right),
\left(K,\Delta_{0},\Delta_{5}\right), 
\left(\Delta_{0},\Delta_{2},\Delta_{4}\right),
\}\end{align}
with the notation taken from Figure \ref{fig:uniform_grid}b.
The set of supercells $\bsupastar$ is:
\begin{align}
\bsupastar= \left\{
  \hat 1, 
\supa[K],
2\hat 1+\hat \sigma_x,
3\hat 1
\right\}
\end{align}
Finally, the $\borbitup[\supa]$ for each $\supa\in\bsupastar$ is:
\begin{align}
\borbitup[\hat 1]&=
\{\left(\Gamma ,\Gamma ,\Gamma \right)\} 
\hspace{3mm}
\borbitup[\supa[K]]=
\{\left(K,K,K\right),
\left(\Gamma ,\bar K,K\right)\}
\nonumber \\
\borbitup[2\hat 1+\hat \sigma_x]&=
\{\left(\Delta_{0},\Delta_{0},\Delta_{0}\right),
\left(\Gamma ,\Delta_{0},\Delta_{3}\right)\}
\nonumber \\
\borbitup[3\hat 1]&=
\{\left(K,\Delta_{0},\Delta_{5}\right),
\left(\Delta_{0},\Delta_{2},\Delta_{4}\right)\}
\end{align}
Here we see that only two out of the seven total $\qset$ need to be computed in the BvK supercell; though it should be noted that those
two have the lowest symmetry.

The next step is to determine the bundled basis for each $\supa\in\bsupastar$.
Therefore, we split $\derivv$ into two separate vectors $\derivvup$ and $\derivvdn$ containing the irreducible derivatives
which do not ($\vee$) and do ($\wedge$) fit into a smaller supercell, respectively. Similarly, the previously defined chain rule matrix $\crulemat$ can be split into
two respective pieces $\crulematup$ and $\crulematdn$. Finally, we can obtain the unknown derivatives which only fit in $\supa$:
$\derivvup= (\crulematup)^{+}(\mderivv - \crulematdn \derivvdn)$, where $(\crulematup)^{+}$ refers to the pseudoinverse.
A necessary condition for the bundled basis is
that $\textrm{rank}(\crulematup)=\nirreptotalup$, and the basis can be chosen using the same schemes as described for SS-BID. 
Once the bundled basis has been chosen for each $\supa\in\bsupastar$, the CFD measurements can be performed, and then the irreducible
derivatives can be extracted from the smallest to largest supercell. It should be noted that all calculations can be performed simultaneously,
given that the bundled basis can be determined \emph{a priori}.

The only remaining idea to be introduced is the notion of ``overbundling"
irreducible derivatives. Given that $\nirreptotalup/\nforce$ is typically
not a round number, it may be possible to obtain irreducible derivatives which
fit in a smaller supercell for free. Specifically, derivatives tallied in
$\nirreptotaldn$ may possibly be added without any increase in $\nmeasure$;
though in general the bundled basis will need to be modified to properly sample
the additional derivatives. 

It is useful to compare the performance of HS-BID with SS-BID, in addition to the competing approaches (see Figure \ref{fig:countruns}b). Assuming 
the first-principles method scales quadratically with system size, HS-BID is more than an order of magnitude faster than all competing approaches
that we examined. The speedup would be far more dramatic for for first-principles methods with poorer scaling, such 
as hybrid functionals. 
It should be emphasized that the speedup of HS-BID compared to SS-BID will be far more dramatic for $\order=2$ as compared to $\order=3$, treated in this example.
Given the efficiency of our new methods, crystals with increasingly complex unit cells may be treated using DFT, and methods which scale
poorly (e.g. hybrid functionals) may now be used to compute phonons and their interactions more regularly.

Finally, we discuss factors related to the quality of the measurements. Given that some measurements may be deficient (i.e. poor quadratic error tails),
it may be easier to simply dispense with them as opposed to fixing them. For example, if one is not overbundling, there may be room to simply remove
a derivative while keeping the chain rule matrix full rank; and we refer to this as ``pruning". If not, one can simply add additional measurements, which we
refer to as ``overmeasuring", and then one can prune away the problem derivatives.

\section{Assessing the results}
\label{sec:assessing}
\subsection{General Considerations}
\label{sec:assessing_general}
No matter what formalism is used to compute the Taylor series of the
Born-Oppenheimer surface, one needs some clear criteria to assess the quality
of the results. Many studies predict some observable and then compare to
experiment. 
This is not an ideal test on its own, even if successful, because it easily allows for a
cancellation of errors and human bias to interact in a dangerous manner; especially so 
when an approach simultaneously fits many derivatives.
Ideally, the test should be purely self-consistent, only answering how well
the Born-Oppenheimer surface of the first-principles method at hand is
captured. In this vein, some studies compare their results to first-principles
molecular dynamics on small supercells. 
We note that first-principles molecular dynamics contains the Taylor
series to infinite order, so failure will not differentiate between a poor expansion and 
activation of higher order terms not included in the expansion being tested.
Furthermore, one can only probe relatively small FTG's in this manner, due to the computational expense of first-principles
approaches. 

Here we consider several different validations for Taylor series, the first being completely generic to any method, and the others being specific to finite displacements. 
The first is the strain derivatives of the phonons, where the $N$-th order strain derivative will
result in an infinite range coupling of the $(N+2)$-th order force tensor. This is an ideal test in that strain derivatives can efficiently be calculated
by simply perturbing the lattice vectors in the context of a phonon calculation; which will not alter the number of atoms within the unit cell in any given calculation. 
A usual scenario is the first volume derivative of the phonons, which,
in conjunction with the phonons, gives rise to the well-known Gr{\"u}neisen parameters\cite{Dove1993book}. Furthermore, the Gr{\"u}neisen parameters are directly connected to thermodynamic
observables, and therefore properly resolving them is physically well justified; which is why Gr{\"u}neisen parameters have often served as a test
of cubic phonon interactions\cite{Lee2014085206}. 
The other two tests we perform are more specific to finite displacement calculations: assessing the quality
of the quadratic error tails and comparing results of BID and LID approaches.
Below we illustrate all three tests.

\subsection{Strain derivatives of the phonons}
\begin{figure}[htbp]
\begin{center}
\includegraphics[width=1.0\linewidth]{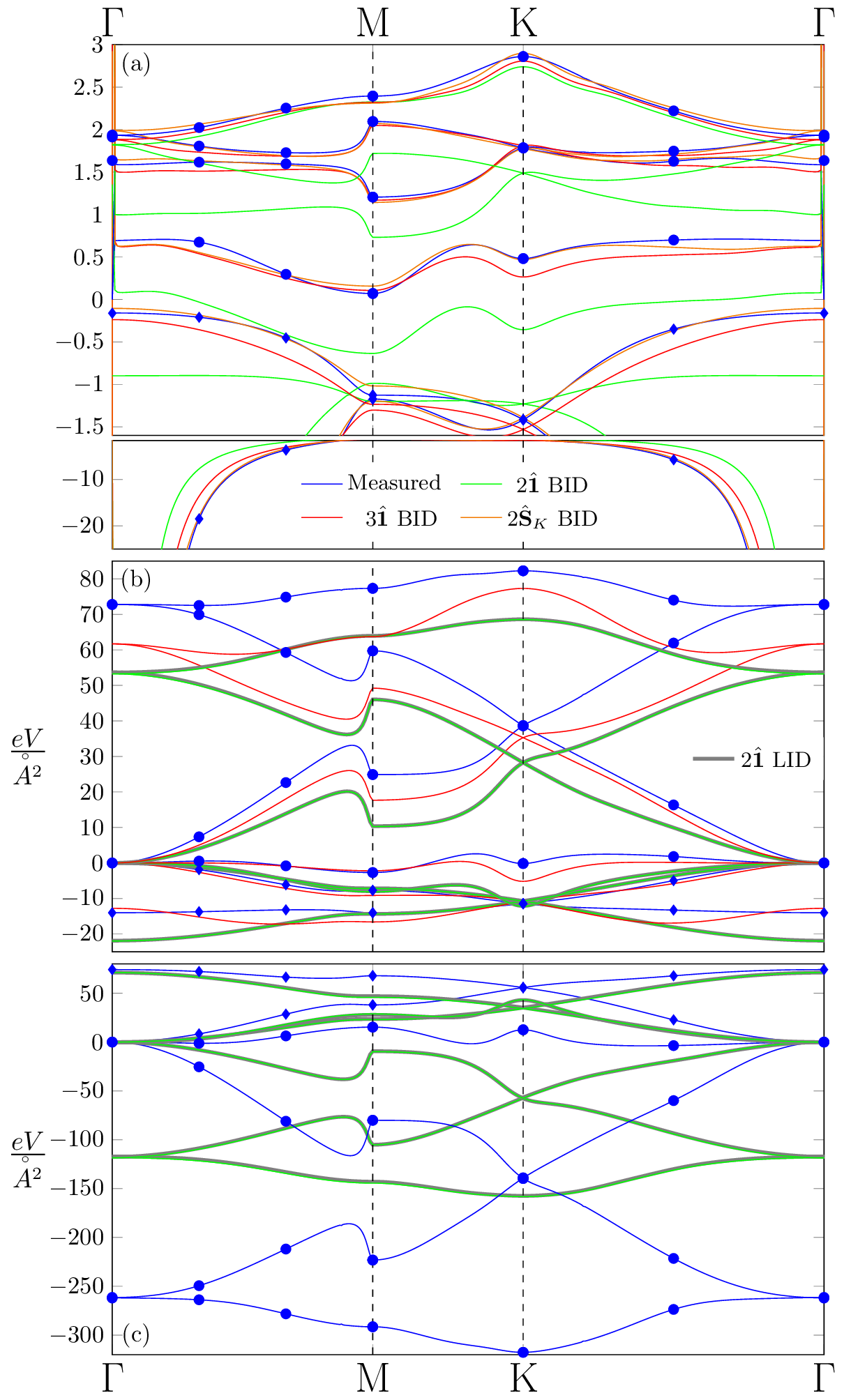}
\end{center}
\caption{
(a) Gr{\"u}neisen parameters of graphene directly measured using identity strain derivatives of the phonons and Fourier
interpolation (blue points and lines; diamonds and circles correspond to
out-of-plane and in-plane modes, respectively) and via Eq. \ref{eq:grun}, which uses the cubic irreducible 
derivatives at various mesh densities. Panels (b) and (c) follow the same conventions, but for the second and third strain derivatives,
respectively; and both panels display the LID results for $2\hat1$ in gray, showing near perfect agreement with BID results.
}
  \label{gruneisen}
\end{figure}
We begin by assessing the order $N$ strain derivatives, and we restrict our attention identity strains (i.e. uniform in all directions) for simplicity. 
Taylor series expanding the dynamical tensor to first order for $N$ selected $\qq[i]$, contracting with the corresponding acoustic displacement vectors to leading order in $\qq$,
taking the small $\qq$ limit of the corresponding displacements,
and taking the identity strain derivatives, we arrive at an
analytic expression for the $N$-th order identity strain derivative of the dynamical matrix. We have restricted ourselves
to crystals which have no internal degrees of freedom, resulting in the following equation:
\begin{equation}
\begin{aligned}
            \frac{ \partial^{N} D\indices*{*_{\vec{\bar q }}^m _{\vec q}^n }   }{\partial \epsilon_{A}^N} &=
            \frac{1}{2^{N }} 
          \sum_{\vec t } 
      e^{-\imag 2\pi \vec q\cdot \vec t  }
            \sum_{\substack{\vec t_1 \dots \vec t_{N} \\ a_1 \dots a_{N}\\ p_1 \dots p_{N}}} 
            \Phi\indices*{*_{\vec 0,}^{m,}   _{\vec t,}^{n,}   _{\vec t_1,}^{(a_1,p_1),} _{\dots,}^{\dots,} _{\vec t_{N}}^{(a_N,p_N)}} 
            \\ &
            \sum_{\alpha_1 \dots \alpha_{N}} 
            \prod_{k=1}^{N}
            \left(\ctrans[k] + \mathcal{A}_{a_k} \right)\cdot\vec e_{\alpha_k}
            \langle u_\Gamma^{(a_k,p_k)}  | \psi_\Gamma^{\alpha_k} \rangle
\end{aligned}
\label{eq:grun}
\end{equation}
where $\epsilon_A$ is the identity strain, $\vec e_i$ is a $d$-dimensional unit vector, and $|\psi_\Gamma^\alpha\rangle$ is an acoustic vector at the zone center. 
For crystals with internal degrees of freedom, or for arbitrary strain states, one must explicitly compute the first order corrections to the $\qq$ dependence
of the acoustic modes; and this requires the phonons. 
The key point is that it is straightforward to directly measure the left hand side of Eq. \ref{eq:grun} by computing the phonons
at a series of different uniform strains, resulting in a set of strain derivative of the phonons defined over some FTG which can then be 
Fourier interpolated. Additionally, the corresponding
quantity can be predicted purely using the $\order=N+2$ irreducible derivatives on the right side of Eq. \ref{eq:grun}; 
and the factor $\prod_{k=1}^{N}\left(\ctrans[k] + \mathcal{A}_{a_k} \right)$ means that
long range terms in the force tensor will be amplified, creating a test that is sensitive to noise in long range terms.

For the case of $N=1$, the Gr{\"u}neisen parameters may be constructed as:
\begin{align}\label{eq:grun1}
\gamma_{\qq}^{i} = 
\left(
\hat U^\dagger_{\qq}
 \hat D_{\bar \qq \qq}^{-1}
\frac{ \partial \hat D_{\bar \qq \qq} }   {\partial \epsilon_{A}}
\hat U_{\qq}
\right)_{ii}
\end{align}
where $\hat U_{\qq}$ is the unitary transformation that diagonalizes the dynamical matrix. Equation \ref{eq:grun} for $N=1$ in conjunction with Equation \ref{eq:grun1} is consistent 
with the equation presented in Ref. \onlinecite{Fabian19971885}.

We begin by comparing our measured and predicted  Gr{\"u}neisen parameters in Figure \ref{gruneisen}a. The direct measurement of the Gr{\"u}neisen parameters
via strain central finite difference are denoted with circles (diamonds) for the in-plane (out-of-plane) modes, while the blue
lines are obtained via Fourier interpolation; and these results are referred to as ``Measured", given that they are numerically exact for the actual points.
The results obtained from using the cubic dynamical tensor in conjunction with Equation \ref{eq:grun} are presented for several 
different FTG's. It should be noted that there are no data points on these curves, as no part of these curves are numerically exact.
The FTG $\supa[BZ]=2\hat 1$ displays relatively poor agreement overall, though the uppermost branch is in good agreement, and several
other branches have a proper shape but are simply shifted 
($\supa[BZ]=\supa[K]$ displayed very similar results, and is not shown for clarity). 
This is consistent with the interpretation that the dynamical tensor is robust, 
but the FTG is simply too small. 
Moving to the next larger FTG, $\supa[BZ]=3\hat 1$, the results markedly improve, with only 
relatively small disagreement; here the discrepancies are likely too large for sensitive quantities like thermal conductivity. 
The next larger FTG, $\supa[BZ]=2\supa[K]$, shows relatively good agreement, with only minor deviations. The  $\order=3$ irreducible derivatives 
for all cases are provided in the Supplementary Information Table S\ref{si:table:irr_deriv_n3}; with $\supa[BZ]=2\supa[K]$ having 215 purely real or imaginary terms.
All the preceding results were obtained using SS-BID, but the results using HS-BID and LID are extremely similar (see Supplementary Information, Fig. S\ref{si:fig:ssvhs}).

The quartic elements of the dynamical tensor can be probed via the second strain derivatives of the phonons, which is shown Figure \ref{gruneisen}b; where the second strain
derivative alone is plotted. For the coarsest FTG, $\supa[BZ]=2\hat 1$, the general shape is smooth and resembles the numerically exact measurements, 
though the deviations are relatively large. However, the near perfect agreement of LID and BID suggests that the derivatives are robust, but a larger FTG is needed. 
Moving to the next larger FTG, $\supa[BZ]=3\hat 1$, the results improve for the out-of-plane modes,
while the in-plane mode results have shifted in the proper direction, but not substantially enough. Even larger FTG's would be needed for a higher
resolution of the results, but we do not proceed further due to computational expense.

The quintic elements of the dynamical tensor can be probed via the third strain derivatives of the phonons, which is shown Figure \ref{gruneisen}c.
In this case, only $\supa[BZ]=2\hat 1$ was attempted. Once again, LID and BID agree extremely well in this case, suggesting that the derivatives are robust.
While the overall shape of the curves are reasonable, it is clear that a larger FTG
would be required to resolve these third strain derivatives.

In summary, strain derivatives can be used as a critical test of the dynamical
tensor, no matter what method is used to compute it.  A more general equation
can be derived for an arbitrary strain derivative, beyond the simple identity
strain considered in Equation \ref{eq:grun}, which would allow for a much more
detailed test; as a larger fraction of the dynamical tensor would be probed. We
leave this to future work.  It also should be noted that the logic of using
strain derivatives of phonons as a test could be inverted to instead use them
as a rich source of information which could be used to assist in extracting the
dynamical tensor, and there are several studies which have begun to pursue
this\cite{Kornbluth2017,Lee2017035105}.

\subsection{Assessing quadratic error tails}
\begin{figure}[htbp]
\begin{center}
\includegraphics[width=1.0\linewidth]{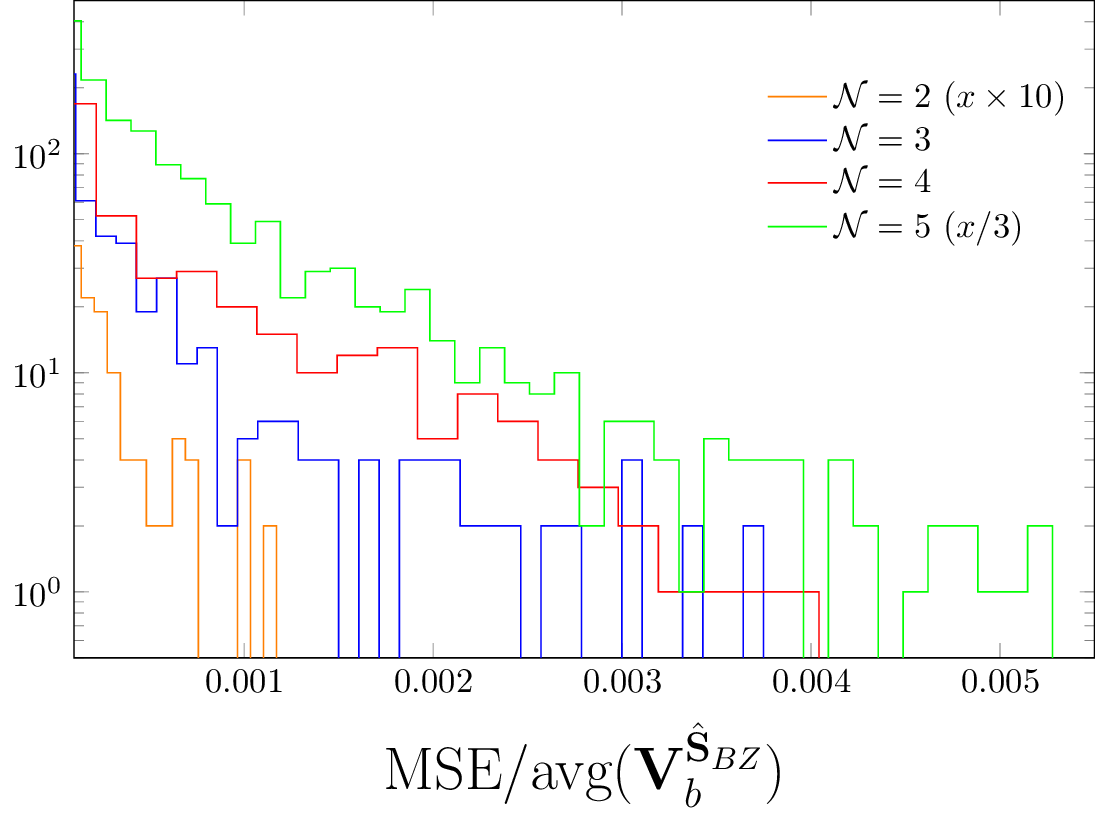}
\end{center}
\caption{
A histogram of mean square error, associated with the quadratic fits to the central finite difference calculations 
as a function of $\Delta$ within the SBB in the SS-BID approach,
\emph{divided} by the average magnitude of the SS-BID SBB derivatives.
Note that values for  $\order=2$ are multiplied by ten while the 
$\order=5$ values are divided by three for ease of viewing. FTG's of $6\hat1$, $2\supa[K]$, $2\hat1$, and $2\hat1$, are used for $\order=2-5$, respectively.
}
\label{fig:err_tail_hist}
\end{figure}

If central finite difference is being used to measure derivatives, then it is critical to assess the quality of the quadratic error tails.
Our algorithm for choosing the set of $\Delta$ used to construct the quadratic error tail is detailed in Section \ref{sec:cfd}. Once 
the $\Delta$ are selected and a least squares fit is performed, there will be a mean square error associated with each quadratic error tail,
and a histogram can be constructed (see Figure \ref{fig:err_tail_hist}). The results are as expected, 
with the error increasing as $\order$ increases from two to five. Furthermore, this should be performed as a diagnostic analysis, and the first evaluation of this 
data did indeed reveal numerous problematic derivatives. The offending derivatives can be inspected to resolve any issues, which usually involves adding 
additional $\Delta$, increasing the convergence parameters of the DFT calculations, or simply pruning the offending derivative (see Section \ref{sec:HSBID}).

\begin{figure}[htpb]
  \centering
  \includegraphics[width=\linewidth]{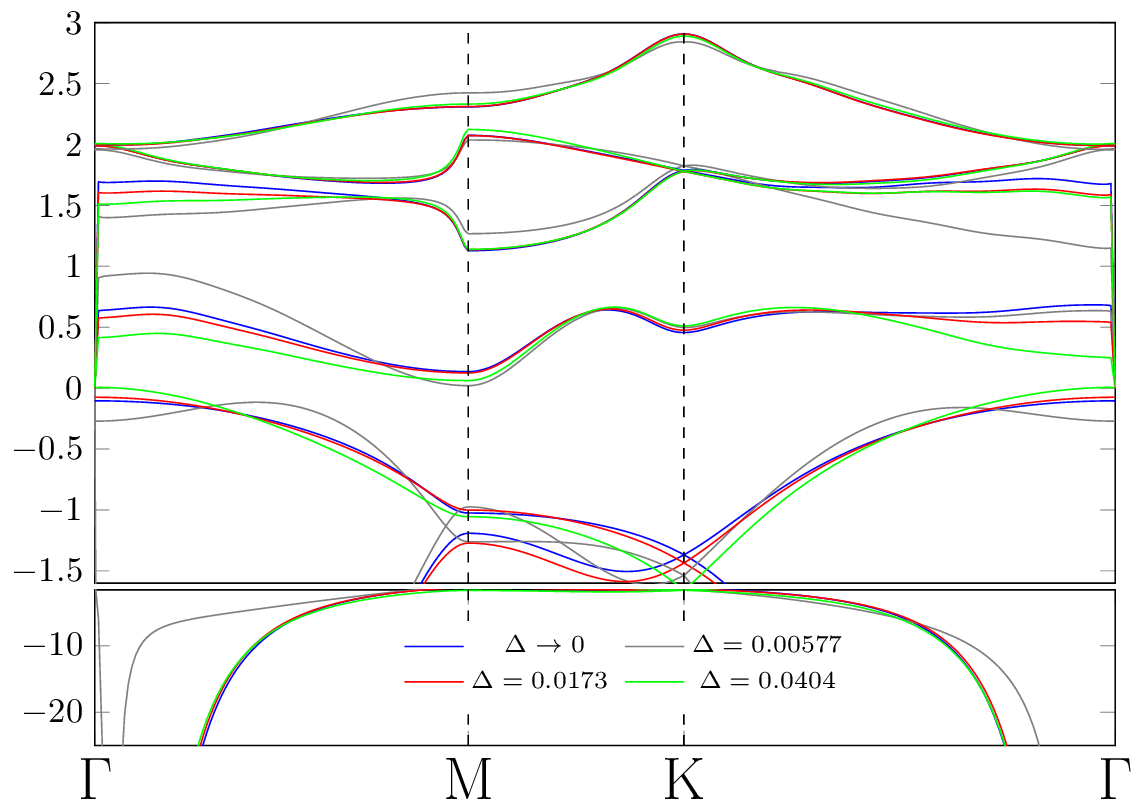}
  \caption{Comparison of the Gr\"uneisen parameters in graphene obtained from cubic irreducible derivatives within $\supa[BZ] = 2 \supa[K]$ using SS-BID. 
  The blue curve uses our algorithm outlined in Section III C to properly extrapolate to $\Delta$ to zero, while the other curves simply 
  use a single value of $\Delta$. 
  } 
  \label{fig:sdg_2K_gru}
\end{figure}

Given the common practice of using a single $\Delta$ to estimate the value of a derivative, as opposed to properly extrapolating to $\Delta=0$, it is interesting to test the efficacy of this on the 
predicted Gr\"uneisen parameters for graphene with $\supa[BZ]=2\supa[K]$ (see Figure \ref{fig:sdg_2K_gru}).
As shown, substantial errors occur if $\Delta$ is too large or too small, though reasonable results
can be obtained with a properly chosen single $\Delta$ in this case; but it can be difficult to choose \emph{a priori}. 
We have observed that the results become more sensitive to a single $\Delta$
as  $\supa[BZ]$ increases (not shown), most likely because more irreducible derivatives are being simultaneously measured.
For a sufficiently large FTG, it is possible that no single $\Delta$ will be effective.

\subsection{Bundled versus Lone derivatives}
\begin{figure}[htbp]
\begin{center}
\includegraphics[width=1.0\linewidth]{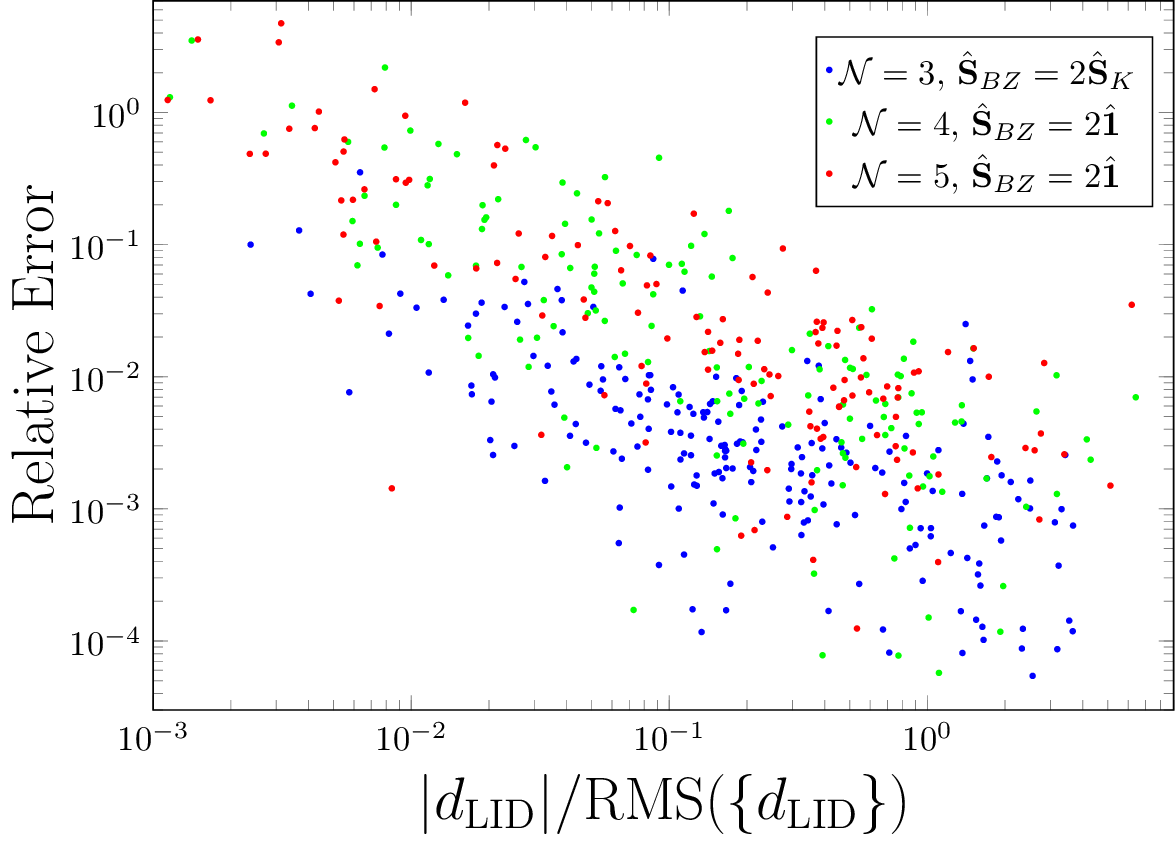}
\end{center}
\caption{
A plot comparing the irreducible derivatives computed using LID and SS-BID.
The $x$-axis denotes relative magnitude $(|d_{\textrm{LID}}|/\textrm{RMS}(\{d_{\textrm{LID}}\}))$
and the $y$-axis denotes relative error $(|d_{\textrm{LID}}-d_{\textrm{BID}}|/|d_{\textrm{LID}}|)$.
For $\order=3$ and $\order=4$, all irreducible derivatives are sampled, 
while only a
subset is provided for $\order=5$. 
}
\label{pack_vs_loneq}
\end{figure}

Another obvious test is to compare BID to LID. Of course, 
if extremely high reliability and precision is needed, and one has the computing resources to execute LID, then LID is the best route.
However, this will not always be possible, and BID will frequently be needed. Bundled derivatives are challenging in the sense that the many irreducible
derivatives that are simultaneously being measured may have starkly different quadratic error tails, which may result in a relatively small region of $\Delta$
which is resolvable as quadratic (see Supplementary Information, Figure S\ref{si:fig:sdg_2K_errortail}a, for a problematic example). As a result, very stringent convergence parameters within DFT may be required to successfully resolve this quadratic region.
Alternatively, LID measures as few irreducible derivatives as possible, and the error tails tend to be much better behaved in this method. 
Therefore, when using BID, one can still compute some fraction of irreducible derivatives using LID as a test; perhaps either a subgroup of the given FTG, or
maybe some random selection.
Figure \ref{pack_vs_loneq} provides a comparison between the irreducible derivatives as computed using LID and BID for $\order=3-5$. 
As expected, the error is smallest for $\order=3$, and increases for $\order=4$ and $\order=5$. 
The terms that have relatively large errors tend to be sufficiently small in magnitude relative to the average magnitude.

\section{Summary and Conclusions}
We have presented a general framework for characterizing and computing phonons and their
interactions.  The first aspect of our work is to write the Taylor series
expansion of the Born-Oppenheimer surface purely in terms of space group
irreducible derivatives. Space group irreducible derivatives guarantee 
invariance to all space group operations, homogeneity of free space, and permutation
symmetry with respect to the order of differentiation; resulting in a Taylor series
that satisfies all the possible symmetries \emph{by construction}.

We demonstrate that it should not be
assumed that numerical implementations of extrinsic symmetry approaches are capturing all symmetry (see Sections \ref{sec:SSBID} and \ref{sec:HSBID} for examples), and such
approaches may produce potentials with broken symmetry that are susceptible
to all sorts of uncontrolled errors. Space group irreducible derivatives not
only guarantee that all symmetry is satisfied by construction, but they also
provide a convenient means for storing and disseminating results; and this will
be critical to data based approaches to the physics of materials. 

The second contribution of this work was to resolve an apparently outstanding mathematical
problem regarding the translation group.
In particular, we resolve the minimum supercell problem, which is to find 
the smallest possible supercell that will accommodate 
$\order$ wavevectors in  a $\cdim$ dimensional crystal. 
We show that this problem is equivalent to constructing
the modulo $\lcmd$ kernel space of the integer matrix $\qset$ under consideration (expressed in lattice coordinates of $\rmat[BZ]$);
which we prove can be achieved using 
the Smith Normal Form, resulting in the Minimum Supercell Multiplicity (MSM) equation (Eq. \ref{eq:ssmult}). 
In practice, this approach
can always be executed with negligible computational cost. Furthermore, the MSM equation
dictates
that for any FTG of an arbitrary $\cdim$-dimensional crystal  at $\order=2$ (i.e. phonons), in addition to any FTG's corresponding to $\supa[BZ]=n\hat1$ %
at arbitrary order $\order$,
the largest necessary supercell multiplicity is $\lcmdm^{\min(\order-1,\cdim)}$.
The implication for
$\order=2$ and $d\le 3$ was only recently realized\cite{Lloyd-williams2015184301}, while the same cannot be said for $\order=\cdim=3$;
which will have a major impact for the computation of cubic interactions using finite displacement approaches.

The third contribution of this work is the formulation of two finite
displacement approaches for computing phonons and their interactions.  First,
we formulate the lone irreducible derivative (LID) approach, which measures a
single irreducible derivative, or as few as possible, at a time in the smallest
possible supercell. The LID approach is the generalization of the
original frozen phonon approach to fully exploit intrinsic symmetrization and minimal supercells at an arbitrary order. While LID does
not efficiently exploit perturbative derivatives, it should be the method of
choice when the most precise results are needed for a finite difference
calculation of a given irreducible derivative.  The second finite displacement
approach we develop is the bundled irreducible derivative (BID) approach; which
maximally exploits perturbative derivatives in order to obtain higher
derivatives via finite difference. BID guarantees that all derivatives are extracted 
in the smallest possible number of calculations. We demonstrate how to implement this BID
approach both using a single supercell approach, and using a hierarchical
supercell approach, which guarantees that all derivatives are executed in the
smallest supercell possible.  

We explicitly execute both LID and BID using the Hellman-Feynman forces (i.e. first
derivatives) for graphene; computing irreducible derivatives at $\order=2-5$.
We explicitly tabulate all irreducible derivatives for $\supa[BZ]=2\supa[K]$ at
$\order=3$; which amounts to 215 purely real or imaginary numbers. We note that 
$\supa[BZ]=2\supa[K]$ will reproduce the numerically exact Gr{\"u}neisen parameters with a 
relatively high fidelity. 
For cubic
interactions in the rock salt structure, we have demonstrated that our Hierarchical Supercell
Bundled Irreducible Derivative (HS-BID)
is more than an  order of magnitude faster than approaches
implemented in the ShengBTE, Phono3py, and AAPL codes. Corresponding speedups at second order
will be even more dramatic.

While perturbation theory should be used
to the highest order possible whenever possible, the many scenarios where it is
not yet available, which range from practical issues in some particular DFT
code or difficult technical issues associated with beyond DFT methods, imply
that finite difference will play a critical role in the foreseeable future. Our
developments will allow finite displacement based methods to be implemented as
efficiently as possible.  Finally, we emphasize that techniques which
use a first-principles molecular dynamics trajectory as data to fit phonon
interactions can also exploit our hierarchical supercell approach.

Our final development relates to assessing the quality of phonon interactions. We build upon the tradition
of using the Gr{\"u}neisen parameters as a test of cubic phonon interactions. We derive an analytic equation to 
compute the $N$-th uniform strain derivative of the phonons, which is a linear combination of the $(N+2)$-th
irreducible derivatives. The strain derivatives of the phonons are straightforward to compute, and provide a stringent,
infinite ranged test of the force tensor, which is constructed from the irreducible derivatives.

The above developments should greatly assist in advancing the computation of
phonons and their interactions, which will impact a broad range of
applications. An important point that has not been addressed in this paper is
that object oriented, modular software has been developed to implement all of
the ideas in this paper at arbitrary order
$\order$.
This 
free, open source software will be disseminated shortly, and described 
in the appropriate forum.

\begin{acknowledgments}
This work was supported by the grant DE-SC0016507 funded by the U.S. Department
of Energy, Office of Science.  This research used resources of the National
Energy Research Scientific Computing Center, a DOE Office of Science User
Facility supported by the Office of Science of the U.S. Department of Energy
under Contract No. DE-AC02-05CH11231.
\end{acknowledgments}

\appendix
\section{Applications to the rocksalt structure}
\label{app:rocksalt}
In this appendix, we consider the rock salt structure and present all space group irreducible derivatives for $\order=3$ in the supercell $\supa[BZ]=2\hat 1$.
The rocksalt structure has space group Fm$\bar3$m, and
the structure can be  defined as
\begin{align}\label{eq:rocksa}
  \lmat =  \frac{a}{2}
    \begin{bmatrix}
      0 & 1 & 1 \\[0.2em]
      1 & 0 & 1 \\[0.2em]
      1 & 1 & 0
    \end{bmatrix} & \hspace{2em} & %
\begin{aligned}
    \cbvec[1] &= \left( 0, 0, 0 \right) \\
    \cbvec[2] &= \frac{a}{2}(1,1,1)
\end{aligned}
\end{align}
where $a$ is the lattice constant. Given the FTG $\supa[BZ] = 2\hat{1} $, we have
\begin{align}
 \bqq[BZ]&= \{\Gamma, L_{a}, L_{b}, L_{c}, L_{d}, X_{x}, X_{y}, X_{z}\}
\\ 
 \bqq[IBZ]&= \{\Gamma, L_{a}, X_{x}\}
\end{align}
where
\begin{align}
  \Gamma &= (0, 0, 0)&
  L_{a}  &= (\frac{1}{2}, 0, 0) &
  L_{b}  &= (0, \frac{1}{2}, 0) \nonumber \\
  L_{c}  &= (0, 0, \frac{1}{2}) &
  L_{d}  &= (\frac{1}{2}, \frac{1}{2}, \frac{1}{2}) &
  X_{x}  &= (0, \frac{1}{2}, \frac{1}{2}) \nonumber \\
  X_{y}  &= (\frac{1}{2}, 0, \frac{1}{2}) &
  X_{z}  &= (\frac{1}{2}, \frac{1}{2}, 0)
\end{align}

The next step is to symmetrize the displacements at each $\qq\in\bqq[IBZ]$,
where $\pg_\Gamma=O_h$, $\pg_L=D_{3d}$, and $\pg_X=D_{4h}$.
Decomposing each representation in terms of irreducible representations (and removing the acoustic modes), we have:
\begin{align}
  \hat \Gamma(\qdisp[\Gamma]) &= T_{1u} \nonumber \\
  \hat \Gamma(\qdisp[L_i]) &= A_{1g} \oplus E_{g} \oplus A_{1u} \oplus E_{u} \nonumber \\
  \hat \Gamma(\qdisp[X_i]) &= 2  A_{2u} \oplus 2 E_{u} 
\end{align}

The induced representations of each $\qdisp[\qq][\alpha]$ must be constructed:
\begin{align}\label{}
\begin{array}{c|ccccc}
\qdisp[L][\alpha]                   &A_{1g} &  E_g    &  A_{2u} &  E_u     \\
\hline
\multirow{3}*{$\dispsym_{\qorbit[L]}^{\alpha_k}$}&A_{1g} &  E_g    &  A_{2u} &  E_u     \\
                                      &T_{2g} &  T_{2g} &  T_{1u} &  T_{1u}  \\
                                      &       &  T_{1g} &         &  T_{2u}  \\
\end{array}
&&
\begin{array}{c|ccccc}
\qdisp[X][\alpha]                   &A_{2u} &  E_u     \\
\hline
\multirow{2}*{$\dispsym_{\qorbit[X]}^{\alpha_k}$}&T_{1u} &  T_{1u}  \\
                                      &       &  T_{2u}  \\
\end{array}
\end{align}
where the induced representations are listed directly below each $\qdisp[\qq][\alpha]$, the index $k$ enumerates them,
and $\alpha_k$ is an irreducible representation of $O_h$. 
For the $\Gamma$ point, $\qdisp[\Gamma][\alpha]=\dispsym_{\qorbit[\Gamma]}^{\alpha}$. We now have all the information
we need to deduce if a star product can be nonzero.

For $\order=3$, the irreducible $\qsetp$ must be constructed:
\begin{align}
  \bqsetpibz = \{ & \left(\Gamma,\Gamma,\Gamma \right), \left(\Gamma,L_{a},L_{a}\right), \left(L_{a},L_{b},X_{z}\right), \nonumber \\%
                  & \left(\Gamma,X_{z},X_{z}\right), \left(X_{x},X_{y},X_{z}\right)\}
\end{align}
Next, each $\qsetp\in\bqsetpibz$ must be examined.
For $\left(\Gamma,\Gamma,\Gamma \right)$, there are only $T_{1u}$ vectors, and 
the symmetric direct product can be constructed,
\begin{align}\label{}
[T_{1u}\otimes T_{1u}\otimes T_{1u}] = A_{2u}\oplus 2T_{1u}\oplus T_{2u}
\end{align}
which does not contain the identity representation. Therefore, there are no cubic
terms contained within the primitive cell.

For $\left(\Gamma,L_{a},L_{a}\right)$, we must execute the symmetric direct product of 
all full space group irreducible representations
associated with each little group irreducible representation for each $\qq\in\qsetp$, which results in
\begin{align}
  \tensor*[^{ }]{d}{}\indices*{*_{\Gamma}^{T_{1u}}_{L_{a}}^{A_{1g}}_{L_{a}}^{A_{2u}}} && %
  \tensor*[^{ }]{d}{}\indices*{*_{\Gamma}^{T_{1u}}_{L_{a}}^{A_{1g}}_{L_{a}}^{E_{u} }} && %
  \tensor*[^{ }]{d}{}\indices*{*_{\Gamma}^{T_{1u}}_{L_{a}}^{A_{2u}}_{L_{a}}^{E_{g} }} \nonumber \\ %
  \tensor*[^{0}]{d}{}\indices*{*_{\Gamma}^{T_{1u}}_{L_{a}}^{E_{g} }_{L_{a}}^{E_{u} }} && %
  \tensor*[^{1}]{d}{}\indices*{*_{\Gamma}^{T_{1u}}_{L_{a}}^{E_{g} }_{L_{a}}^{E_{u} }}  %
\end{align}
where the left superscript indicates that multiple identity representations are produced in that product.

For $\left(L_{a},L_{b},X_{z}\right)$, we follow the same procedure, obtaining
\begin{align}
  \tensor*[^{ }]{d}{}\indices*{*_{L_{b}}^{A_{1g}}_{L_{a}}^{A_{2u}}_{X_{z}}^{A_{2u}                                    }}   &&
  \tensor*[^{ }]{d}{}\indices*{*_{L_{b}}^{A_{1g}}_{L_{a}}^{A_{2u}}_{X_{z}}^{\tensor*[^{1}]{\hspace{-0.2em}A}{}_{2u}^{}}}   &&
  \tensor*[^{ }]{d}{}\indices*{*_{L_{b}}^{A_{1g}}_{L_{a}}^{A_{2u}}_{X_{z}}^{E_{u }                                    }}   &&
  \tensor*[^{ }]{d}{}\indices*{*_{L_{b}}^{A_{1g}}_{L_{a}}^{A_{2u}}_{X_{z}}^{\tensor*[^{1}]{\hspace{-0.2em}E}{}_{u }^{}}}   \nonumber \\
  \tensor*[^{ }]{d}{}\indices*{*_{L_{b}}^{A_{1g}}_{L_{a}}^{E_{u }}_{X_{z}}^{A_{2u}                                    }}   &&
  \tensor*[^{ }]{d}{}\indices*{*_{L_{b}}^{A_{1g}}_{L_{a}}^{E_{u }}_{X_{z}}^{\tensor*[^{1}]{\hspace{-0.2em}A}{}_{2u}^{}}}   &&
  \tensor*[^{0}]{d}{}\indices*{*_{L_{b}}^{A_{1g}}_{L_{a}}^{E_{u }}_{X_{z}}^{E_{u }                                    }}   &&
  \tensor*[^{1}]{d}{}\indices*{*_{L_{b}}^{A_{1g}}_{L_{a}}^{E_{u }}_{X_{z}}^{E_{u }                                    }}   \nonumber \\
  \tensor*[^{0}]{d}{}\indices*{*_{L_{b}}^{A_{1g}}_{L_{a}}^{E_{u }}_{X_{z}}^{\tensor*[^{1}]{\hspace{-0.2em}E}{}_{u }^{}}}   &&
  \tensor*[^{1}]{d}{}\indices*{*_{L_{b}}^{A_{1g}}_{L_{a}}^{E_{u }}_{X_{z}}^{\tensor*[^{1}]{\hspace{-0.2em}E}{}_{u }^{}}}   &&
  \tensor*[^{0}]{d}{}\indices*{*_{L_{b}}^{E_{g }}_{L_{a}}^{A_{2u}}_{X_{z}}^{E_{u }                                    }}   &&
  \tensor*[^{1}]{d}{}\indices*{*_{L_{b}}^{E_{g }}_{L_{a}}^{A_{2u}}_{X_{z}}^{E_{u }                                    }}   \nonumber \\
  \tensor*[^{0}]{d}{}\indices*{*_{L_{b}}^{E_{g }}_{L_{a}}^{A_{2u}}_{X_{z}}^{\tensor*[^{1}]{\hspace{-0.2em}E}{}_{u }^{}}}   &&
  \tensor*[^{1}]{d}{}\indices*{*_{L_{b}}^{E_{g }}_{L_{a}}^{A_{2u}}_{X_{z}}^{\tensor*[^{1}]{\hspace{-0.2em}E}{}_{u }^{}}}   &&
  \tensor*[^{0}]{d}{}\indices*{*_{L_{b}}^{E_{g }}_{L_{a}}^{E_{u }}_{X_{z}}^{A_{2u}                                    }}   &&
  \tensor*[^{1}]{d}{}\indices*{*_{L_{b}}^{E_{g }}_{L_{a}}^{E_{u }}_{X_{z}}^{A_{2u}                                    }}   \nonumber \\
  \tensor*[^{0}]{d}{}\indices*{*_{L_{b}}^{E_{g }}_{L_{a}}^{E_{u }}_{X_{z}}^{E_{u }                                    }}   &&
  \tensor*[^{1}]{d}{}\indices*{*_{L_{b}}^{E_{g }}_{L_{a}}^{E_{u }}_{X_{z}}^{E_{u }                                    }}   &&
  \tensor*[^{2}]{d}{}\indices*{*_{L_{b}}^{E_{g }}_{L_{a}}^{E_{u }}_{X_{z}}^{E_{u }                                    }}   &&
  \tensor*[^{3}]{d}{}\indices*{*_{L_{b}}^{E_{g }}_{L_{a}}^{E_{u }}_{X_{z}}^{E_{u }                                    }}   \nonumber \\
  \tensor*[^{0}]{d}{}\indices*{*_{L_{b}}^{E_{g }}_{L_{a}}^{E_{u }}_{X_{z}}^{\tensor*[^{1}]{\hspace{-0.2em}E}{}_{u }^{}}}   &&
  \tensor*[^{1}]{d}{}\indices*{*_{L_{b}}^{E_{g }}_{L_{a}}^{E_{u }}_{X_{z}}^{\tensor*[^{1}]{\hspace{-0.2em}E}{}_{u }^{}}}   &&
  \tensor*[^{2}]{d}{}\indices*{*_{L_{b}}^{E_{g }}_{L_{a}}^{E_{u }}_{X_{z}}^{\tensor*[^{1}]{\hspace{-0.2em}E}{}_{u }^{}}}   &&
  \tensor*[^{3}]{d}{}\indices*{*_{L_{b}}^{E_{g }}_{L_{a}}^{E_{u }}_{X_{z}}^{\tensor*[^{1}]{\hspace{-0.2em}E}{}_{u }^{}}}   \nonumber \\
  \tensor*[^{0}]{d}{}\indices*{*_{L_{b}}^{E_{g }}_{L_{a}}^{E_{u }}_{X_{z}}^{\tensor*[^{1}]{\hspace{-0.2em}A}{}_{2u}^{}}}   &&
  \tensor*[^{1}]{d}{}\indices*{*_{L_{b}}^{E_{g }}_{L_{a}}^{E_{u }}_{X_{z}}^{\tensor*[^{1}]{\hspace{-0.2em}A}{}_{2u}^{}}}   &&
  \tensor*[^{ }]{d}{}\indices*{*_{L_{b}}^{E_{g }}_{L_{a}}^{A_{2u}}_{X_{z}}^{A_{2u}                                    }}   &&
  \tensor*[^{ }]{d}{}\indices*{*_{L_{b}}^{E_{g }}_{L_{a}}^{A_{2u}}_{X_{z}}^{\tensor*[^{1}]{\hspace{-0.2em}A}{}_{2u}^{}}}   \nonumber \\
\end{align}

The same analysis for $\left(\Gamma,X_{z},X_{z}\right)$ and $\left(X_{x},X_{y},X_{z}\right)$ proves that there are no allowed derivatives
in those two $\qsetp$.
In conclusion, there are a total of 33 space group irreducible derivatives. One can reach the same conclusion by inspecting 
the product and symmetric product tables, which were constructed to third order, for Fm$\bar 3$m by Birman \emph{et. al}\cite{Chen1968639}.

\subsection{SS-BID and HS-BID approach with $\pd[1]$ for $\order=3$ and $\supa[BZ]=2\hat 1$}

We begin by evaluating the SS-BID approach, where the maximum number of irreducible derivatives are measured simultaneously in 
the BvK supercell. In this case, we have $\nirreptotal[2\hat1]=33$ and $\nforce[2\hat 1]=2^3\cdot6-3=45$, and therefore
$\nmeasure[2\hat 1]=1$; meaning that all irreducible derivatives can be obtained in a single measurement.

However, it is clearly more efficient to avoid the BvK supercell altogether using the HS-BID approach.
Using the Smith Normal Form of the two allowed $\qsetp$, we can find the smallest supercells for each case:
\begin{align}\label{}
\supa[\left(\Gamma,L_{a},L_{a}\right)] =
\begin{bmatrix} 2 & 0 & 0 \\ 0 & 1 & 0 \\ 0 & 0 & 1 \end{bmatrix}
&&
\supa[\left(L_{a},L_{b},X_{z}\right)]=
\begin{bmatrix} 2 & 0 & 0 \\ 0 & 2 & 0 \\ 0 & 0 & 1 \end{bmatrix}
\end{align}

For $\supa[\left(\Gamma,L_{a},L_{a}\right)]$, there are $\nirreptotalup[\vee\supa[\left(\Gamma,L_{a},L_{a}\right)]]=5$ irreducible derivatives 
and there are $\nforce[\vee\supa[\left(\Gamma,L_{a},L_{a}\right)]]=2\cdot6-3=9$ nonzero force equations, so we see that
all irreducible derivatives can be obtained in a single measurement $\nmeasure[\vee\supa[\left(\Gamma,L_{a},L_{a}\right)]]=1$.
For $\left(L_{a},L_{b},X_{z}\right)$, there are $\nirreptotalup[\vee\supa[\left(L_{a},L_{b},X_{z}\right)]]=28$ irreducible derivatives and there
are $\nforce[\vee\supa[\left(L_{a},L_{b},X_{z}\right)]]=3\cdot6=18$ nonzero force equations, where we do not count the $\Gamma$-point optical modes as there are no derivatives
with respect to the $\Gamma$ being computed in this supercell; so we see that $\nmeasure[\vee\supa[\left(L_{a},L_{b},X_{z}\right)]]=2$.

Finally, we note that we can exploit overbundling in this situation, given that $\supa[\left(L_{a},L_{b},X_{z}\right)]$ also accommodates $\left(\Gamma,L_{a},L_{a}\right)$.
In this scenario, we would have $\nirreptotalup[\supa[\left(L_{a},L_{b},X_{z}\right)]]=33$ and $\nforce[\supa[\left(L_{a},L_{b},X_{z}\right)]]=4\cdot6-3=21$;
so $\nmeasure[\supa[\left(L_{a},L_{b},X_{z}\right)]]=2$  and all 33 irreducible derivatives can be obtained in two measurements. Therefore, 
we obtained all 5 irreducible derivatives from $\left(\Gamma,L_{a},L_{a}\right)$ at no cost.

\section{Phonons of ZrO$_2$}
\label{app:zro2}
In this section, we execute our method in the case of phonons of ZrO$_2$, comparing to the previously published work\cite{Parlinski19974063}.
ZrO$_2$ has space group symmetry  $Fm\bar{3}m$, 
and we study the FTG corresponding to the  $2\times2\times2$ supercell of the conventional cubic unit cell.
The primitive lattice cell vectors have the same form as rock salt, Eq. \ref{eq:rocksa}, while the conventional
cubic cell is
\begin{align}
\supa[C] = \begin{bmatrix}
\bar 1 &  1 &  1 \\
 1 & \bar 1 &  1 \\
 1 &  1 & \bar 1 \\
\end{bmatrix}
\end{align}
Therefore, the BvK supercell is $\supa[BZ]=2\supa[C]$, and we have $\nq=\det(\supa[BZ])=32$.
All of the conventions defined in rock salt will follow throughout, though we do not need subscripts to label a given $\qq\in\qorbit[\qq]$ as this is second order.
The irreducible Brillouin zone is given by
\begin{align}
\bqq[IBZ] = \{\Gamma, L, X, A, \Delta, W\}
\end{align}
where
\begin{align}
\Gamma &= \left(0, 0, 0\right) &
L  &= \left(\frac{1}{2}, 0, 0\right) &
X      &= \left(\frac{1}{2}, \frac{1}{2}, 0 \right) \nonumber \\
A      &= \left(\frac{1}{4}, \frac{3}{4}, 0\right)  &
\Delta &= \left(\frac{1}{4}, \frac{1}{4}, 0\right) &
W      &= \left(\frac{1}{4}, \frac{3}{4}, \frac{1}{2}\right) 
\end{align}

The irreducible representations of the displacements according to $\pg_{\qq}$ are:
\begin{align}\label{eq:zro2symirreps}
\hat\Gamma(\qdisp[\Gamma]) =& \left( T_{2g} \right) \oplus T_{1u} \nonumber \\
\hat\Gamma(\qdisp[L     ]) =& \left( A_{1g} \oplus {}^{1}\hspace{-0.2em}A_{2u} \oplus E_{g} \oplus {}^{1}\hspace{-0.2em}E_{u} \right) \oplus \nonumber \\
                            & \left( A_{2u} \oplus E_{u} \right) \nonumber \\
\hat\Gamma(\qdisp[X     ]) =& \left( A_{1g} \oplus B_{1u} \oplus E_{g} \oplus {}^{1}\hspace{-0.2em}E_{u} \right) \oplus \nonumber \\
                            & \left( A_{2u} \oplus E_{u} \right) \nonumber \\
\hat\Gamma(\qdisp[A     ]) =& \left( {}^{1}\hspace{-0.2em}A_{1} \oplus {}^{2}\hspace{-0.2em}A_{1} \oplus A_{2} \oplus {}^{1}\hspace{-0.2em}B_{1} \oplus %
                              {}^{2}\hspace{-0.2em}B_{1} \oplus {}^{1}\hspace{-0.2em}B_{2} \right) \oplus \nonumber \\
                            & \left( A_{1} \oplus B_{1} \oplus B_{2} \right) \nonumber \\
\hat\Gamma(\qdisp[\Delta]) =& \left( {}^{1}\hspace{-0.2em}A_{1} \oplus B_{2} \oplus {}^{1}\hspace{-0.2em}E \oplus {}^{2}\hspace{-0.2em}E \right) \oplus \nonumber \\
                            & \left( A_{1} \oplus E \right) \nonumber \\
\hat\Gamma(\qdisp[W     ]) =& \left( A_{1} \oplus A_{2} \oplus {}^{1}\hspace{-0.2em}B_{1} \oplus B_{2} \oplus {}^{1}\hspace{-0.2em}E \right) \oplus \nonumber \\
                            & \left( B_{1} \oplus E \right)
\end{align}
where the first set of parenthesis enclose irreducible representations purely associated with O atoms, while the second set correspond purely to Zr.
The $T_{1u}$ mode, which is not enclosed in any parenthesis, is a mixture of Zr and O atoms.

The number of irreducible derivatives can be determined by inspecting Eq. \ref{eq:zro2symirreps}, counting once for each irreducible
representation and once for each pair of repeating irreducible representations at a given $\qq$.
All 52 space group irreducible derivatives are listed in Table \ref{tab:zro2_irrterms}, and they may be chosen to be real given the presence of inversion and
time reversal symmetry. 
This proves that the analysis in Ref. \onlinecite{Parlinski19974063} did not properly account for all symmetry, as they arrived
at 59 nonzero parameters. %

\begin{figure}[htpb]
\centering
\includegraphics[width=0.99\linewidth]{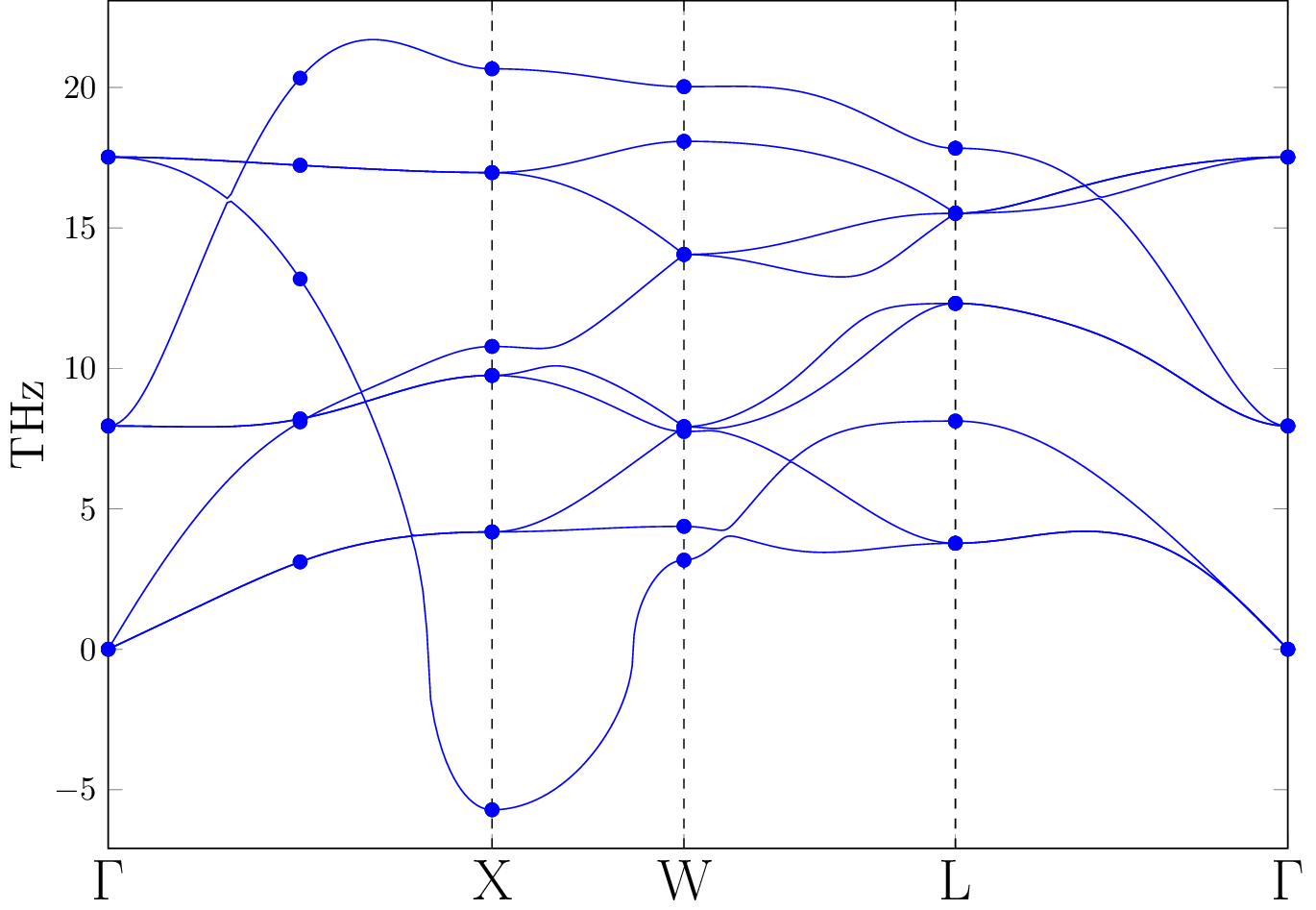}
\caption{
  Phonons of ZrO${}_2$ within DFT for $\supa[BZ]=2\supa[C]$, where data points are direct computational measurements and lines are Fourier interpolation of the 
  measurements. The irreducible derivatives for $\supa[BZ]$ are shown in Table \ref{tab:zro2_irrterms}. 
  When irreducible representations do not repeat at a given $\qq$, the phonon frequency is given by
  $\omega_{\qq}^{\alpha}=\sqrt{\dqnirrsym\indices*{*_{\bar \qq}^{\alpha}_{\qq}^{\alpha}}/m}$, 
  where $m=m_{i}\cdot 1.0364\times 10^{-28}\textrm{eV}\cdot \textrm{s}^2/\AA^2 $, with either
  $m_{O} = 15.9994$ or $m_{Zr} = 91.224$. The $y$-axis plots $\omega_{\qq}^{\alpha}\cdot10^{-12}/(2\pi)$, giving units of THz.
  LO-TO splitting has not been incorporated.
}
\label{fig:zro2_phonon}
\end{figure}

We now turn to extracting these 52 irreducible derivatives using SS-BID, and we can use the specific second order equation for the number of measurements in Eq. \ref{eq:nmphonons}.
The result is that $\nmeasure[2\supa[C]]=1$, and all derivatives can be extracted from a single measurement; as compared to the two measurements (i.e. four calculations)
in the original study\cite{Parlinski19974063}. We demonstrate the result of this single measurement, providing all space group irreducible derivatives in Table \ref{tab:zro2_irrterms},
along with a plot of the phonons in Figure \ref{fig:zro2_phonon}.

The execution of the SS-BID is only a proof of principle, as in practice one would always perform HS-BID as it far more efficient.
Indeed,  HS-BID completely avoids the BvK supercell $\supa[BZ]=2\supa[C]$, 
extracting all irreducible derivatives from smaller supercells. In this case, we have  
$\bsupastar=\{\supa[\Gamma],\supa[L],\supa[X],\supa[A],\supa[\Delta],\supa[W] \}$, where
\begin{align}
&\supa[\Gamma] = \hat 1 &
\supa[L] = \begin{bmatrix} 2 & 0 & 0 \\ 0 & 1 & 0 \\ 0 & 0 & 1\end{bmatrix} &
\supa[X] = \begin{bmatrix} 1 & 1 & 0 \\ 0 & 2 & 0 \\ 0 & 0 & 1\end{bmatrix} \nonumber \\
& \supa[A] = \begin{bmatrix} 1 & 1 & 0 \\ 0 & 4 & 0 \\ 0 & 0 & 1\end{bmatrix} &
\supa[\Delta] = \begin{bmatrix} 2 & 2 & 0 \\ 1 & 3 & 0 \\ 0 & 0 & 1\end{bmatrix} &
\supa[W] = \begin{bmatrix} 2 & 0 & 1 \\ 1 & 1 & 0 \\ 0 & 0 & 2\end{bmatrix}
\label{eq:zro2:supas}
\end{align}
The number of calculations required in each supercell is 1, 2, 1, 2, 1, and 1, respectively.
The gain in time complexity as is described in Section \ref{sec:HSBID} is an order of magnitude.
Furthermore, overbundling can be exploited with only computing $\supa[A],\supa[\Delta],\supa[W]$
supercells with 2, 1, and 1 calculations respectively, increasing additional efficiency by avoiding smaller supercells.

Density Functional Theory (DFT) calculations within the local density approximation (LDA) \cite{Perdew19815048}%
were performed using the
Projector Augmented Wave (PAW) method \cite{Blochl199417953,Kresse19991758},
as implemented in the Vienna Ab-initio Simulation Package (VASP) \cite{Kresse1993558,Kresse199414251,Kresse199615,Kresse199611169}. 
A plane wave basis with a kinetic energy cutoff of 700 eV was employed. We used a $\Gamma$-centered \textbf{k}-point 
mesh of 4$\times$4$\times$4. 
All $k$-point integrations were done using tetrahedron method with Bl\"ochl corrections\cite{Blochl199416223}.
The crystal structure was relaxed, yielding a lattice parameter of 5.0303\AA.

\begin{table}
\centering
\caption{Irreducible derivatives of ZrO$_2$ for $\order=2$ and  $\supa[BZ]=2\supa[C]$ in units of $eV/\AA^2$.}
\label{tab:zro2_irrterms}
\begin{tabular}{rrrr}
\hline\hline
Derivative & Value & Derivative & Value \\
\hline\\[-0.9em]
                                                                        $d\indices*{*_{\Gamma}^{T_{2g}^{}}_{\Gamma}^{T_{2g}^{}}}$ &   $20.105$ & %
                                                                        $d\indices*{*_{\Gamma}^{T_{1u}^{}}_{\Gamma}^{T_{1u}^{}}}$ &    $9.189$ \\[0.4em]
\hline\\[-0.9em]                                                                                                                                             
                                                                          $d\indices*{*_{L}^{A_{1g}^{}}_{L}^{A_{1g}^{}}}$ &   $20.828$ & %
                                                                            $d\indices*{*_{L}^{E_{g}^{}}_{L}^{E_{g}^{}}}$ &   $15.766$ \\[0.4em]
                                                                          $d\indices*{*_{L}^{A_{2u}^{}}_{L}^{A_{2u}^{}}}$ &   $33.220$ & %
                                         $d\indices*{*_{L}^{A_{2u}^{}}_{L}^{\tensor*[^{1}]{\hspace{-0.2em}A}{}_{2u}^{}}}$ &  $- 9.242$ \\[0.4em]
        $d\indices*{*_{L}^{\tensor*[^{1}]{\hspace{-0.2em}A}{}_{2u}^{}}_{L}^{\tensor*[^{1}]{\hspace{-0.2em}A}{}_{2u}^{}}}$ &   $14.278$ & %
                                                                            $d\indices*{*_{L}^{E_{u}^{}}_{L}^{E_{u}^{}}}$ &    $5.937$ \\[0.4em]
                                           $d\indices*{*_{L}^{E_{u}^{}}_{L}^{\tensor*[^{1}]{\hspace{-0.2em}E}{}_{u}^{}}}$ &  $- 2.346$ & %
          $d\indices*{*_{L}^{\tensor*[^{1}]{\hspace{-0.2em}E}{}_{u}^{}}_{L}^{\tensor*[^{1}]{\hspace{-0.2em}E}{}_{u}^{}}}$ &    $9.819$ \\[0.4em]
\hline\\[-0.9em]                                                                                                                                             
                                                                                  $d\indices*{*_{X}^{A_{1g}^{}}_{X}^{A_{1g}^{}}}$ &   $27.957$ & %
                                                                                    $d\indices*{*_{X}^{E_{g}^{}}_{X}^{E_{g}^{}}}$ &    $6.221$ \\[0.4em]
                                                                                  $d\indices*{*_{X}^{A_{2u}^{}}_{X}^{A_{2u}^{}}}$ &   $43.374$ & %
                                                                                  $d\indices*{*_{X}^{B_{1u}^{}}_{X}^{B_{1u}^{}}}$ &  $- 2.139$ \\[0.4em]
                                                                                    $d\indices*{*_{X}^{E_{u}^{}}_{X}^{E_{u}^{}}}$ &    $6.614$ & %
                                                   $d\indices*{*_{X}^{E_{u}^{}}_{X}^{\tensor*[^{1}]{\hspace{-0.2em}E}{}_{u}^{}}}$ &  $- 1.339$ \\[0.4em]
                  $d\indices*{*_{X}^{\tensor*[^{1}]{\hspace{-0.2em}E}{}_{u}^{}}_{X}^{\tensor*[^{1}]{\hspace{-0.2em}E}{}_{u}^{}}}$ &   $18.834$ & \\[0.4em]
\hline\\[-0.9em]                                                                                                                                             
                                                                              $d\indices*{*_{\bar{A}}^{A_{1}^{}}_{A}^{A_{1}^{}}}$ &   $30.934$ & %
                                             $d\indices*{*_{\bar{A}}^{A_{1}^{}}_{A}^{\tensor*[^{1}]{\hspace{-0.2em}A}{}_{1}^{}}}$ &    $4.000$ \\[0.4em]
                                             $d\indices*{*_{\bar{A}}^{A_{1}^{}}_{A}^{\tensor*[^{2}]{\hspace{-0.2em}A}{}_{1}^{}}}$ & $- 10.720$ & %
            $d\indices*{*_{\bar{A}}^{\tensor*[^{1}]{\hspace{-0.2em}A}{}_{1}^{}}_{A}^{\tensor*[^{1}]{\hspace{-0.2em}A}{}_{1}^{}}}$ &   $22.021$ \\[0.4em]
            $d\indices*{*_{\bar{A}}^{\tensor*[^{1}]{\hspace{-0.2em}A}{}_{1}^{}}_{A}^{\tensor*[^{2}]{\hspace{-0.2em}A}{}_{1}^{}}}$ &  $- 1.691$ & %
            $d\indices*{*_{\bar{A}}^{\tensor*[^{2}]{\hspace{-0.2em}A}{}_{1}^{}}_{A}^{\tensor*[^{2}]{\hspace{-0.2em}A}{}_{1}^{}}}$ &   $21.250$ \\[0.4em]
                                                                              $d\indices*{*_{\bar{A}}^{B_{2}^{}}_{A}^{B_{2}^{}}}$ &    $6.135$ & %
                                             $d\indices*{*_{\bar{A}}^{B_{2}^{}}_{A}^{\tensor*[^{1}]{\hspace{-0.2em}B}{}_{2}^{}}}$ &  $- 2.558$ \\[0.4em]
                                             $d\indices*{*_{\bar{A}}^{B_{1}^{}}_{A}^{\tensor*[^{2}]{\hspace{-0.2em}B}{}_{1}^{}}}$ &    $2.580$ & %
            $d\indices*{*_{\bar{A}}^{\tensor*[^{1}]{\hspace{-0.2em}B}{}_{2}^{}}_{A}^{\tensor*[^{1}]{\hspace{-0.2em}B}{}_{2}^{}}}$ &   $12.342$ \\[0.4em]
            $d\indices*{*_{\bar{A}}^{\tensor*[^{1}]{\hspace{-0.2em}B}{}_{1}^{}}_{A}^{\tensor*[^{2}]{\hspace{-0.2em}B}{}_{1}^{}}}$ &  $- 1.515$ & %
            $d\indices*{*_{\bar{A}}^{\tensor*[^{2}]{\hspace{-0.2em}B}{}_{1}^{}}_{A}^{\tensor*[^{2}]{\hspace{-0.2em}B}{}_{1}^{}}}$ &   $10.480$ \\[0.4em]
                                                                              $d\indices*{*_{\bar{A}}^{A_{2}^{}}_{A}^{A_{2}^{}}}$ &   $13.783$ & %
                                                                              $d\indices*{*_{\bar{A}}^{B_{1}^{}}_{A}^{B_{1}^{}}}$ &   $13.532$ \\[0.4em]
                                             $d\indices*{*_{\bar{A}}^{B_{1}^{}}_{A}^{\tensor*[^{1}]{\hspace{-0.2em}B}{}_{1}^{}}}$ &  $- 2.990$ & %
            $d\indices*{*_{\bar{A}}^{\tensor*[^{1}]{\hspace{-0.2em}B}{}_{1}^{}}_{A}^{\tensor*[^{1}]{\hspace{-0.2em}B}{}_{1}^{}}}$ &    $3.481$ \\[0.4em]
\hline\\[-0.9em]                                                                                                                                             
                                                                    $d\indices*{*_{\bar{\Delta}}^{A_{1}^{}}_{\Delta}^{A_{1}^{}}}$ &   $41.658$ & %
                                   $d\indices*{*_{\bar{\Delta}}^{A_{1}^{}}_{\Delta}^{\tensor*[^{1}]{\hspace{-0.2em}A}{}_{1}^{}}}$ & $- 18.449$ \\[0.4em]
  $d\indices*{*_{\bar{\Delta}}^{\tensor*[^{1}]{\hspace{-0.2em}A}{}_{1}^{}}_{\Delta}^{\tensor*[^{1}]{\hspace{-0.2em}A}{}_{1}^{}}}$ &   $24.050$ & %
                                                                    $d\indices*{*_{\bar{\Delta}}^{B_{2}^{}}_{\Delta}^{B_{2}^{}}}$ &   $11.378$ \\[0.4em]
                                                                      $d\indices*{*_{\bar{\Delta}}^{E_{}^{}}_{\Delta}^{E_{}^{}}}$ &    $6.380$ & %
                                     $d\indices*{*_{\bar{\Delta}}^{E_{}^{}}_{\Delta}^{\tensor*[^{1}]{\hspace{-0.2em}E}{}_{}^{}}}$ &  $- 2.896$ \\[0.4em]
                                     $d\indices*{*_{\bar{\Delta}}^{E_{}^{}}_{\Delta}^{\tensor*[^{2}]{\hspace{-0.2em}E}{}_{}^{}}}$ &    $0.918$ & %
    $d\indices*{*_{\bar{\Delta}}^{\tensor*[^{1}]{\hspace{-0.2em}E}{}_{}^{}}_{\Delta}^{\tensor*[^{1}]{\hspace{-0.2em}E}{}_{}^{}}}$ &    $4.395$ \\[0.4em]
    $d\indices*{*_{\bar{\Delta}}^{\tensor*[^{1}]{\hspace{-0.2em}E}{}_{}^{}}_{\Delta}^{\tensor*[^{2}]{\hspace{-0.2em}E}{}_{}^{}}}$ &    $2.673$ & %
    $d\indices*{*_{\bar{\Delta}}^{\tensor*[^{2}]{\hspace{-0.2em}E}{}_{}^{}}_{\Delta}^{\tensor*[^{2}]{\hspace{-0.2em}E}{}_{}^{}}}$ &   $18.958$ \\[0.4em]
\hline\\[-0.9em]                                                                                                                                             
                                                                              $d\indices*{*_{\bar{W}}^{A_{1}^{}}_{W}^{A_{1}^{}}}$ &   $26.263$ & %
                                                                              $d\indices*{*_{\bar{W}}^{B_{1}^{}}_{W}^{B_{1}^{}}}$ &    $7.193$ \\[0.4em]
                                             $d\indices*{*_{\bar{W}}^{B_{1}^{}}_{W}^{\tensor*[^{1}]{\hspace{-0.2em}B}{}_{1}^{}}}$ &    $0.973$ & %
            $d\indices*{*_{\bar{W}}^{\tensor*[^{1}]{\hspace{-0.2em}B}{}_{1}^{}}_{W}^{\tensor*[^{1}]{\hspace{-0.2em}B}{}_{1}^{}}}$ &   $21.398$ \\[0.4em]
                                                                              $d\indices*{*_{\bar{W}}^{A_{2}^{}}_{W}^{A_{2}^{}}}$ &    $0.659$ & %
                                                                              $d\indices*{*_{\bar{W}}^{B_{2}^{}}_{W}^{B_{2}^{}}}$ &    $3.936$ \\[0.4em]
                                                                                $d\indices*{*_{\bar{W}}^{E_{}^{}}_{W}^{E_{}^{}}}$ &   $23.596$ & %
                                               $d\indices*{*_{\bar{W}}^{E_{}^{}}_{W}^{\tensor*[^{1}]{\hspace{-0.2em}E}{}_{}^{}}}$ &    $1.211$ \\[0.4em]
              $d\indices*{*_{\bar{W}}^{\tensor*[^{1}]{\hspace{-0.2em}E}{}_{}^{}}_{W}^{\tensor*[^{1}]{\hspace{-0.2em}E}{}_{}^{}}}$ &   $12.905$ &  \\[0.2em]
\hline\hline
\end{tabular}
 \end{table}

\section{Example of minimum cell for given $\qset$}
\label{app:SNF}
Here we give an example illustrating how to find the minimum supercell that accommodates a given $\qset$.
There is no need to specify a crystal structure given that this problem is only specific to the translation group.
Let us consider an example for $\order=3$:
\begin{align}
\qset = \left\{ \left( \frac{1}{4}, \frac{3}{4}, \frac{1}{2} \right), \left( \frac{1}{4}, \frac{1}{4}, 0 \right) %
               \left( \frac{1}{2}, 0, \frac{1}{2} \right) \right\}
\end{align}

Here $\lcmd = 4$, and we can now drop the third row, for example, and rewrite in terms of lattice coordinates of $\rmat[BZ]$:
\begin{align}
  \qset[]['] = %
  \begin{bmatrix}
    1 & 3 & 2 \\
    1 & 1 & 0 
  \end{bmatrix}
\end{align}
We can apply row and column operations $\hat R$  and $\hat C$ in order to achieve the Smith Normal Form $\snf=  \hat R \qset[][\prime] \hat C$:
\begin{align}
  \snf= 
  \begin{bmatrix}
    1 & 0 & 0 \\
    0 & 2 & 0 \\
  \end{bmatrix}
  &&
  \hat R = %
  \begin{bmatrix}
    0 &  1  \\
    1 & -1  \\
  \end{bmatrix}
  &&
  \hat C = %
  \begin{bmatrix}
    1 & -1 &  1 \\
    0 &  1 & -1 \\
    0 &  0 &  1 \\
  \end{bmatrix}
\end{align}

It is straightforward to write the kernel of $\snf$:
\begin{align}
  \ker(\snf) = %
  \begin{bmatrix}
    4 &  0 & 0 \\
    0 &  2 & 0 \\
    0 &  0 & 1 \\
  \end{bmatrix}
\end{align}
Finally, the kernel of $\ker(\qset[]['])$ can be easily constructed, in addition to a minimal supercell which accommodates $\qset$:
\begin{align}
  \ker(\qset[][']) = \hat C\ker( \snf) = %
  \begin{bmatrix}
    4 & -2 &  1 \\
    0 &  2 & -1 \\
    0 &  0 &  1 \\
  \end{bmatrix} =\supa[\qset]^\tp
\end{align}
We emphasize that this particular choice of supercell is not unique, and may be reshaped.

\bibliography{main}

\end{document}


\title{A group theoretical approach to computing phonons and their interactions}
\author{Lyuwen Fu}
\email{lyuwen.fu@columbia.edu}
\author{Mordechai Kornbluth}
\email{mckornbluth@gmail.com}
\author{Zhengqian Cheng}
\email{chengzhengqian@gmail.com}
\author{Chris A. Marianetti}
\email{chris.marianetti@columbia.edu}
\affiliation{Department of Applied Physics and Applied Mathematics, Columbia University, New York, NY 10027}

\section{Supplementary Information}

\begin{table}[htpb]
\caption{A Glossary of the key variables used throughout the manuscript.}
\label{si:table:glossary}
\begin{tabular}{c|m{15cm}}
  $\cdim$         & Dimensionality of the crystal, given by the number of generators of the translation group. \\
  $\natom$        & The number of atoms in the primitive unit cell. \\
  $\npol$         & The number of displacement polarizations for each atom. \\
  $\nq$         & The number of $q$-points within the FBZ, or $t$-points in BvK supercell: $\nq=\det(\supa[BZ])$. \\
  $\pg$         & The point group of the space group.  \\
  $\pgorder$    & The order of the point group of the space group. \\
  $\lvec[i]$      & A primitive, real space lattice vector. The index $i$ is enumerated over $\cdim$ different vectors. \\
  $\lmat$         & A rank-$\cdim$ matrix of row-stacked $\lvec$. \\
  $\cbvec[i]$      & A Cartesian vector specifying a basis atom, with $i\in[1,\natom]$.  \\
  $\rvec$         & A primitive, reciprocal space lattice vector, with $i\in[1,\cdim]$.  \\
  $\rmat$         & A column-stacked matrix of $\rvec$, defined from $\lmat \rmat =2\pi \hat 1$. \\
  $\ctrans$       & A  vector of integers, $\ctrans \in \mathbb{Z}^d$, defining a lattice translation $\ctrans\lmat$ or a reciprocal lattice 
                    translation $\rmat[BZ]\ctrans$. \\
  $\qq$        & A  $\cdim$ dimensional vector of rational fractions, $0\le \qqc[i]<1$, defining a reciprocal lattice point $\rmat\qq$. \\
  $\supa$      & A supercell matrix, which is a rank-$\cdim$ matrix of integers with $\det(\supa)\ne0$. \\
  $\supa[BZ]$  & The supercell matrix which creates the BvK supercell; must retain point symmetry of lattice.   \\
  $\lmat[BZ]$  & Matrix of lattice vectors of  BvK supercell $\lmat[BZ]=\supa[BZ]\lmat$.  \\
  $\rmat[BZ]$  & Matrix of reciprocal lattice vectors of Brillouin zone subcell $\rmat[BZ]=\rmat\supa[BZ]^{-1}$.   \\
  $\bctrans[BZ]$        & A set of length $\nq$ containing each $\ctrans$ in the BvK supercell. \\
  $\bqq[BZ]$        & A set of length $\nq$ containing each $\qq$ in the first Brillouin zone. \\
  $\bqq[IBZ]$      & A set containing each $\qq$ in the irreducible Brillouin zone. \\
  $\qorbit$      & A set composed of point equivalent $\qq$, known as a ``star". The total number of stars in a $\supa[BZ]$ is $|\bqq[IBZ]|$. \\
  $\qorbit[BZ]$      & A set composed of all stars;  $|\qorbit[BZ]|=|\bqq[IBZ]|$. \\
  $\rdispvec[\ctrans][a][\alpha]$ 
                  & Real space displacement amplitude from lattice cell $\ctrans$ and nucleus $a$, along direction $\alpha$. \\
  $\qdispvec$     & Reciprocal space displacement amplitude of mode $\qq$ and nucleus $a$, along direction $\alpha$. \\
  $\order$        & Order of the derivative of the function defined over the lattice displacements. \\
  $\qset$         & A  $\order\times\cdim$ matrix of $\order$ row stacked $\qq[i]$; and $(\qset)$  implies the $\order$-tuple $(\qq[i],\dots,\qq[\order])$. \\
  $\qsetp$         & The \emph{multiset}  $\qsetp=\{\qq|\qq\in(\qset)\}$. \\
  $\orbit$        & The star of  $\qsetp$. The total number of $\orbit$ in a $\supa[BZ]$ is $|\bqsetpibz|$. \\
  $\borbit$       & The set of all $\{\orbit\}$ where $\supa$ accommodates at least one member $\qsetp\in\orbit$. \\
  $\borbitup$    & The set of all $\{\orbit\}$ where $\supa$ has the smallest $\det(\supa)$ that accommodates at least one member $\qset\in\orbit$, and that member will not fit in any smaller supercell. \\
  $\bqsetp[\supa]$ & The set of all $\qsetp$ within the supercell $\supa$. The case $\bqsetp[\supa[BZ]]$ may be abbreviated as $\bqsetpbz$.\\
  $\bqsetpibz$   & A set of all point irreducible $\qsetp$ in the first Brillouin zone. \\[0.5em]
  $\tset$         & A  $\order\times\cdim$ matrix of $\order$ row stacked lattice vectors $\ctrans\in\bctrans[BZ]$. \\
  $\btset[\supa]$   & The set  of all $\tset$ within the supercell $\supa[]$. $\btset[\supa[BZ]]$ may be abbreviated as $\btset[BZ]$. \\
  $\bsupastar$       & 
  A smallest set of $\supa$ matrices which can accommodate at least one $\qsetp\in\orbit$ for all $\orbit\in\borbit[BZ]$, with the constraint that each $\supa$ has the minimum $|\det(\supa)|$ allowed by the translation group. \\
  $\nirrep$   & The number of irreducible derivatives at order $\order$ for any $\bqset\in\orbit$. \\
  $\nirreptotal$   & The total number of irreducible derivatives at order $\order$ and supercell $\supa$. \\
  $\nforce$     & The number of nonzero force equations in $\supa$. \\
  $\dqn[\qset]$   & Dynamical tensor, which is the derivative of the energy with respect displacements that transform as space group irreducible representations. \\
  $\dqnirrsym\indices*{*_{\qq[1]}^{\alpha_1}_{\cdots}^{\cdots}_{\qq[\order]}^{\alpha_\order}}$  & Space group irreducible derivative, where $\alpha_i$ labels a given irreducible representations, independent of row.   \\
  $\phin[\tset]$  & Force tensor, which is the derivative of the energy with respect to real space displacements.  \\
\end{tabular}
\end{table}

\begin{figure}[htpb]
  \centering
  \includegraphics[width=0.75\linewidth]{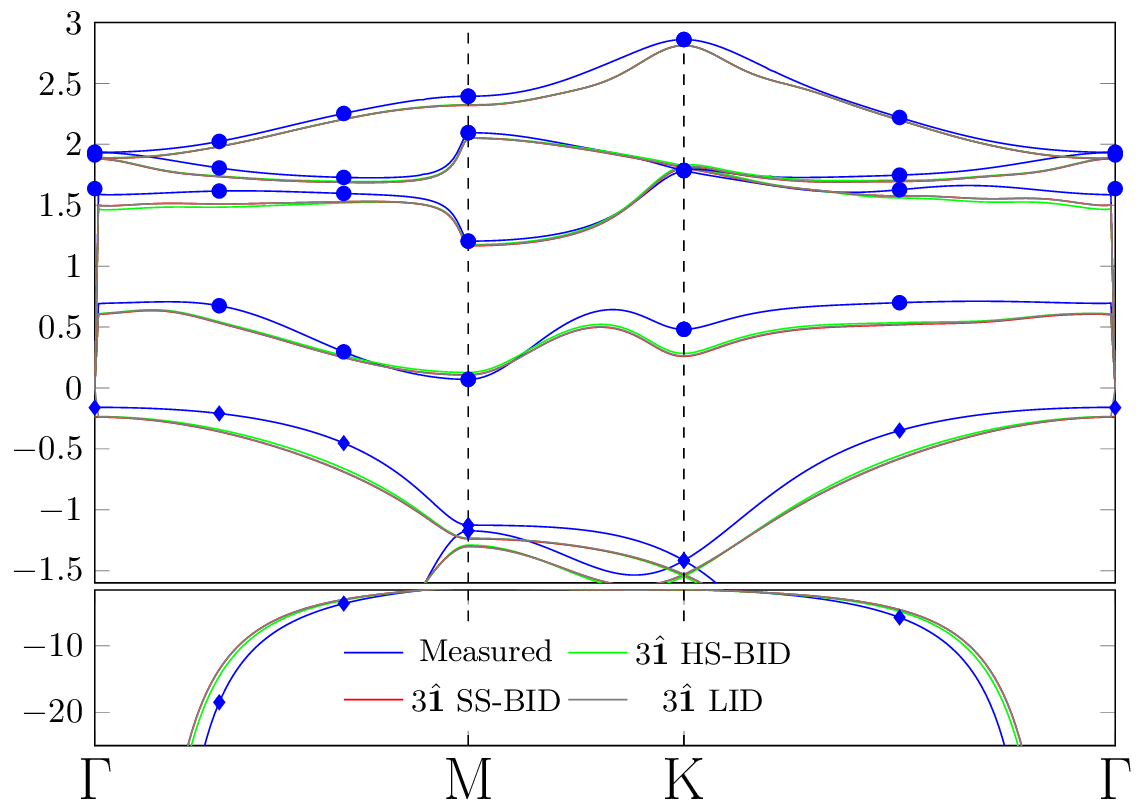}
  \caption{Gr{\"u}neisen parameters for graphene, following the same conventions as Fig. 6a in the main text. This figure compares the LID, HS-BID, and SS-BID method for $\supa[BZ] = 3 \hat 1$,
  showing very good agreement. The LID and HS-BID are so close that they are difficult to distinguish.}
  \label{si:fig:ssvhs}
\end{figure}

\begin{figure}[htpb]
  \centering
  \includegraphics[width=0.75\linewidth]{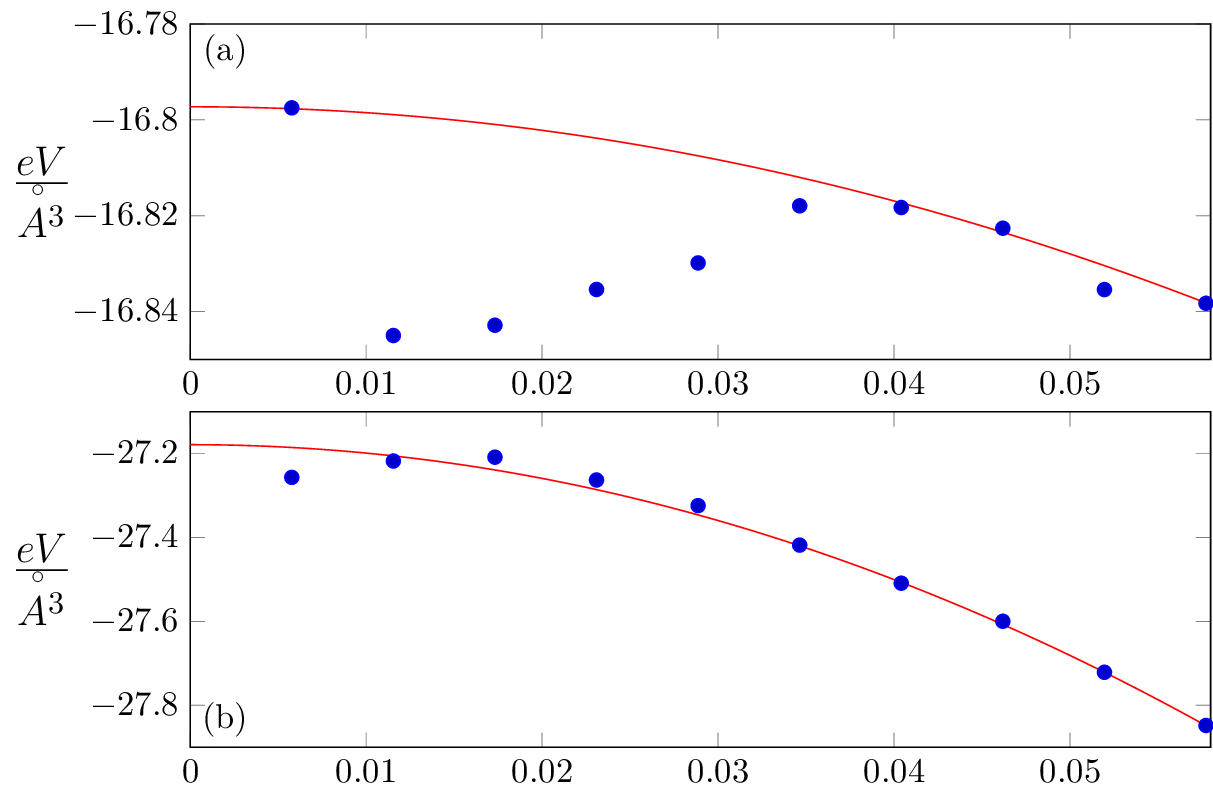}
  \caption{Selected error tails from calculations using the SBB basis in the SS-BID method for $\supa[BZ] = 2 \supa[K]$ in graphene.
  Panel (a) displays one of many types of subtleties that may occur when attempting to simultaneously measure many derivatives within the SS-BID method.
  It is possible that our algorithm for fitting quadratics (see Section III C) has made a poor choice for selecting a subset of points, as
  points 2 through 6 may comprise the true error tail; though the relative error would be less than a half of a percent. Requiring selected points
  to be consecutive would circumvent this issue, but we have found nonconsecutive points to be helpful in other scenarios. Panel (b) is 
  relatively well behaved.
  }
  \label{si:fig:sdg_2K_errortail}
\end{figure}

\begin{figure}[htpb]
  \centering
  \includegraphics[width=0.75\linewidth]{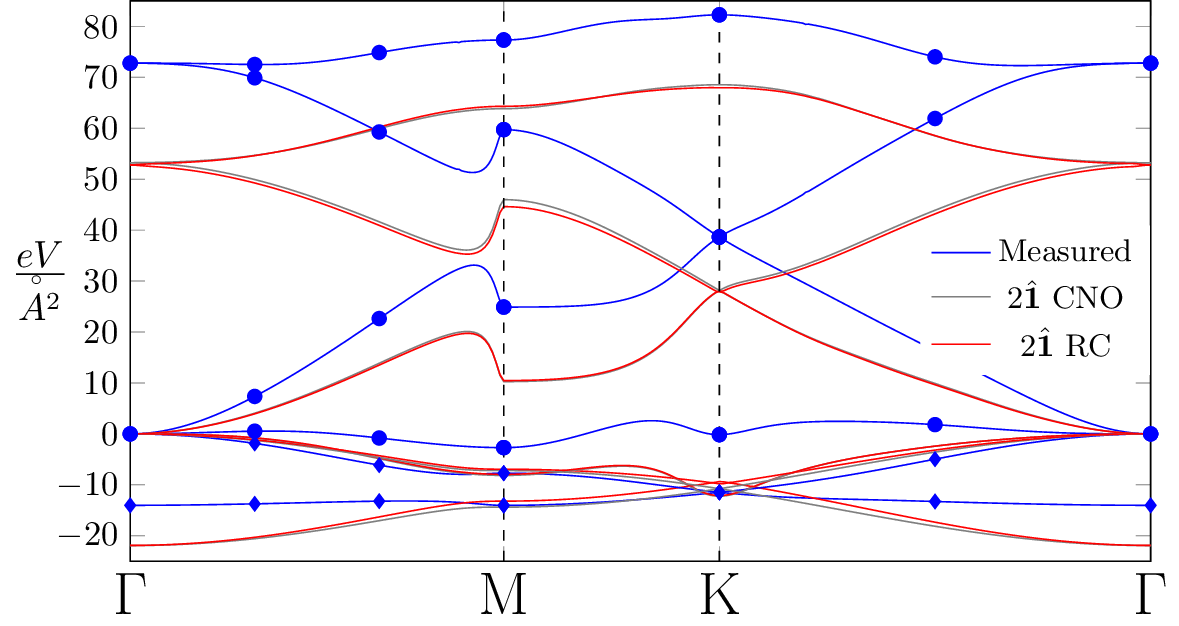}
  \caption{A plot of the second identity strain derivative of the phonons, following the same conventions as Fig. 6b in the main text. 
  Red and gray curves compare the result of using the CNO and SBB bundled basis within SS-BID to extract irreducible derivative from $\supa[BZ] = 2 \hat 1$
  at $\order=4$. The two approaches are in good agreement, though small differences can be observed.
  }
  \label{si:fig:cno_v_rc}
\end{figure}

\clearpage
\begin{longtable*}{l|rrrrrr}
\caption{Symmetrized displacement amplitudes $\dispsym_{\qq}^\alpha$ for all $\qq\in\bqq[BZ]$. The two carbon atoms in the primitive unit cell are denoted $\textrm{C}_1$ and
$\textrm{C}_2$.
         \label{si:table:qbasis}
         }\\
\hline\\[-0.8em]
 & $\rdispvec[\qq][\textrm{C}_1][x]$ & $\rdispvec[\qq][\textrm{C}_1][y]$ & $\rdispvec[\qq][\textrm{C}_1][z]$  & $\rdispvec[\qq][\textrm{C}_2][x]$  & $\rdispvec[\qq][\textrm{C}_2][y]$ & $\rdispvec[\qq][\textrm{C}_2][z]$  \\[0.2em]
\hline
\\[\dimexpr-\normalbaselineskip+0.4em]
\endfirsthead
\hline\\[-0.8em]
 & $\rdispvec[\qq][\textrm{C}_1][x]$ & $\rdispvec[\qq][\textrm{C}_1][y]$ & $\rdispvec[\qq][\textrm{C}_1][z]$  & $\rdispvec[\qq][\textrm{C}_2][x]$  & $\rdispvec[\qq][\textrm{C}_2][y]$ & $\rdispvec[\qq][\textrm{C}_2][z]$  \\[0.2em]
\hline
\\[\dimexpr-\normalbaselineskip+0.4em]
\endhead
\hline\endfoot
${u_{\Gamma}^{B_{2g}^{}}}$ & $  0 $ & $  0 $ & $  \frac{ \sqrt{2}}{2} $ & $  0 $ & $  0 $ & $ -\frac{ \sqrt{2}}{2} $ \\[0.4em]
\multirow{2}{*}{${u_{\Gamma}^{E_{2g}^{}}}$} & $  0 $ & $  \frac{ \sqrt{2}}{2} $ & $  0 $ & $  0 $ & $ -\frac{ \sqrt{2}}{2} $ & $  0 $ \\[0.4em]
                                            & $ -\frac{ \sqrt{2}}{2} $ & $  0 $ & $  0 $ & $  \frac{ \sqrt{2}}{2} $ & $  0 $ & $  0 $ \\[0.4em]
${u_{M_{x}}^{A_{1g}^{}}}$ & $  \frac{ \sqrt{2}}{2} $ & $  0 $ & $  0 $ & $ -\frac{ \sqrt{2}}{2} $ & $  0 $ & $  0 $ \\[0.4em]
${u_{M_{x}}^{B_{3g}^{}}}$ & $  0 $ & $ -\frac{ \sqrt{2}}{2} $ & $  0 $ & $  0 $ & $  \frac{ \sqrt{2}}{2} $ & $  0 $ \\[0.4em]
${u_{M_{x}}^{B_{2g}^{}}}$ & $  0 $ & $  0 $ & $ -\frac{ \sqrt{2}}{2} $ & $  0 $ & $  0 $ & $  \frac{ \sqrt{2}}{2} $ \\[0.4em]
${u_{M_{x}}^{B_{1u}^{}}}$ & $  \frac{ \sqrt{2}}{2} $ & $  0 $ & $  0 $ & $  \frac{ \sqrt{2}}{2} $ & $  0 $ & $  0 $ \\[0.4em]
${u_{M_{x}}^{B_{2u}^{}}}$ & $  0 $ & $ -\frac{ \sqrt{2}}{2} $ & $  0 $ & $  0 $ & $ -\frac{ \sqrt{2}}{2} $ & $  0 $ \\
\hline
 \end{longtable*}

\clearpage
\begin{longtable*}{rrrrrrrrrr}
\caption{All space group irreducible derivatives for graphene at $\order=3$ for $\supa[BZ]=\hat 1$, $\supa[K]$, $2\hat 1$, $3\hat 1$, and $2\supa[K]$.
         Horizontal lines separate irreducible derivatives associated with $\qsetp[\hat1]$, $\qsetp[\supa[K]]\cap\qsetp[\hat1]$, $\qsetp[2\hat1]\cap\qsetp[\hat1]$,
         $\qsetp[3\hat1]\cap(\qsetp[\supa[K]]\cup\qsetp[\hat1])$, and $\qsetp[\supa[2K]]\cap(\qsetp[2\hat1]\cup\qsetp[\supa[K]]\cup\qsetp[\hat1])$. 
         All results were computed using the LID approach.
         \label{si:table:irr_deriv_n3}
         }\\
\hline\hline
Derivative & Value & Derivative & Value & Derivative & Value & Derivative & Value & Derivative & Value \\
\hline\\[-0.9em]
\endfirsthead
\hline\hline
Derivative & Value & Derivative & Value & Derivative & Value & Derivative & Value & Derivative & Value \\
\hline\\[-0.9em]
\endhead
\hline\hline
\endfoot
  $d\indices*{*_{\Gamma}^{E_{2g}^{}}_{\Gamma}^{E_{2g}^{}}_{\Gamma}^{E_{2g}^{}}}$ & 
$425.751$ & 
  & 
  & 
  & 
  & 
  & 
  & 
  & 
   \\[0.2em]

 \hline\\[-0.9em]
                  $d\indices*{*_{K}^{A_{1}^{\prime}}_{K}^{A_{1}^{\prime}}_{K}^{A_{1}^{\prime}}}$ & 
 $440.064$ & 
          $d\indices*{*_{K}^{A_{1}^{\prime}}_{K}^{A_{2}^{\prime}}_{K}^{A_{2}^{\prime}}}$ & 
  $11.138$ & 
       $d\indices*{*_{K}^{A_{1}^{\prime}}_{K}^{E_{}^{\prime}}_{K}^{E_{}^{\prime}}}$ & 
  $289.379$ & 
$d\indices*{*_{K}^{A_{1}^{\prime}}_{K}^{E_{}^{\prime\prime}}_{K}^{E_{}^{\prime\prime}}}$ & 
$- 53.543$ & 
       $d\indices*{*_{K}^{E_{}^{\prime}}_{K}^{E_{}^{\prime}}_{K}^{E_{}^{\prime}}}$ & 
$- 239.640$ \\[0.4em]
         $d\indices*{*_{K}^{E_{}^{\prime\prime}}_{K}^{E_{}^{\prime\prime}}_{K}^{E_{}^{\prime}}}$ & 
  $24.732$ & 
$d\indices*{*_{\Gamma}^{B_{2g}^{}}_{\bar{K}}^{E_{}^{\prime}}_{K}^{E_{}^{\prime\prime}}}$ & 
$- 41.699$ & 
$d\indices*{*_{\Gamma}^{E_{2g}^{}}_{\bar{K}}^{A_{1}^{\prime}}_{K}^{E_{}^{\prime}}}$ & 
$- 455.829$ & 
     $d\indices*{*_{\Gamma}^{E_{2g}^{}}_{\bar{K}}^{A_{2}^{\prime}}_{K}^{E_{}^{\prime}}}$ & 
$- 50.356$ & 
$d\indices*{*_{\Gamma}^{E_{2g}^{}}_{\bar{K}}^{E_{}^{\prime}}_{K}^{E_{}^{\prime}}}$ & 
  $204.211$ \\[0.4em]
  $d\indices*{*_{\Gamma}^{E_{2g}^{}}_{\bar{K}}^{E_{}^{\prime\prime}}_{K}^{E_{}^{\prime\prime}}}$ & 
$- 32.416$ & 
                                                                                         & 
           & 
                                                                                    & 
            & 
                                                                                         & 
           & 
                                                                                   & 
             \\[0.2em]
 \hline\\
\end{longtable*}
